\documentclass[twocolumn]{pasj01}
\usepackage{natbib,bm,url}
\usepackage{color}
\begin{document} 
\Received{2021/03/11}
\Accepted{2021/04/17}
\Published{}

\title{SIRIUS project. I. Star formation models for star-by-star simulations of star clusters and galaxy formation}

\author{Yutaka \textsc{Hirai}\altaffilmark{1, 2}\thanks{JSPS Research Fellow}}
\altaffiltext{1}{Astronomical Institute, Tohoku University, 6-3 Aramaki, Aoba-ku, Sendai, Miyagi 980-8578, Japan}
\altaffiltext{2}{RIKEN Center for Computational Science, 7-1-26 Minatojima-minami-machi, Chuo-ku,
Kobe, Hyogo 650-0047, Japan}
\email{yutaka.hirai@astr.tohoku.ac.jp}

\author{Michiko S. \textsc{Fujii}\altaffilmark{3}}%
\altaffiltext{2}{Department of Astronomy, Graduate School of Science, The University of Tokyo, 7-3-1 Hongo, Bunkyo-ku, Tokyo 113-0033, Japan}
\email{fujii@astron.s.u-tokyo.ac.jp}

\author{Takayuki R. \textsc{Saitoh}\altaffilmark{4, 5}}
\altaffiltext{4}{Department of Planetology, Graduate School of Science, Kobe University, 1-1 Rokkodai-cho, Nada-ku, Kobe, Hyogo 657-8501, Japan}
\altaffiltext{5}{Earth-Life Science Institute, Tokyo Institute of Technology, 2-12-1 Ookayama, Meguro-ku, Tokyo 152-8551, Japan}
\email{saitoh@people.kobe-u.ac.jp}

\KeyWords{methods: numerical --- ISM: clouds --- open clusters and associations: general --- galaxies: formation --- galaxies: star clusters: general} 

\maketitle

\begin{abstract}
Most stars are formed as star clusters in galaxies, which then disperse into galactic disks. Upcoming exascale supercomputational facilities will enable performing simulations of galaxies and their formation by resolving individual stars (star-by-star simulations). This will substantially advance our understanding of star formation in galaxies, star cluster formation, and assembly histories of galaxies. In previous galaxy simulations, a simple stellar population approximation was used. It is, however, difficult to improve the mass resolution with this approximation. Therefore, a model for forming individual stars that can be used in simulations of galaxies must be established. In this first paper of a series of the SIRIUS (SImulations Resolving IndividUal Stars) project, we demonstrate a stochastic star formation model for star-by-star simulations. An assumed stellar initial mass function (IMF) is randomly assigned to newly formed stars in this model. We introduce a maximum search radius to assemble the mass from surrounding gas particles to form star particles. In this study, we perform a series of $N$-body/smoothed particle hydrodynamics simulations of star cluster formations from turbulent molecular clouds and ultra-faint dwarf galaxies as test cases. The IMF can be correctly sampled if a maximum search radius that is larger than the value estimated from the threshold density for star formation is adopted. In small clouds, the formation of massive stars is highly stochastic because of the small number of stars. We confirm that the star formation efficiency and threshold density do not strongly affect the results. We find that our model can naturally reproduce the relationship between the most massive stars and the total stellar mass of star clusters. Herein, we demonstrate that our models can be applied to simulations varying from star clusters to galaxies for a wide range of resolutions. 
\end{abstract}

\section{Introduction}
Our goal is to {gain a comprehensive picture of the formation and evolution of star clusters and galaxies}.  {Simulations that can resolve individual stars (hereafter, star-by-star simulations)} of galaxies are expected to provide a breakthrough in studies of galaxy formation. These simulations can assess the formation of star clusters and their evolution across the cosmic time \citep{2019ARA&A..57..227K}. Humanity's understanding of the assembly histories of galaxies will be considerably improved by star-by-star comparisons with the chemo-dynamical properties of stars obtained from the astrometric satellite Gaia \citep{2018A&A...616A...1G}, spectroscopic observations with astronomical telescopes, and simulations. Feedback from supernovae is independent of the models in these high-resolution simulations because the latter can detail the evolution of supernova remnants \citep[e.g.,][]{2012MNRAS.426..140D, 2018MNRAS.477.1578H,2019MNRAS.483.3363H}.

Galaxies consist of objects with a broad mass range. The largest objects in the Local Group are M31 and the Milky Way, as they have a total stellar mass of $\sim$ 10$^{11}\,M_{\odot}$. Conversely, recently discovered ultra-faint dwarf galaxies have a total stellar mass of only $\lesssim$ 10$^5\,M_{\odot}$ \citep[e.g.,][]{2019ARA&A..57..375S}. Globular clusters and open star clusters also have an extensive mass range, from 10$^2$ to 10$^7\,M_{\odot}$ \citep[e.g.,][]{2010ARA&A..48..431P}. These objects are formed within the broader events of galaxy formation. \citet{2009PASJ...61..481S} have shown that mergers of galaxies induce the formation of star clusters. \citet{2018MNRAS.474.4232K} have also identified that the mergers of high-redshift galaxies form globular cluster-like objects \citep[see also,][]{2020MNRAS.493.4315M}. To comprehensively understand their formation and relationship to the building blocks of galaxies, it is necessary to evaluate small star clusters and ultra-faint dwarf galaxies within the formation of more massive galaxies.

In the last decade, the mass resolution in simulations of galaxies has greatly improved \citep[and references therein]{2020NatRP...2...42V}. \citet{2014hpcn.conf...54B} performed an $N$-body simulation of a Milky Way mass galaxy using 10$^{11}$ particles. Current state-of-the-art hydrodynamic simulations of Milky Way mass galaxies have reached a mass resolution of less than $10^{4}\,M_{\odot}$ \citep[e.g.,][]{Auriga, FIRE2, 2020MNRAS.498.1765F, 2020arXiv200606008A,2020arXiv200811207A}. A considerably higher resolution is possible in simulations of dwarf galaxies \citep[e.g.,][]{2015ApJ...814...41H, 2017MNRAS.466.2474H, 2019ApJ...886L...3R, 2019MNRAS.490.4447W, 2019ApJ...879L..18L, 2020ApJ...891....2L, 2020MNRAS.491.1656A,2020arXiv201007311G, 2020arXiv201010533S}. Recently, \citet{2019MNRAS.483.3363H} performed a series of simulations of isolated dwarf galaxies with a mass resolution of $\sim$ 1\,$M_{\odot}$. They showed that properties of supernova driven winds converged within the simulations, with one gas-particle mass of less than 5\,$M_{\odot}$. \citet{2019MNRAS.482.1304E} also computed isolated dwarf galaxies with star-by-star yields of supernovae. They have shown that the outflows caused by supernova feedback have a larger metallicity than that of the interstellar medium (ISM). 

Exascale computational facilities will make it possible to perform star-by-star simulations up to the Milky Way mass galaxy scale within the next decade. These facilities are planned in different institutions. The supercomputer Fugaku in RIKEN {has commenced} operation. Oak Ridge National Laboratory plans to operate the exascale supercomputer Frontier in 2021. {China plans three projects for exascale computing.} By using such facilities, star-by-star simulations with 10$^{11}$ particles are expected to be possible if code with high-scalability can be developed.

{A sink particle approach has often been used in relatively small-scale simulations \citep[e.g.,][]{1995MNRAS.277..362B, 2003MNRAS.343..413B, 2004MNRAS.349..735B, 2004ApJ...611..399K, 2005MNRAS.356.1201B, 2005A&A...435..611J, 2005MNRAS.359..809C, 2010ApJ...713..269F,2013MNRAS.430.3261H, 2014MNRAS.445.4015B, 2016ApJ...823...28K, 2017MNRAS.466.1903G,2018PASJ...70S..54S,2018ApJ...859...68K,2020MNRAS.497.3830F}. 
\citet{2003MNRAS.343..413B} performed a series of $N$-body/smoothed particle hydrodynamics (SPH) star cluster formation simulations from turbulent molecular clouds. The mass of one gas particle of their simulation was 0.002\,$M_{\odot}$, and their results showed that the hierarchical fragmentation of turbulent molecular clouds helped form small star clusters. The merging of these objects formed the final star clusters. \citet{2019MNRAS.489.1880H} performed a series of radiation-magneto-hydrodynamic simulations of star clusters with a spatial resolution of 200 to 2000\,au. They found that the IMF's observed power-law slope could be reproduced if they assumed that 40\% of a star-forming gas clump was converted into the most massive stars, and others were distributed to the smaller mass stars.}

{For more massive clusters, `cluster particle' approach is used \citep{2012MNRAS.424..377D,2014MNRAS.442..694D,2017MNRAS.466..407S, 2018ApJ...859...68K,2018NatAs...2..725H,2019ApJ...887...62W, 2019MNRAS.489.1880H,2020MNRAS.497.3830F}. This is similar to the simple stellar population (SSP) approximation in galaxy simulations and cluster particles, containing a bunch of stars following a given mass function. The masses of cluster particles depend on the simulation scale and the resolution, but they are typically orders of ten to a hundred.}

{Sink particle approach is difficult to apply for simulations in a scale of galaxies. Simulations} cannot resolve the formation of the lowest-mass stars even if exascale supercomputers are used. At least 10$^{15}$ particles are required to resolve the Jeans mass of 0.1\,$M_{\odot}$ with 100 particles, corresponding to the formation region for the stars with the lowest mass in the simulations of Milky Way mass galaxies. There are no computational resources that can compute such simulations. {If we adopt the sink particle approach to galaxy formation simulations with a mass resolution of $\gtrsim$ 10\,$M_{\odot}$, a large amount of gas (typically $\gtrsim$ 500\,$M_{\odot}$) is locked up in a sink particle. In the case of poor resolution, not all gas particles are going to form stars, resulting in locking too much non-star-forming gas in a sink particle. \citet{2017ApJ...846..133K} have shown that the ISM properties such as vertical velocity dispersion and hot gas fraction do not converge in simulations with the grid resolution larger than 16\,pc because supernovae are clustered in the large sink particles. These consequences mean that we cannot apply the sink particle approach to galaxy formation simulations. }

In almost all galaxy formation simulations, the SSP approximation, which considers a stellar component as a cluster of stars sharing the same age and metallicity with a given stellar initial mass function (IMF),
is used to model star formation.
With this approximation, once a gas particle satisfies a set of conditions imitating real star forming regions,
(a part of) its mass converts into a collision-less star particle by following Schmidt's law \citep{1959ApJ...129..243S}:
\begin{equation}\label{eq:schmidt}
\frac{d\rho_*}{dt}\,=\,
-\frac{d\rho_{\mathrm{gas}}}{dt}\,\equiv\,c_{*} \frac{\rho_{\mathrm{gas}}}{t_{\mathrm{dyn}}},
\end{equation}
where $\rho_*$ and $\rho_{\mathrm{gas}}$ exhibit stellar and gas densities, respectively, $t_{\mathrm{dyn}}$ is the local dynamical time, and $c_*$ is a dimensionless parameter ranging from $0.01-0.1$ \citep{1992ApJ...391..502K}.
Although there are some variations in modeling star formation and its conditions \citep[e.g.,][]{1993MNRAS.265..271N,1994A&A...281L..97S,
2006MNRAS.373.1074S, 2008PASJ...60..667S, 2011MNRAS.417..950H}, the star formation models used in galaxy formation simulations are essentially the same.

These star formation models cannot be easily applied to star-by-star simulations because of the breakdown of the SSP approximation. \citet{2016A&A...588A..21R} have shown that the SSP approximation cannot correctly sample the IMF in the simulations of mass resolution of $\lesssim$ $10^{3}M_{\odot}$. {Since we have not yet understood what conditions derive IMFs, we need to rely on the stochastic sampling of IMFs \citep[e.g.,][]{2014MNRAS.438.1305H,2017MNRAS.471.2151H,2019MNRAS.483.3363H}. This case requires star particles with different masses. If the mass of a star particle is larger than the mass of a gas particle, the masses from surrounding gas particles must be accounted for. Moreover, the IMF should be sampled correctly in sufficiently large systems. However, there are no systematic studies for modeling star-by-star simulations. It is necessary to confirm that the model can correctly sample the IMFs and compute properties of star clusters and galaxies.}

This study is the first in a series of the SIRIUS (SImulations Resolving IndividUal Stars) project, which seeks to understand the chemo-dynamical evolution of star clusters and galaxies with high-resolution simulations. {This project consists of three code papers: star formation model (this study), ASURA+BRIDGE code \citep{FujiiASURABRIDGE}, and feedback \citep{2021arXiv210302829F} and subsequent science papers.} The purpose of this study is to construct a star formation model for star-by-star simulations and clarify the effects of parameters of the model in the simulations. In this study, we perform a series of star cluster formation simulations from turbulent molecular clouds to test the newly developed models. We study the condition to sample the IMF in this model and the influence of the parameters in star-by-star simulations. 

This paper is organized as follows. The next section describes the implementation of the star formation models for star-by-star simulations. {Section \ref{sec:simulations} shows the code and initial conditions.} Section \ref{sec:results} systematically studies the effects of parameters on the sampling of the assumed IMF in star clusters. Section \ref{sec:UFD} discusses the formation of an ultra-faint dwarf galaxy (UFD). In section \ref{sec:discussion}, we discuss the applicability of our model. Section \ref{sec:conclusions} summarizes the main results. 

\section{Star formation scheme}\label{sec:method}
\subsection{Procedure for star formation}\label{sec:sf}
{The models of star formation developed for simulations with SSP approximation \citep[e.g.,][]{1992ApJ...391..502K,2003MNRAS.345..429O,2006MNRAS.373.1074S,2008PASJ...60..667S} must be modified for star-by-star simulations of star clusters and galaxies. The intended mass resolution is $m_{\rm{gas}}\,\leq\,m_{\rm{max,\,IMF}}$ in this study. We also assumed that the stellar mass from the adopted IMF was assigned to each star particle.}

In this section, we describe the procedure for the proposed star formation model. The first step was to check the conditions for star formation. Gas particles became eligible for star formation when they were conversing ($\nabla\cdot\bm{v}<0$) in a higher density region than the threshold density ($n_{\rm{th}}$) and in a colder region than the threshold temperature ($T_{\rm{th}}$). Gas particles that formed stars during the given time interval $\Delta t$ were stochastically selected by the following equation:
\begin{equation}
p\,=\,\frac{m_{\mathrm{gas}}}{\langle{m_{*}}\rangle}
\left\{1-\exp\left(-c_{*}
\frac{\Delta t}{t_{\mathrm{dyn}}}\right)\right\},\label{eq:sf_probability}
\end{equation}
where $m_{\rm{gas}}$, $\langle{m_{*}}\rangle$, $c_{*}$, and $t_{\mathrm{dyn}}$ were the mass of one gas particle, the average value of stellar mass in the assumed IMF, the dimensionless star formation efficiency, and the local dynamical time, respectively.\footnote{
{We can rewrite equation (\ref{eq:sf_probability}) as follows if $c_* \Delta t / t_{\mathrm{dyn}} \ll 1$:
\begin{equation}
p\,=\,
\left\{1-\exp\left(-c_{*}\frac{m_{\mathrm{gas}}}{\langle{m_{*}}\rangle}
\frac{\Delta t}{t_{\mathrm{dyn}}}\right)\right\}.\label{eq:sf_probability_new}
\end{equation}
This expression is harmonized to star formation's probabilistic manner because its range is from 0 to unity. In our numerical experiments, both expressions provided almost the same results, 
which indicated that the condition $c_* \Delta t / t_{\mathrm{dyn}} \ll 1$ was satisfied in our simulations.  In this study, we used equation \ref{eq:sf_probability}.}
}
We set the dimensionless star formation efficiency as 0.02 and 0.1 following its observed constraints per free-fall time \cite[and references therein]{2019ARA&A..57..227K}.

We introduced the coefficient, $m_{\mathrm{gas}}/\langle{m_{*}}\rangle$. This expression was adopted to scale the number of newly formed stars to the mass resolution. Note that the denominator of the coefficient was not the mass of the star particle ($m_{\rm{*}}$), which was adopted in models of \citet{2003MNRAS.345..429O, 2006MNRAS.373.1074S}; instead, it was the average stellar mass computed from the adopted IMF ($\langle{m_{*}}\rangle$). This difference came from the mass of each star particle. The masses of each star particle in the SSP approximation were almost constant, whereas the masses were different among star particles in our case.

The second step was to compare the value of $p$ to the random number ($\mathcal{R}$) from 0 to 1. If $p\,>\,\mathcal{R}$, we assigned stellar mass ($m_{*}$) from the minimum ($m_{\rm{min,\, IMF}}$) to the maximum ($m_{\rm{max,\,IMF}}$) mass of the IMF using the Chemical Evolution Library  \citep[CELib,][]{2017AJ....153...85S}. The detailed implementation of CELib is described in section \ref{sec:IMF}.

\begin{figure}[htbp]
\begin{center}
\includegraphics[width = 8cm]{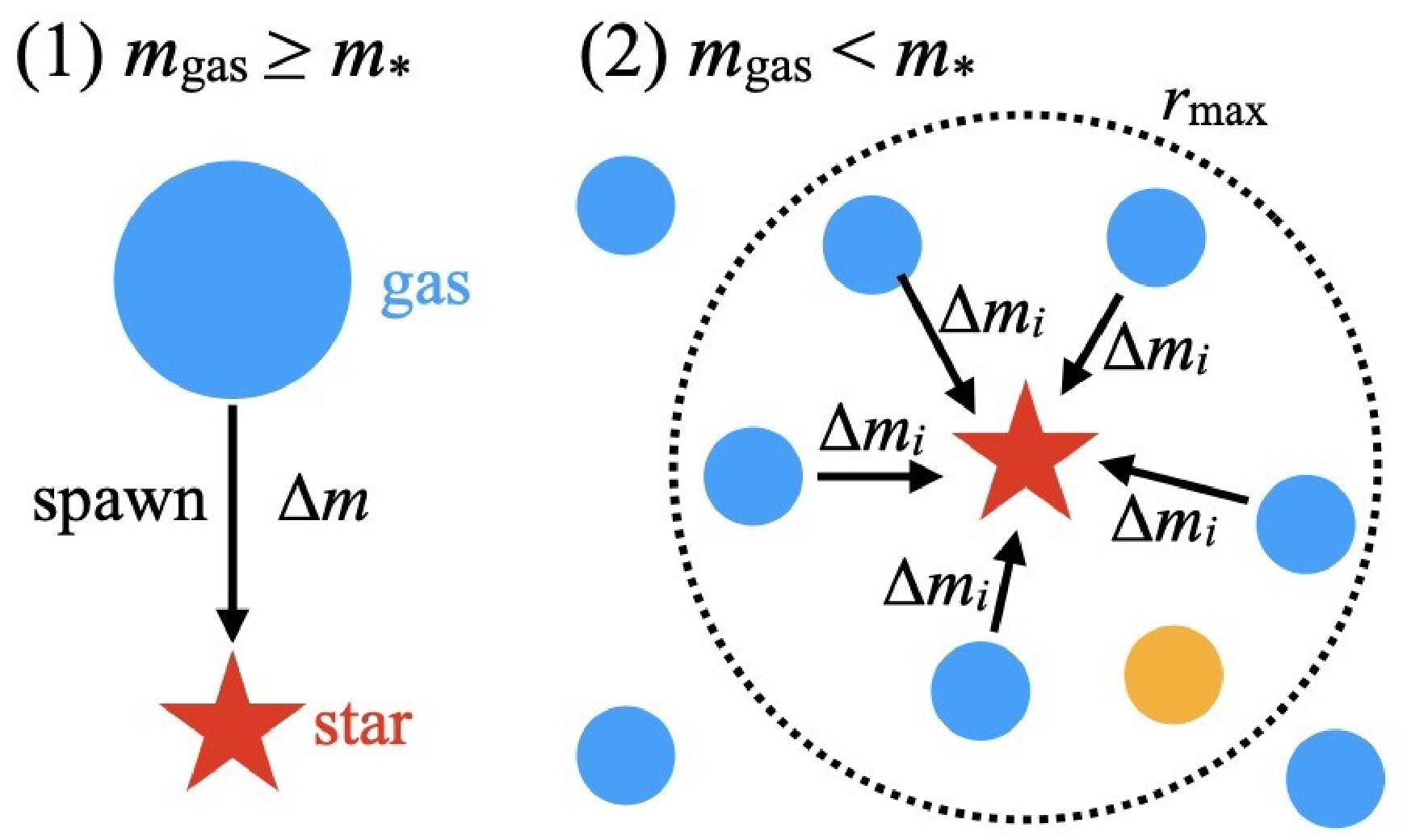} 
\end{center}
\caption{Illustration of our star formation scheme. Gas particles that satisfied all conditions of star formation were eliminated from the source of mass to form a new star particle (yellow plot). The black-dashed circle represents the maximum search radius ($r_{\rm{max}}$, color online).}\label{fig:SF_scheme}
\end{figure}

In the final step, gas particles that satisfied all conditions of star formation were converted into star particles through one of two methods depending on whether the mass of a gas particle ($m_{\rm{gas}}$) was larger than $m_{*}$ or not. Figure \ref{fig:SF_scheme} shows the schematic for converting gas particles into a star particle. In the case of $m_{\rm{gas}} \geq m_{*}$, a gas particle was spawned to form a star particle (case 1 in figure \ref{fig:SF_scheme}). The mass of the gas particle was reduced by $\Delta{m}\,=\,m_{\rm{gas}}\,-\,m_{*}$. {Positions and velocities of the parent gas particles are inherited to the newly formed stars. When $m_{\rm{gas}}\,>\,\langle{m_{*}}\rangle$, mass resolution of the simulation was not enough to explicitly sample all mass ranges of stars in the IMFs. Several ways were proposed to assign properties of stars to star particles \citep{2013MNRAS.435.1701C, 2017MNRAS.471.2151H,2019MNRAS.483.3363H,2020MNRAS.492....8A}. Since all models in this study had $m_{\rm{min,\, IMF}}$ $\leq$ 1.5\,$M_{\odot}$, lifetimes of un-sampled stars were much longer than the total time of the performed simulations. We did not put the effect of stellar evolution of low mass stars in this study.}

If $m_{\rm{gas}} < m_{*}$, a star particle was generated by assembling masses of surrounding gas particles (case 2 in figure \ref{fig:SF_scheme}). In this case, we first determined the region that contained a mass of 5--10\,$m_{*}$ ($\equiv{m_{\rm{inc}}}$). We then adopted the maximum search radius ($r_{\rm{max}}$) to gather gas mass to form stars and prevent an assemblage of this mass in an unrealistically large region. If the required radius to assemble the gas mass in the region containing the mass of $m_{\rm{inc}}$ exceeded $r_{\rm{max}}$, we forced the search radius to be $r_{\rm{max}}$.

We estimated the required search radius ($r_{\rm{th}}$) to form a star with a mass $m_{*}$ by the following equation: 
\begin{equation}\label{eq:radius}
r_{\rm{th}}\,=\,\left(\frac{3m_{*}}{4\pi{n_{\rm{H}}m_{\rm{H}}}}\right)^{\frac{1}{3}},
\end{equation}
where $n_{\rm{H}}$ and $m_{\rm{H}}$ were the number density and mass of hydrogen, respectively. To form a star with a mass of 100\,$M_{\odot}$ in a region of $n_{\rm{H}}\,=\,1.2\,\times\,10^5\>$cm$^{-3}$ and $n_{\rm{H}}\,=\,1.2\,\times\,10^7\>$cm$^{-3}$, the maximum search radius must be larger than 0.21$\>$pc and 0.04$\>$pc, respectively. If the gas mass within $r_{\rm{max}}$ was less than 2\,$m_{*}$, we randomly re-assign the smaller stellar mass for a star particle. If there were gas particles that satisfied all conditions of star formation, they were excluded from the mass transfer. {Gas particles with temperature higher than 10$^3$ K were also excluded to prevent assembling of mass from hot gas.} 

We then converted the gas particle at the center of this region into a star particle. {Positions ($\bm{x}_{*,\rm{new}}$) and velocities ($\bm{v}_{*,\rm{new}}$) of a newly star particles were re-assigned to ensure the momentum conservation. If the positions and velocities of the parent gas particle were $\bm{x}_{\rm{gas, p}}$ and $\bm{v}_{\rm{gas, p}}$, $\bm{x}_{*,\rm{new}}$ and $\bm{v}_{*,\rm{new}}$ were reassigned as follows:}
\begin{equation}\label{eq:starpos}
    \bm{x}_{*,\rm{new}}\,=\,\frac{m_{*}\bm{x}_{\rm{gas, p}}+f\Sigma^{N_{\rm{list}}}_{i=0}m_{{\rm{gas},}i}\bm{x}_{{\rm{gas, ,}}i}}{m_{*}+f\Sigma^{N_{\rm{list}}}_{i=0}m_{{\rm{gas},}i}},
\end{equation}

\begin{equation}\label{eq:starvel}
    \bm{v}_{*,\rm{new}}\,=\,\frac{m_{*}\bm{v}_{\rm{gas, p}}+f\Sigma^{N_{\rm{list}}}_{i=0}m_{{\rm{gas},}i}\bm{v}_{{\rm{gas, ,}}i}}{m_{*}+f\Sigma^{N_{\rm{list}}}_{i=0}m_{{\rm{gas},}i}},
\end{equation}
{where $\bm{x}_{{\rm{gas}},i}$, $\bm{v}_{{\rm{gas}},i}$, $m_{{\rm{gas},}i}$ were positions, velocities, and masses of assembled gas particles, respectively. The amount of reduced gas mass was set as $f\equiv{(m_{*}-m_{{\rm{gas}})}/m_{\rm{inc}}}$.}

Next, we reduced the masses of surrounding gas particles. The mass of gas particles after mass conversion was $(1-f)\,m_{\rm{gas}}$ to satisfy the mass conservation. {After the star formation, gas particles with ten times less massive than the average gas particle mass were merged to the nearest neighbor gas particle. When two particles were merged, the positions and velocities of merged particles were the centers of mass of two particles to ensure momentum conservation. This scheme was introduced to prevent gas particles that have significantly less massive than gas particles in its neighborhood.}

\subsection{Sampling of the IMF by CELib}\label{sec:IMF}
\begin{figure}[htbp]
 \begin{center}
  \includegraphics[width=8cm]{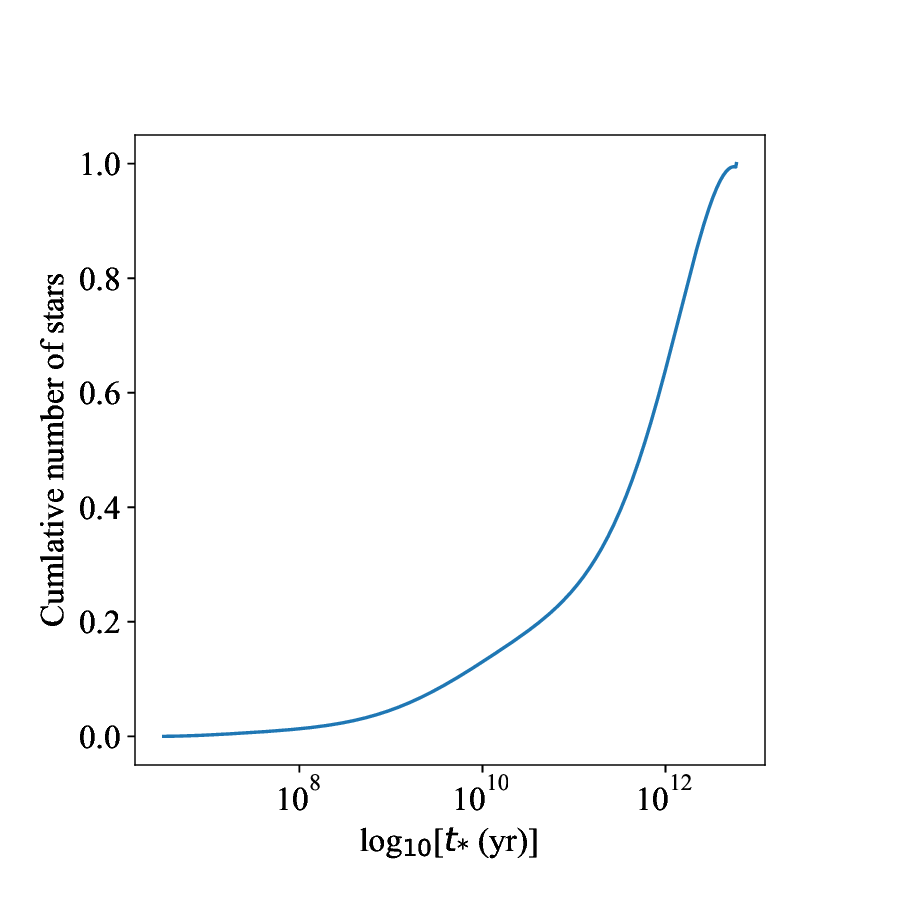}
 \end{center}
\caption{Cumulative number normalized from 0 to 1 of stars with solar metallicity as a function of the lifetime ($t_{*}$).}\label{fig:lifetime}
\end{figure}

We updated CELib to assign stellar masses to newly formed star particles. CELib first converted lifetimes to a table numbered from 0 to 1 as weighted by the IMF (figure \ref{fig:lifetime}). It then assigned the stellar mass and the lifetime to the new star particle.

Stellar lifetimes from lifetime tables were interpolated via polynomial function fitted using the least-squares fitting method \citep{2017AJ....153...85S}. We used the stellar lifetime table from \citet{1998A&A...334..505P} for 0.6 to 100\,$M_{\odot}$. This lifetime represented the sum of the timescales of hydrogen- and helium-burning computed in the Padova stellar evolution library \citep{1993A&AS..100..647B,1994A&AS..105...29F,1994A&AS..104..365F}. For lifetimes ($t_{*}$) of stars with less than 0.6\,$M_{\odot}$, we extrapolated the lifetime table using the stellar mass-luminosity ($L$) relationship ($L\,\propto\,m_{*}^4$, $t_{*}\,\propto\,m_{*}L^{-1}$, i.e., $t_{*}\,\propto\,m_{*}^{-3}$). Lifetimes of stars from 150 to 300\,$M_{\odot}$ were taken from \citet{2002A&A...382...28S}. 

We adopted the IMF of \citet{2003PASP..115..763C} in the models, except for M40ks and M40kt. Recent observations of low mass stars suggested that the IMF in the low mass range shows a flatter index of the power law than $-$1.35 \citep{2001MNRAS.322..231K}. The functional form of the adopted Chabrier IMF is as follows:
\begin{eqnarray}
\frac{\mathrm{d}N}{\mathrm{d}\log_{10}m} \propto \left\{
\begin{array}{ll}
\exp\left[-\frac{\left\{\log_{10}(\frac{m}{0.079})\right\}^2}{2(0.69)^2}\right], & \\
\hspace{2.0cm}(0.1\,M_{\odot} \leq m \leq 1\,M_{\odot}),&\\
m^{-1.3}, &\\
\hspace{2.0cm}(1\,M_{\odot} < m \leq 100\,M_{\odot}),&\\
\end{array}\right.
\end{eqnarray}
where $m$ and $N$ were the mass and the number of stars, respectively. We also adopted the classical Salpeter IMF \citep{1955ApJ...121..161S} from 0.1 to 100\,$M_{\odot}$:
\begin{equation}
\frac{\mathrm{d}N}{\mathrm{d}\log_{10}m} \propto m^{-1.35},
\end{equation}
and the IMF suggested from the simulation of the Population III star formation \citep[hereafter, we refer to this IMF as Susa IMF;][]{2014ApJ...792...32S}. We defined the Susa IMF from 0.7 to 300\,$M_{\odot}$ as follows:
\begin{equation}
\frac{\mathrm{d}N}{\mathrm{d}\log_{10}m} \propto \exp\left\{-\frac{\left[\log_{10}\left(\frac{m}{22.0}\right)\right]^2}{2(0.5)^2}\right\}.
\end{equation}
This IMF was characterized by a {large fraction of massive stars}.

\begin{figure*}[htbp]
 \begin{center}
   \includegraphics[width=5cm]{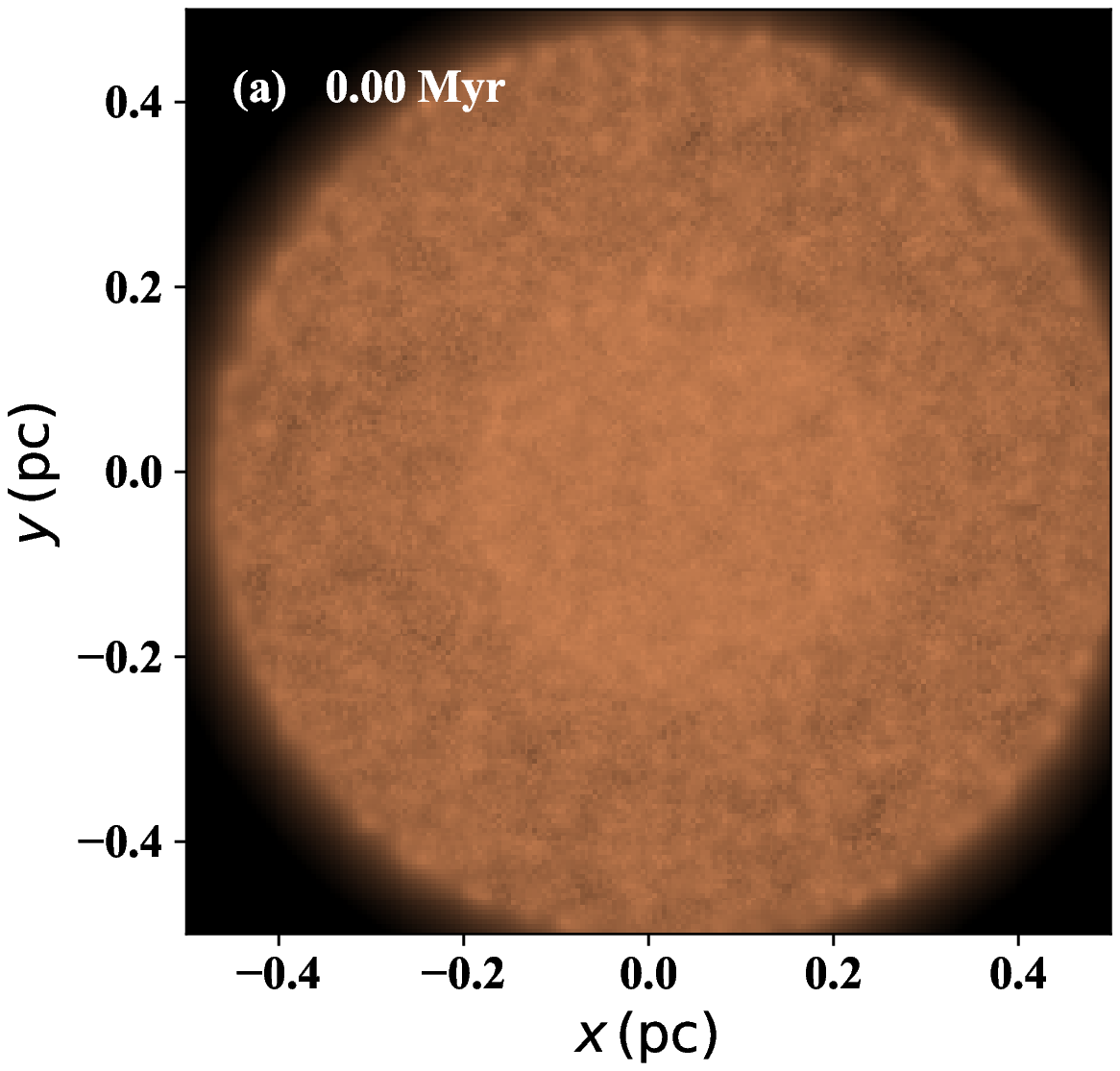}
   \includegraphics[width=5cm]{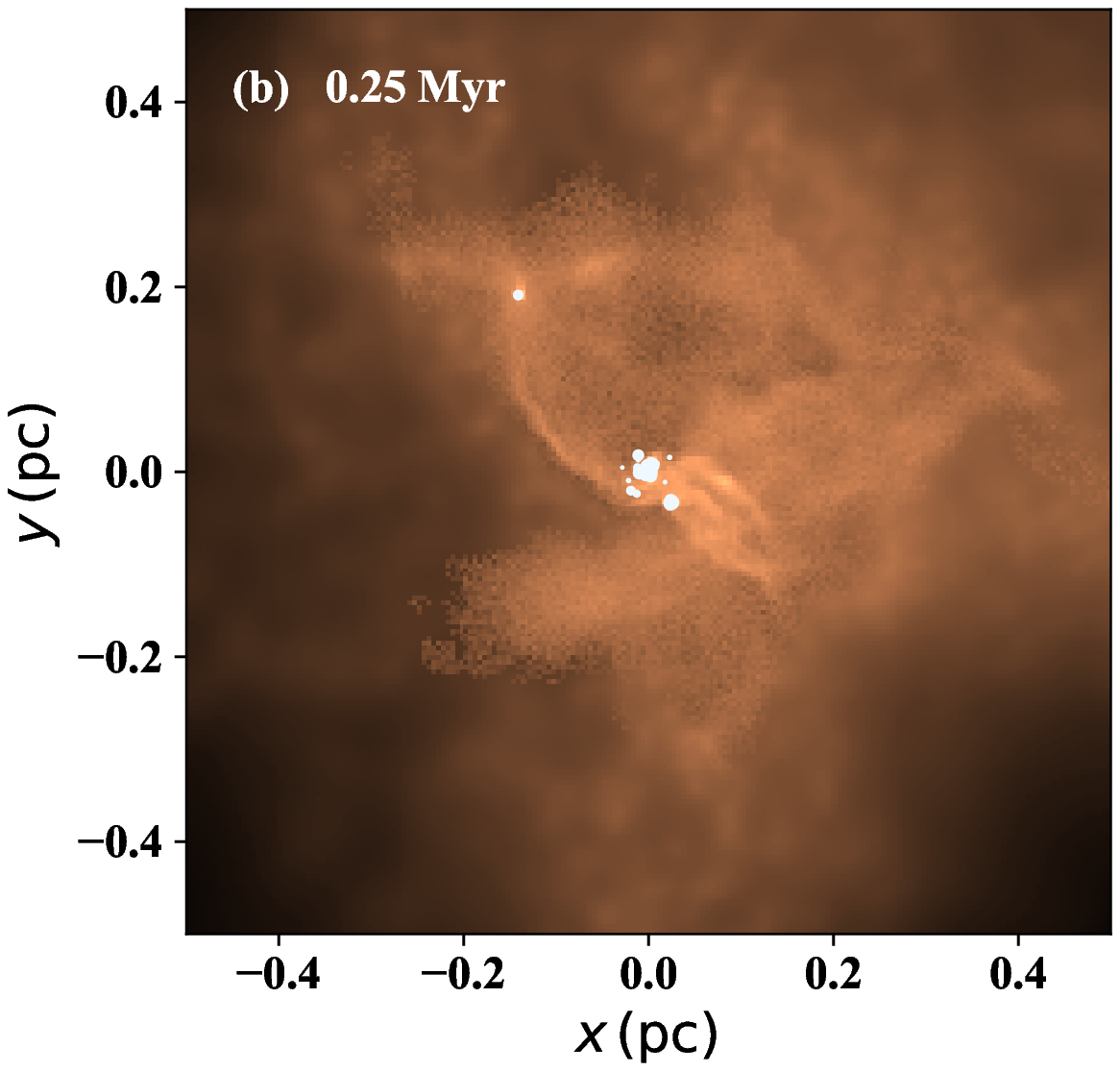}
   \includegraphics[width=5cm]{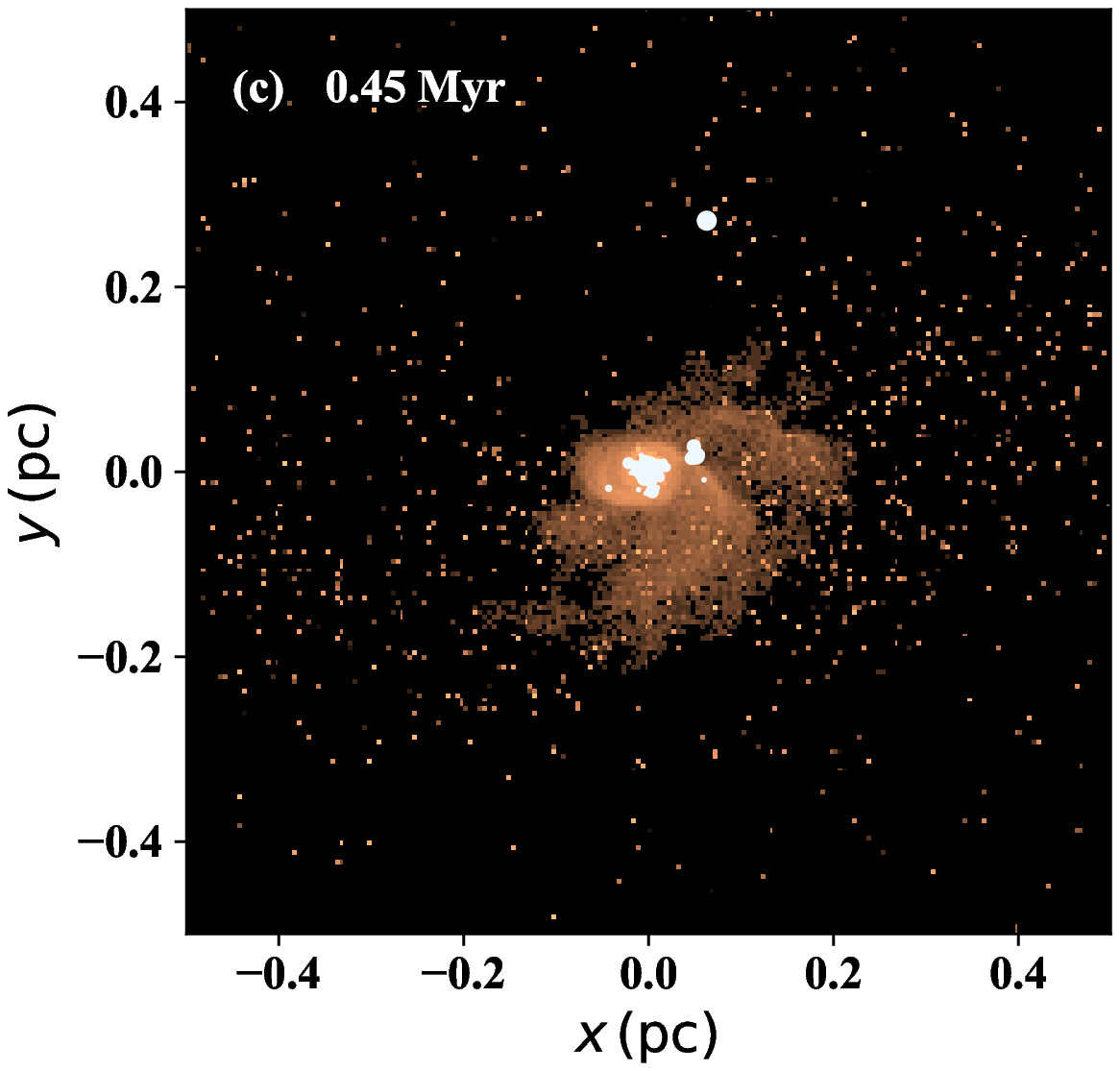}
 \end{center}
\caption{The IMFs generated by CELib (red-solid curve). The blue-dashed curve represents (a) Chabrier IMF, (b) Salpeter IMF, and (c) Susa IMF (color online).}\label{fig:IMF}
\end{figure*}

Figure \ref{fig:IMF} shows the mass function {stochastically} generated by CELib using 10$^6$ samples. We generated 10$^6$ random numbers from 0 to 1 by CELib and then assigned stellar mass following the lifetime table weighted by the IMF. Note that the IMF was generated not by $N$-body/SPH simulations but by only using CELib. As shown in this figure, CELib can sample the assumed IMFs. Deviations seen in higher mass stars were caused by the small number of samples. 

\section{Simulations}\label{sec:simulations}
\subsection{Code}\label{sec:code}
We adopted an $N$-body/SPH simulation code, ASURA \citep{2008PASJ...60..667S,2009PASJ...61..481S}. Gravity was computed using the tree method \citep{1986Natur.324..446B} with the tolerance parameter $\theta\,=\,0.5$. Because star clusters were collisional systems, these objects should be treated as direct $N$-body \citep[e.g.,][]{2007PASJ...59.1095F, 2013MNRAS.430.1599H, 2019ApJ...887...62W} to force the accuracy to a sufficiently high level. We did not, however, apply a direct $N$-body computation in this study; this was to avoid increasing uncertain parameters. Our next paper will show the implementation of the direct $N$-body and its effects on the properties of star clusters \citep{FujiiASURABRIDGE}.\\
\indent Hydrodynamics in ASURA were computed with the density-independent SPH method \citep{2013ApJ...768...44S}. 
{An artificial viscosity was introduced to handle shocks. We adopted a variable viscosity model proposed by \citet{1997JCoPh.136...41M} with a slight modification of \citet{2009NewAR..53...78R}.}
We implemented a timestep limiter for the supernova shocked region \citep{2009ApJ...697L..99S} and a Fully Asynchronous Split Time-Integrator \citep[FAST,][]{2010PASJ...62..301S} to accelerate the computation. We applied the cooling and heating function from 10 to 10$^9$ K, as generated by Cloudy ver.13.05 \citep{1998PASP..110..761F, 2013RMxAA..49..137F, 2017RMxAA..53..385F}. {At the end of the lifetime, stars with 13 to 40\,$M_{\odot}$ exploded as core-collapse supernovae. We assumed that each supernova distributes thermal energy of 10$^{51}$ erg and elements to surrounding gas particles with the yield of \citet{2013ARA&A..51..457N}. We did not implement other types of nucleosynthetic events.} Metal diffusion was also computed based on the turbulence-motivated model \citep{2010MNRAS.407.1581S, 2017AJ....153...85S,2017ApJ...838L..23H}. {We set the scaling factor for metal diffusion as 0.01 following \citet{2017ApJ...838L..23H}. For cosmological zoom-in simulations, we also implemented the effects of ultra-violet background field \citep{2012ApJ...746..125H} and self-shielding \citep{2013MNRAS.430.2427R}.}
\subsection{Initial conditions}\label{sec:IC}
\subsubsection{Star clusters from turbulent molecular clouds}
We adopted the turbulence-molecular cloud model \citep{2003MNRAS.343..413B, 2015PASJ...67...59F, 2015MNRAS.449..726F, 2016ApJ...817....4F} using initial conditions generated by the Astronomical Multipurpose Software Environment \citep[AMUSE,][]{2009NewA...14..369P, 2013CoPhC.184..456P, 2013A&A...557A..84P, 2018araa.book.....P}. The supersonic turbulent motion of gas was modeled as a divergence-free random Gaussian velocity field proportional to the wave number of velocity perturbations with a power law index of $-$4 \citep{2001ApJ...546..980O}. The initial total gas masses of clouds were set as 1\,$\times$\,10$^3$\,$M_{\odot}$ (models B03) and 4\,$\times$\,10$^4$\,$M_{\odot}$ (models M40k). The mass resolution of model B03h was the same as the model adopted in \citet{2003MNRAS.343..413B}. The radii of the clouds were 0.5$\>$pc and 10.0$\>$pc for models B03 and M40k, respectively. The free-fall times for the clouds of B03 and M40k were 0.19 and 0.83 Myr, respectively. {We set the gravitational softening length and the threshold density for star formation following the Jeans length, assuming that the Jeans mass was resolved by 100 gas particles and the temperature was 20 K. For models with mass resolutions of 0.001, 0.002, 0.01, 0.02, and 0.1\,$M_{\odot}$, the softening lengths corresponded to 3.2 \,$\times\,10^2$, 6.5\,$\times\,10^2$, 3.2\,$\times\,10^3$, 6.5\,$\times\,10^3$, and 3.2\,$\times\,10^4$\,au, respectively.} We set the metallicity as $Z$\,=\,0.013 \citep{2009ARA&A..47..481A}. Table \ref{tab:model} lists the models adopted in this study.
\begin{table*}[htbp]
  \tbl{List of models.\footnotemark[$*$]}{
  \begin{tabular}{llllllllll}
  \hline              
   Name & $M_{\rm{tot}}$ ($M_{\odot}$)&$r_{\rm{t}}$ (pc) &$N_{\rm g}$ & $m_{\rm g}$ $(M_{\odot})$ & $\epsilon_{\rm g}$ (au) & $n_{\rm th}$ (cm$^{-3}$) & $r_{\rm{max}}$ (pc) &$c_{*}$&IMFs\\ 
  \hline
      B03vh& $1\,\times\,10^3$& 0.5 & $1\,\times\,10^6$ & 0.001 & $3.2\,\times\,10^2$&$1.2\,\times\,10^9$ &0.2&0.02&\citet{2003PASP..115..763C}\\
      B03h& $1\,\times\,10^3$& 0.5 & $5\,\times\,10^5$ & 0.002 & $6.5\,\times\,10^2$&$3.0\,\times\,10^8$ &0.2&0.02&\citet{2003PASP..115..763C}\\
      B03m & $1\,\times\,10^3$& 0.5 & $1\,\times\,10^5$ & 0.01 & $3.2\,\times\,10^3$&$1.2\,\times\,10^7$&0.2&0.02&\citet{2003PASP..115..763C}\\       
      B03l & $1\,\times\,10^3$& 0.5 & $5\,\times\,10^4$ & 0.02 & $6.5\,\times\,10^3$&$3.0\,\times\,10^6$ &0.2 &0.02&\citet{2003PASP..115..763C}\\
      B03vl & $1\,\times\,10^3$& 0.5 & $1\,\times\,10^4$ & 0.1 & $3.2\,\times\,10^4$&$1.2\,\times\,10^5$ &0.2 &0.02&\citet{2003PASP..115..763C}\\
      B03e & $1\,\times\,10^3$& 0.5 & $1\,\times\,10^5$ & 0.01 & $3.2\,\times\,10^4$&$1.2\,\times\,10^5$&0.2&0.02&\citet{2003PASP..115..763C}\\
      B03n & $1\,\times\,10^3$& 0.5 & $1\,\times\, 10^5$ & 0.01 & $3.2\,\times\,10^3$&$1.2\,\times\,10^5$&0.2&0.02&\citet{2003PASP..115..763C}\\        
      B03c & $1\,\times\,10^3$& 0.5 & $1\,\times\, 10^5$ & 0.01 & $3.2\,\times\,10^3$&$1.2\,\times\,10^7$&0.2&0.2&\citet{2003PASP..115..763C}\\ 
      B03sr & $1\,\times\,10^3$& 0.5 & $1\,\times\,10^5$ & 0.01 & $3.2\,\times\,10^3$&$1.2\,\times\,10^7$&0.02&0.02&\citet{2003PASP..115..763C}\\
      B03lr & $1\,\times\,10^3$& 0.5 & $1\,\times\,10^5$ & 0.01 & $3.2\,\times\, 10^3$&$1.2\,\times\,10^7$&2.0&0.02&\citet{2003PASP..115..763C}\\      
      M40km & $4\,\times\,10^4$& 10.0& $4\,\times\,10^6$ & 0.01 & $3.2\,\times\, 10^3$&$1.2\,\times\,10^7$&0.2&0.02&\citet{2003PASP..115..763C}\\   
      M40kl & $4\,\times\,10^4$& 10.0& $4\,\times\,10^5$ & 0.1 & $3.2\,\times\,10^4$&$1.2\,\times\,10^5$&0.2&0.02&\citet{2003PASP..115..763C}\\
      M40ke & $4\,\times\,10^4$& 10.0& $4\,\times\,10^6$ & 0.01 & $3.2\,\times\,10^4$&$1.2\,\times\,10^5$&0.2&0.02&\citet{2003PASP..115..763C}\\
      M40ksr & $4\,\times\,10^4$& 10.0& $4\,\times\,10^5$ & 0.1 & $3.2\,\times\,10^4$&$1.2\,\times\,10^5$&0.02&0.02&\citet{2003PASP..115..763C}\\
      M40klr & $4\,\times\,10^4$& 10.0& $4\,\times\,10^5$ & 0.1 & $3.2\,\times\,10^4$&$1.2\,\times\,10^5$&2.0&0.02&\citet{2003PASP..115..763C}\\
      M40ks & $4\,\times\,10^4$& 10.0& $4\,\times\,10^5$ & 0.1 & $3.2\,\times\,10^4$&$1.2\,\times\,10^5$&0.2&0.02&\citet{1955ApJ...121..161S}\\ M40kt & $4\,\times\,10^4$& 10.0& $4\,\times\,10^5$ & 0.1 & $3.2\,\times\,10^4$&$1.2\,\times\,10^5$&0.5&0.02&\citet{2014ApJ...792...32S}\\
      UFD& $1\,\times\,10^7$& 770 & $2\,\times\,10^7$ & 18.4 & $1.9\,\times\,10^6$&$1.0\,\times\,10^2$ &2.2&0.02&\citet{2003PASP..115..763C}\\
      \hline
      \end{tabular}}\label{tab:model}
\begin{tabnote}
\footnotemark[$*$] From left to right, the columns show the names of the models, the initial total gas mass ($M_{\rm{tot}}$), the initial truncation radius ($r_{\rm{t}}$), the initial number of gas particles ($N_{\rm{g}}$), the mass of one gas particle ($m_{\rm{g}}$), the gravitational softening length ($\epsilon_{\rm{g}}$), the threshold density for star formation ($n_{\rm{th}}$), the maximum search radius ($r_{\rm{max}}$), the dimensionless star formation efficiency parameter ($c_{*}$), and adopted IMFs.
\end{tabnote}
\end{table*}
\subsubsection{Cosmological zoom-in simulations of ultra-faint dwarf galaxies}
{We performed a cosmological zoom-in simulation of an UFD. The initial condition was generated by \textsc{music} \citep{2011MNRAS.415.2101H}. A pre-flight $N$-body simulation was performed using \textsc{Gadget-2} \citep{2005MNRAS.364.1105S} with a box size of (4 Mpc$\>h^{-1}$)$^3$. A halo for a zoom-in simulation was selected using AMIGA halo finder \citep{2004MNRAS.351..399G, 2009ApJS..182..608K}. We selected a halo with a mass of 4.9\,$\times\,10^8\,M_{\odot}$ at the redshift $z$\,=\,3. We confirmed that there were no halos over 10$^{12}\,M_{\odot}$ within 1$\>$Mpc from the zoomed-in halo.} 

{A zoom-in hydrodynamic simulation was performed using \textsc{ASURA}. The total number of particles in the zoomed-in initial condition was 3.8$\times\,10^7$. Masses of a dark matter particle and a gas particle were 99.0\,$M_{\odot}$ and 18.5\,$M_{\odot}$, respectively. To avoid making stars with masses largely different from gas particles, we set the Chabrier IMF from 1.5\,$M_{\odot}$ to 100\,$M_{\odot}$. This assumption let average star particle mass 4.7\,$M_{\odot}$, corresponding to 0.25 times the mass of gas particles. This value led to a good balance between time resolution and CPU time \citep{2003MNRAS.339..289S, 2012A&A...538A..82R}. We set the gravitational softening lengths of dark matter particles as 19.8$\>$pc following \citet{FIRE2}. The softening lengths of gas and star particles were set to contain 100 gas particles in the threshold density. In this simulation, this value corresponded to 9.1$\>$pc.}

\section{Star cluster formation}\label{sec:results}
\subsection{Formation of star clusters from the turbulent molecular clouds with 10$^3\,M_{\odot}$}\label{sec:B03}

\begin{figure*}[htbp]
 \begin{center}
  \includegraphics[width=5cm]{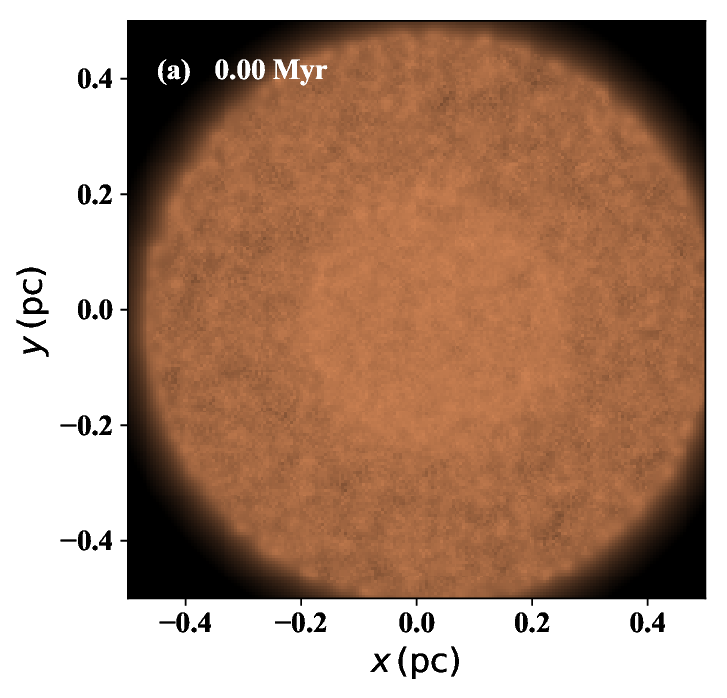}
  \includegraphics[width=5cm]{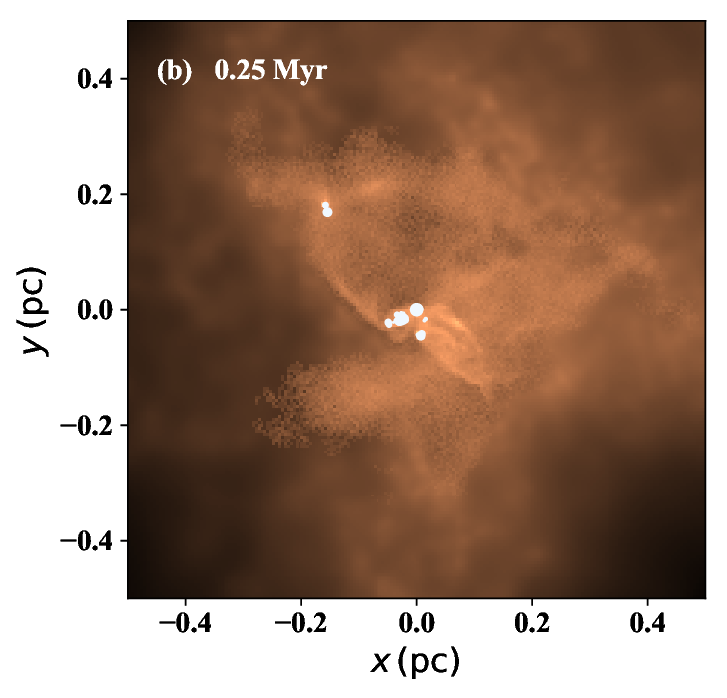}
  \includegraphics[width=5cm]{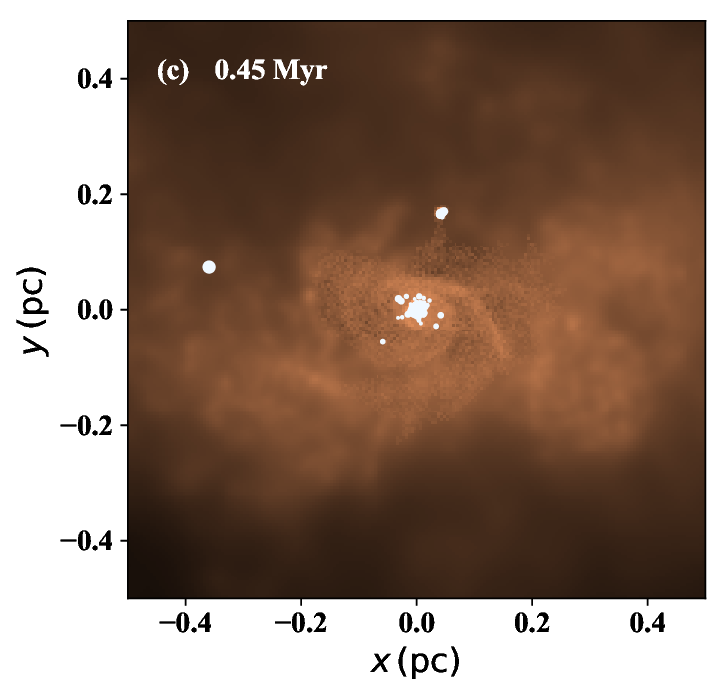}
 \end{center}
\caption{Gas column density and stellar distributions in model B03h. Panels (a), (b), and (c) represent snapshots at 0.00, 0.25, and 0.45 Myr from the beginning of the simulation, respectively. The color gradation shows the logarithm of the column density from 10$^{21}\>$cm$^{-2}$ (black) to 10$^{25}\>$cm$^{-2}$ (yellow). White dots depict stars. Larger sizes of dots represent more massive stars (color online).}\label{fig:snapshot}
\end{figure*}

The turbulent motion of the gas induced the evolution of the molecular cloud. Figure \ref{fig:snapshot} shows snapshots of gas and stellar density distributions in model B03h. We placed the cloud with the random Gaussian velocity field as described in section \ref{sec:IC} (figure \ref{fig:snapshot}a). Supersonic turbulent motion in this model produced shocks leading to the formation of filamentary structures. Shocks also expelled the kinetic energy of the gas. This effect locally reduced the support of turbulence. Once high density regions in the filament self-gravitate, they collapse and can form stars (figure \ref{fig:snapshot}b). After the star formation begins, gases are consumed (figure \ref{fig:snapshot}c).

\begin{figure}[htbp]
 \begin{center}
  \includegraphics[width=8cm]{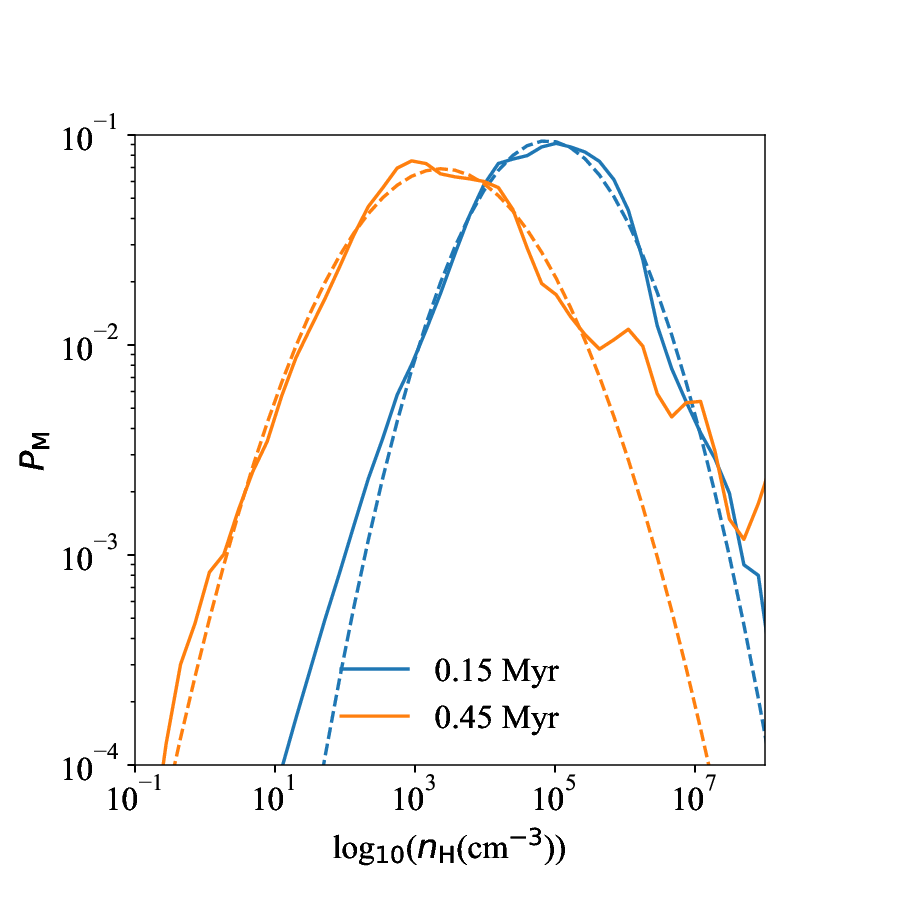}
 \end{center}
\caption{The density distribution of gas in model B03h at 0.15\,Myr (the orange-solid curve) and 0.45\,Myr (the blue-solid curve) from the beginning of the simulation. Orange- and blue-dashed curves represent the best-fit log-normal curves at 0.15\,Myr and 0.45\,Myr, respectively (color online).}\label{fig:DensityPDF}
\end{figure}

{Figure \ref{fig:DensityPDF} shows gas density probability distribution function (PDF) in model B03h. This gas density PDF can be well fitted with a log-normal PDF. The mean density ($\langle$log$_{10}n_{\rm{H}}\rangle$) and the standard deviation ($\sigma$) are $\langle$log$_{10}n_{\rm{H}}\rangle$\,=\,4.89, $\sigma$\,=\,0.86 at 0.15 Myr and $\langle$log$_{10}n_{\rm{H}}\rangle$\,=\,3.38, $\sigma$\,=\,1.06 at 0.45 Myr. Decrease of the mean density is owing to the star formation. At 0.45 Myr, the cloud develops the power-law tail in a high-density region (see section \ref{sec:comparison} for the discussion).} 

\subsubsection{Mass resolution}\label{sec:B03_resolution}
\begin{figure}[htbp]
 \begin{center}
  \includegraphics[width=8cm]{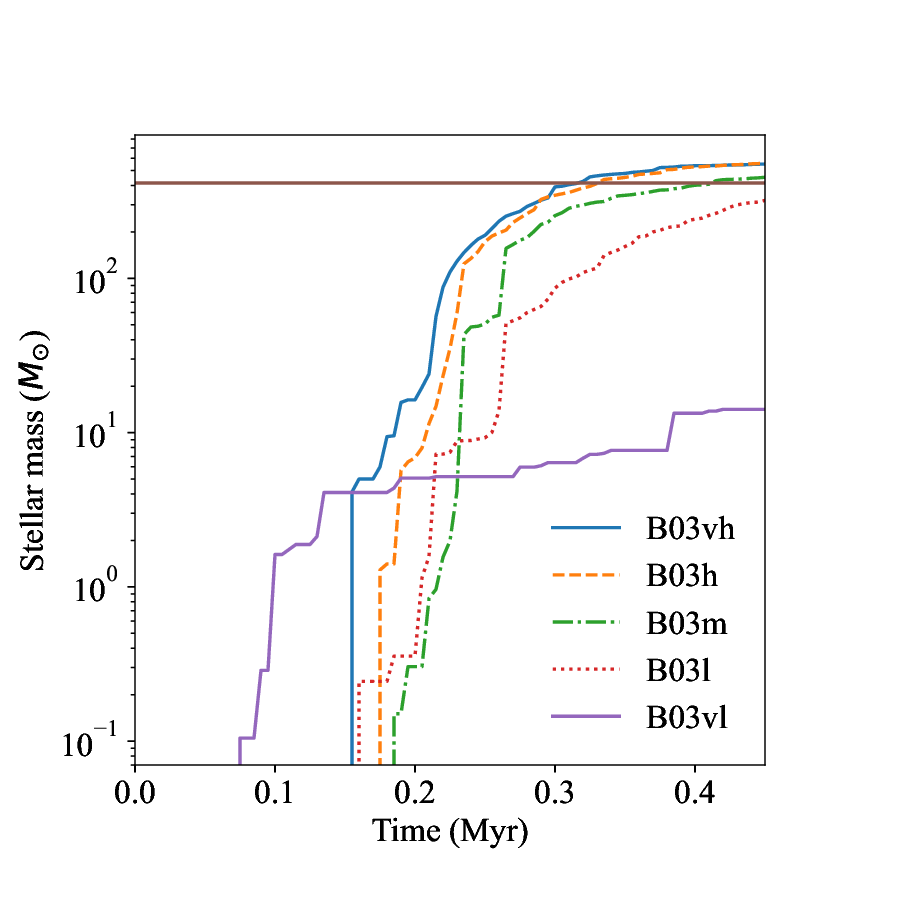}
 \end{center}
\caption{Total stellar mass as a function of time from the beginning of the simulation (the solid blue curve: B03vh, the dashed orange curve: B03h, the dash-dotted green curve: B03m, the dotted red curve: B03l, and the solid purple curve: B03vl; color online). The dotted brown line depicts the total stellar mass (415\,$M_{\odot}$) at 0.45 Myr in the model computed in \citet{2003MNRAS.343..413B}.}\label{fig:B03massvstime_resolution}
\end{figure}

Figure \ref{fig:B03massvstime_resolution} shows the total stellar mass as a function of time in models of different mass resolutions. The total stellar masses in models B03vh, B03h, B03m, and B03l are 551, 556, 452, and 320\,$M_{\odot}$, respectively, at 0.45 Myr. These masses are similar to those in \citet[see section \ref{sec:comparison}]{2003MNRAS.343..413B}. As shown in section \ref{sec:IC}, the Jeans lengths in the star forming regions in these models are less than $6.5\,\times\,10^3$\,au (= $3.2\,\times\,10^{-2}$$\>$pc). These sizes are significantly smaller than that of the system (= $5.0\,\times\,10^{-1}$$\>$pc), allowing them to detail the star formation process. They can therefore convert over 30\% of gas to stars. However, model B03vl only has a total stellar mass of 14\,$M_{\odot}$ at 0.45 Myr. Because the Jeans length $3.2\,\times\,10^4$\,au (= $1.6\,\times\,10^{-1}$$\>$pc) is a similar size as that of the system, this model cannot emulate the star formation process correctly.

\subsubsection{Gravitational softening length and threshold density for star formation}
\begin{figure*}[htbp]
 \begin{center}
  \includegraphics[width=6cm]{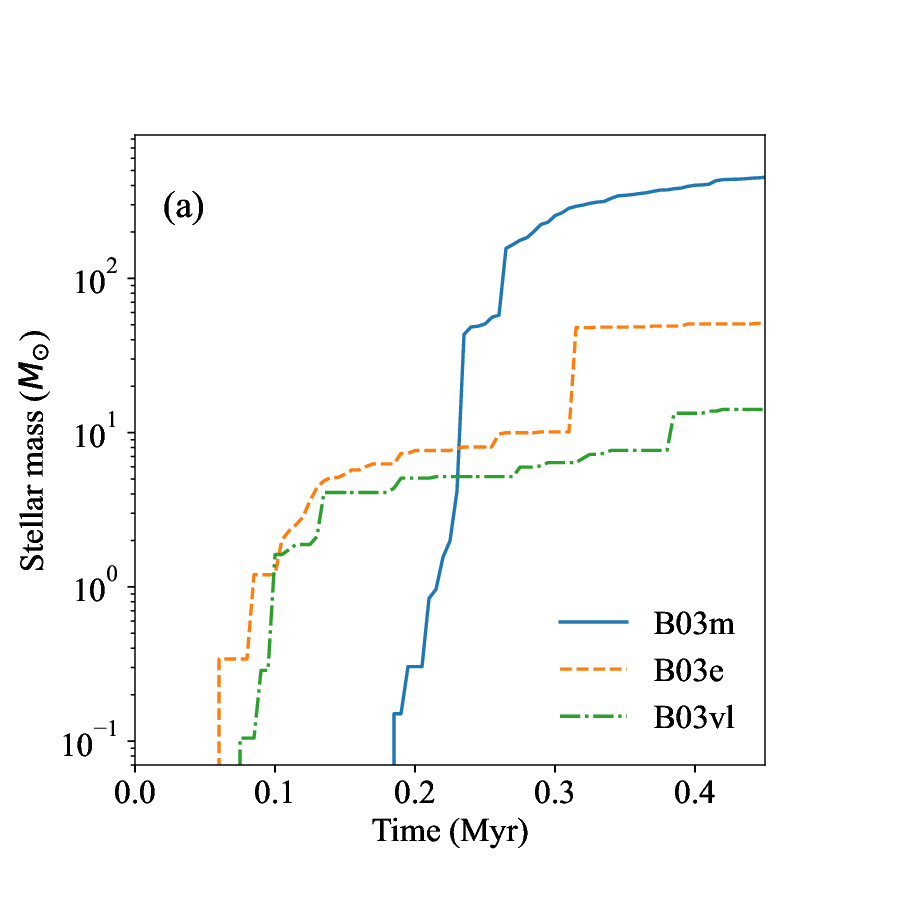}
  \hspace{-1cm}
  \includegraphics[width=6cm]{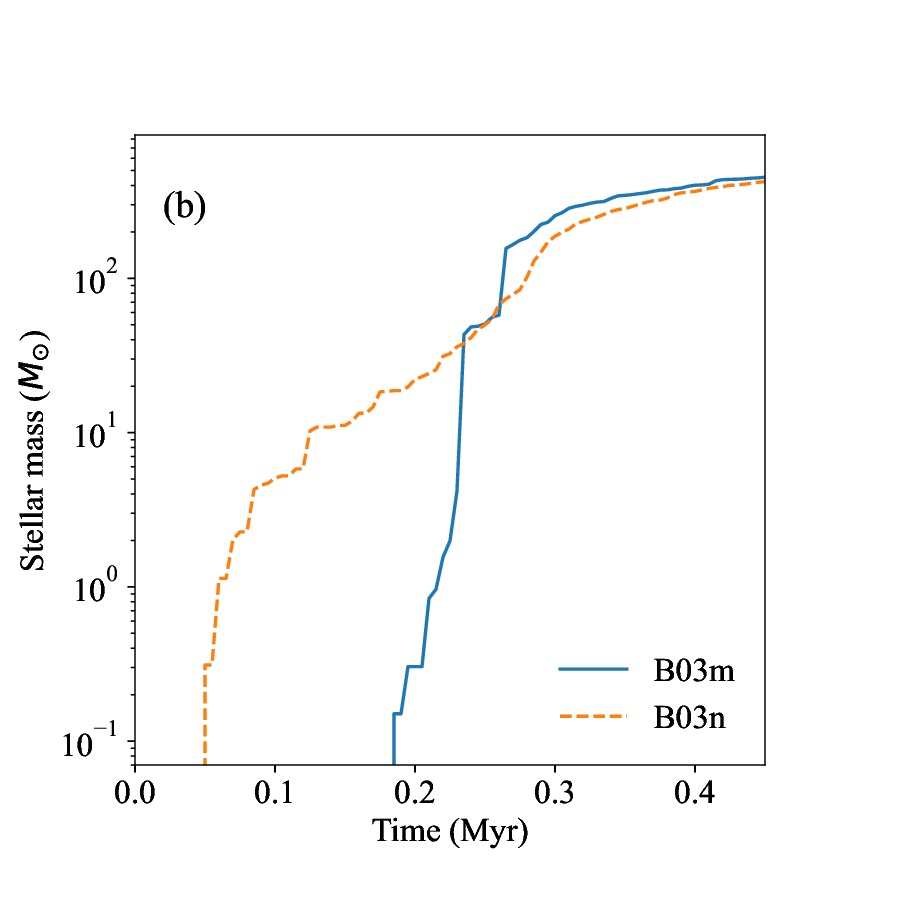}
  \hspace{-1cm}  
  \includegraphics[width=6cm]{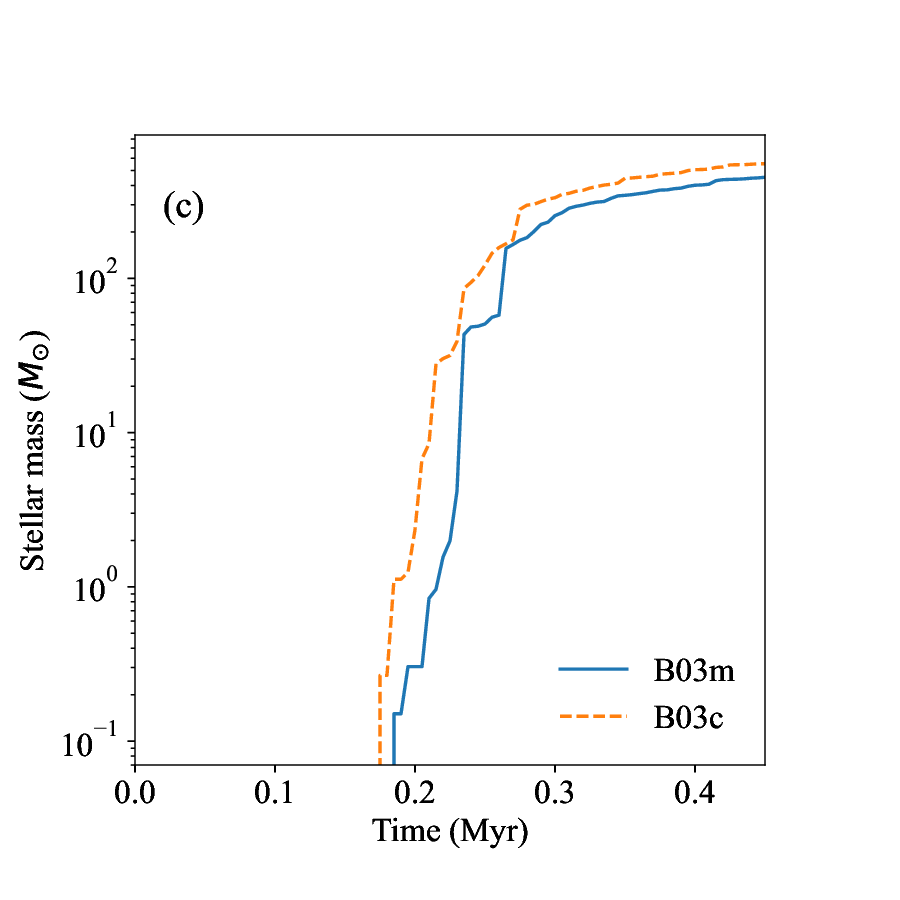}
 \end{center}
\caption{Similar to figure \ref{fig:B03massvstime_resolution}, but for models (a) B03m (solid blue curve), B03e (dashed orange curve), B03vl (dash-dotted green curve), (b) B03m (solid blue curve), B03n (dashed orange curve), (c) B03m (solid blue curve), and B03c (dashed orange curve, color online).}\label{fig:B03massvstime_softening}
\end{figure*}

We have varied the gravitational softening length and threshold density for star formation to clarify the parameters that show the largest impacts on the total stellar mass of the system. Both parameters are related to the resolution of the simulation. The effect of gravitational softening is evident in figure \ref{fig:B03massvstime_softening}a. Model B03e adopts the same parameters of gravitational softening length and threshold density for star formation as model B03vl ($\epsilon_{\rm{g}}$\,=\,3.2\,$\times$\,10$^4$\,au and $n_{\rm{th}}$\,=\,1.2\,$\times$\,$10^5\>$cm$^{-3}$) but has the same mass resolution as model B03m ($m_{\rm{g}}$\,=\,0.01\,$M_{\odot}$). As shown in this figure, the star formation in model B03e is suppressed at the similar level as model B03vl. This result means that resolving the self-gravitating clumps is one of the keys for forming stars in these models. Models B03vl and B03e do not have a sufficiently high resolution to resolve star-forming clumps in the molecular cloud.

\begin{figure}[htbp]
 \begin{center}
  \includegraphics[width=8cm]{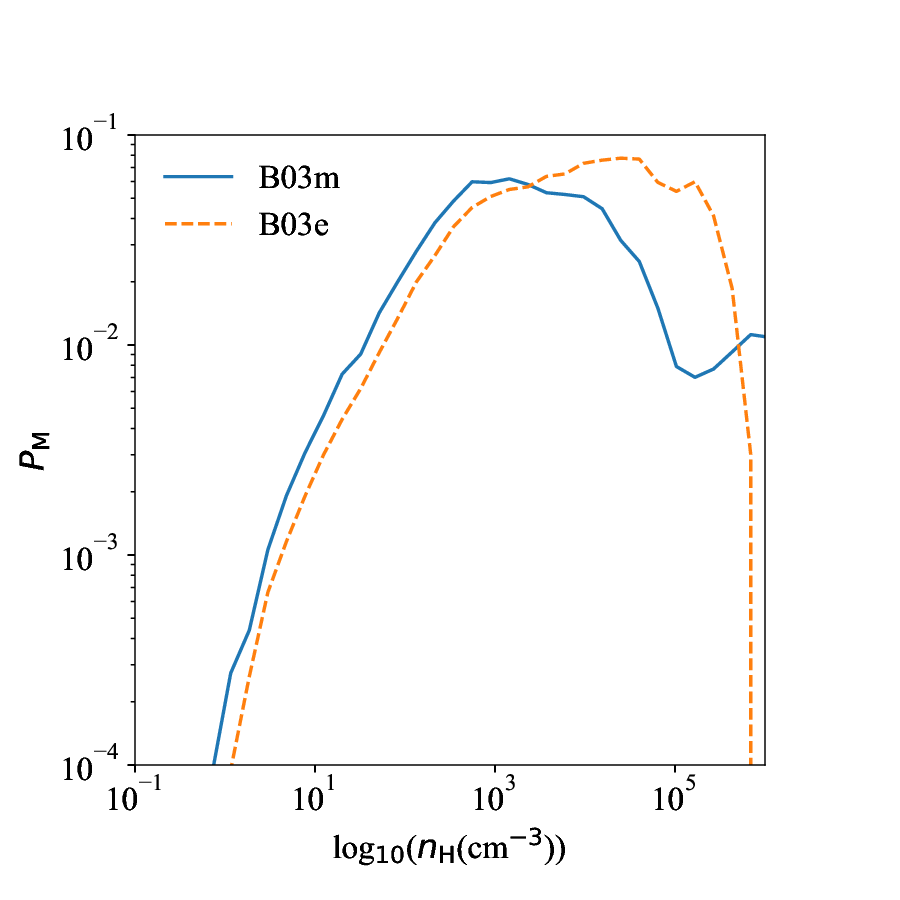}
 \end{center}
\caption{Similar to figure \ref{fig:DensityPDF}, but for models B03m (blue-solid curve) and B03e (orange-dashed curve, color online).}\label{fig:DensityPDFB03me}
\end{figure}

{Lack of sufficient spatial resolution prevents resolving high-density gas. Figure \ref{fig:DensityPDFB03me} shows gas density PDFs in models B03m and B03e. As shown in this figure, model B03e lacks gas with $\gtrsim\,10^{6}$\,cm$^{-3}$, which cannot be resolved in this model. These results mean} that if the gravitational softening length is excessively large compared to the size of the system, the star formation cannot be computed correctly.

The value of the threshold density for star formation does not substantially affect the total stellar mass of the system. {Figure \ref{fig:B03massvstime_softening}b compares the time evolution of the stellar masses in models B03n ($n_{\rm{th}}$\,=\,$1.2\,\times\,10^5\>$cm$^{-3}$) and B03m ($n_{\rm{th}}$\,=\,$1.2\,\times\,10^7\>$cm$^{-3}$).} In model B03n, the stellar mass at 0.45 Myr is 423\,$M_{\odot}$. This value is similar to that of model B03m (452\,$M_{\odot}$). Regardless of the value of $n_{\rm{th}}$, most of the stars are formed in a region with a significantly higher density than the threshold for star formation. The average star formation density at 0.25 Myr in models B03m and B03n are $1.2\,\times\,10^{8}\>$cm$^{-3}$ and $4.9\,\times\,10^{7}\>$cm$^{-3}$, respectively. The total stellar masses at 0.45 Myr in both models are similar, owing to the limited initial total gas mass of the cloud (= 1000\,$M_{\odot}$). 

Model B03n begins star formation earlier than that of model B03m (figure \ref{fig:B03massvstime_softening}b). The conditions of star formation are more easily satisfied in models with a lower value of $n_{\rm{th}}$. These results suggest that the choice of $n_{\rm{th}}$ does not considerably affect the formation of stars in this model. 

\subsubsection{Star formation efficiency}\label{sec:B03_SFE}
The value of $c_{*}$ does not strongly affect the total stellar mass of the system. Model B03c ($c_{*}$\,=\,0.1) has a stellar mass of 554\,$M_{\odot}$. This mass is only 1.2 times larger stellar mass than that of model B03m (452\,$M_{\odot}$, see figure \ref{fig:B03massvstime_softening}c), whereas model B03c has a value of $c_{*}$ that is five times larger than that of model B03m. The threshold density for star formation ($n_{\rm{th}}\,=\,1.2\,\times\,10^7\>$cm$^{-3}$) is 2 dex larger than the mean density of the cloud ($\sim\,10^5\>$cm$^{-3}$). Because we have adopted the Schmidt law (equation \ref{eq:schmidt}), the timescale of the star formation is short enough to diminish the effect of $c_{*}$ in this case. Therefore, the value of $c_{*}$ does not substantially affect the total stellar mass of the system. 

\subsubsection{Maximum search radius}\label{sec:B03_rmax}
\begin{figure*}[htbp]
 \begin{center}
  \includegraphics[width=6cm]{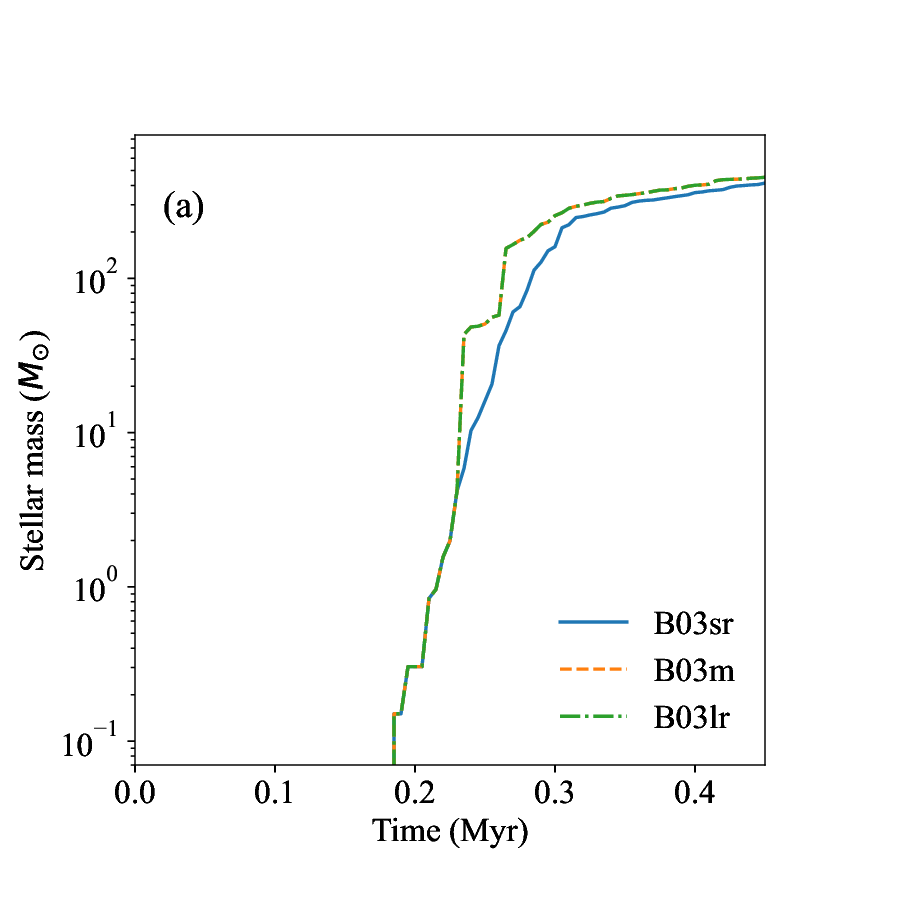}
  \hspace{-1cm}
  \includegraphics[width=6cm]{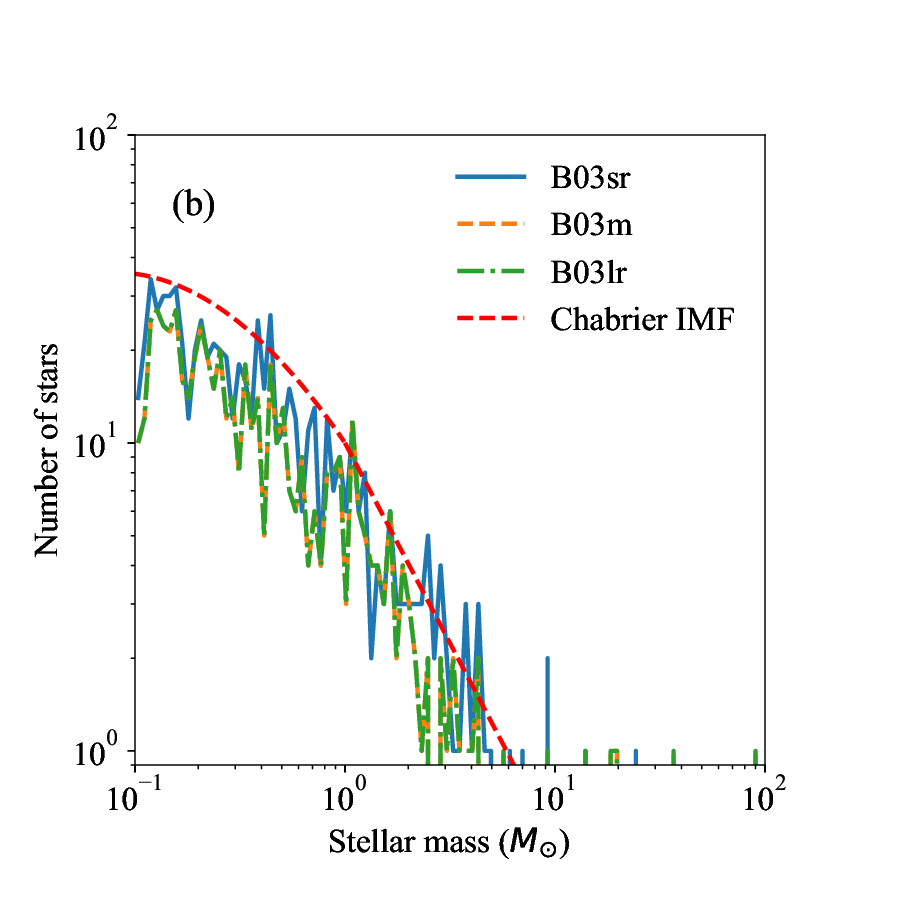}
  \hspace{-1cm}  
  \includegraphics[width=6cm]{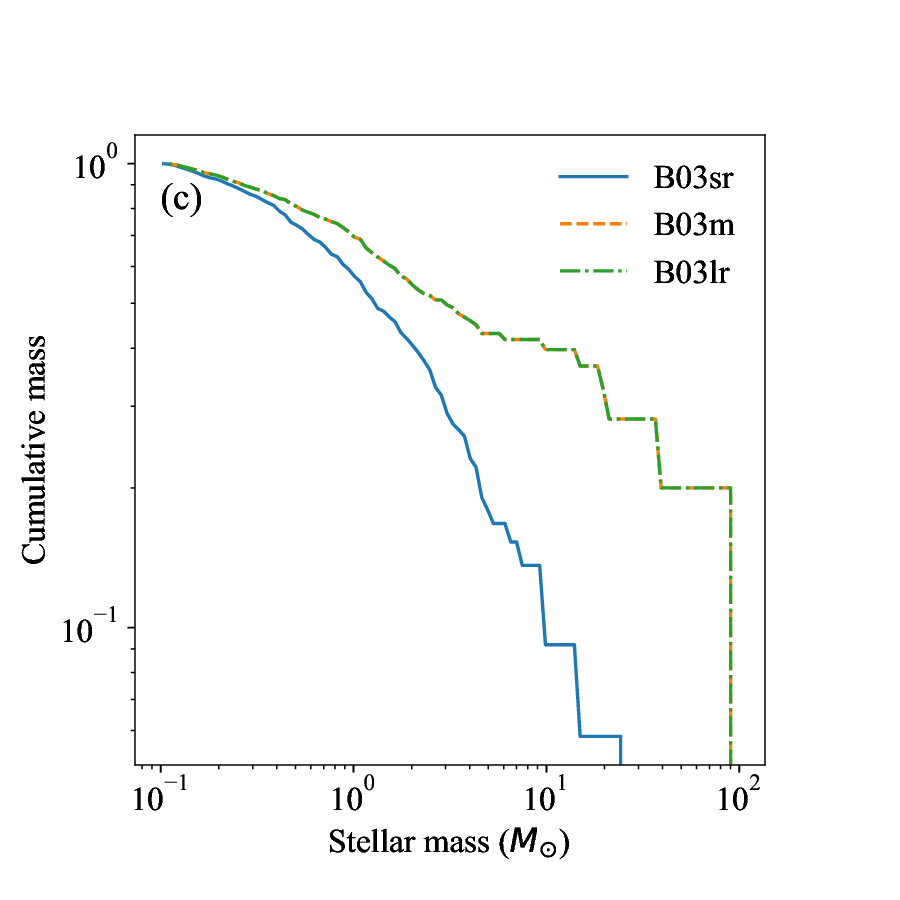}
 \end{center}
\caption{The effect of the maximum search radius in models B03sr ($r_{\rm{max}}$\,=\,0.02$\>$pc, solid blue curves), B03m ($r_{\rm{max}}$\,=\,0.2$\>$pc, dashed orange curves), and B03lr ($r_{\rm{max}}$\,=\,2.0$\>$pc, dash-dotted green curves). (a) Total stellar mass as a function of time, (b) the number of stars as a function of the mass of each star particle, and (c) cumulative mass functions. Panels (b) and (c) are plotted at 0.45 Myr from the beginning of the simulation. The red dotted curve in panel (b) represents the Chabrier IMF (color online).}\label{fig:B03m_rmax}
\end{figure*}

The maximum search radius does not affect the time evolution of the total stellar mass. Figure \ref{fig:B03m_rmax}a shows the total stellar mass as a function of time in models B03sr ($r_{\rm{max}}$\,=\,0.02$\>$pc), B03m ($r_{\rm{max}}$\,=\,0.2$\>$pc), and B03lr ($r_{\rm{max}}$\,=\,2.0$\>$pc). As shown in this figure, there is no significant difference among the models. 

Regardless of the value of the maximum search radius, the assumed IMF for stars with a mass lower than 10\,$M_{\odot}$ is reproduced in model B03. Figure \ref{fig:B03m_rmax}b represents the stellar mass functions computed in models B03sr, B03m, and B03lr. For stars larger than 1\,$M_{\odot}$, the mass function follows the power-law distribution with an index of $-$1.3. The flattening shape in lower mass stars is caused by adopting the log-normal distribution in the Chabrier IMF.

Figure \ref{fig:B03m_rmax}c denotes the cumulative mass of stars as a function of the stellar mass. According to this figure, the cumulative masses of models B03m and B03lr are the same. These models adopt the same parameters, except for the maximum search radius. The required search radius to form a star with 100\,$M_{\odot}$ in these models is $r_{\rm{th}}$\,=\,0.04$\>$pc (equation \ref{eq:radius}). Models B03m and B03lr have larger values of $r_{\rm{max}}$ than $r_{\rm{th}}$. This result implies that the maximum search radius does not affect the stellar mass function as far as its size is larger than that expected from the density of the star-forming region.

The formation of the most massive stars appears to be suppressed in model B03sr. However, it is difficult to evaluate the effects of the maximum search radius on the formation of massive stars in this model. Because the mass of the cloud has only 1000\,$M_{\odot}$ {and the typical conversion fraction of gas to stars is $\sim$ 0.4}, a few massive stars are formed in these models. In section \ref{M40k}, we discuss the effects of using a maximum search radius with more massive clouds. 

\subsubsection{Run-to-run variations}\label{sec:B03_rand}
\begin{figure}[htbp]
 \begin{center}
  \includegraphics[width=8cm]{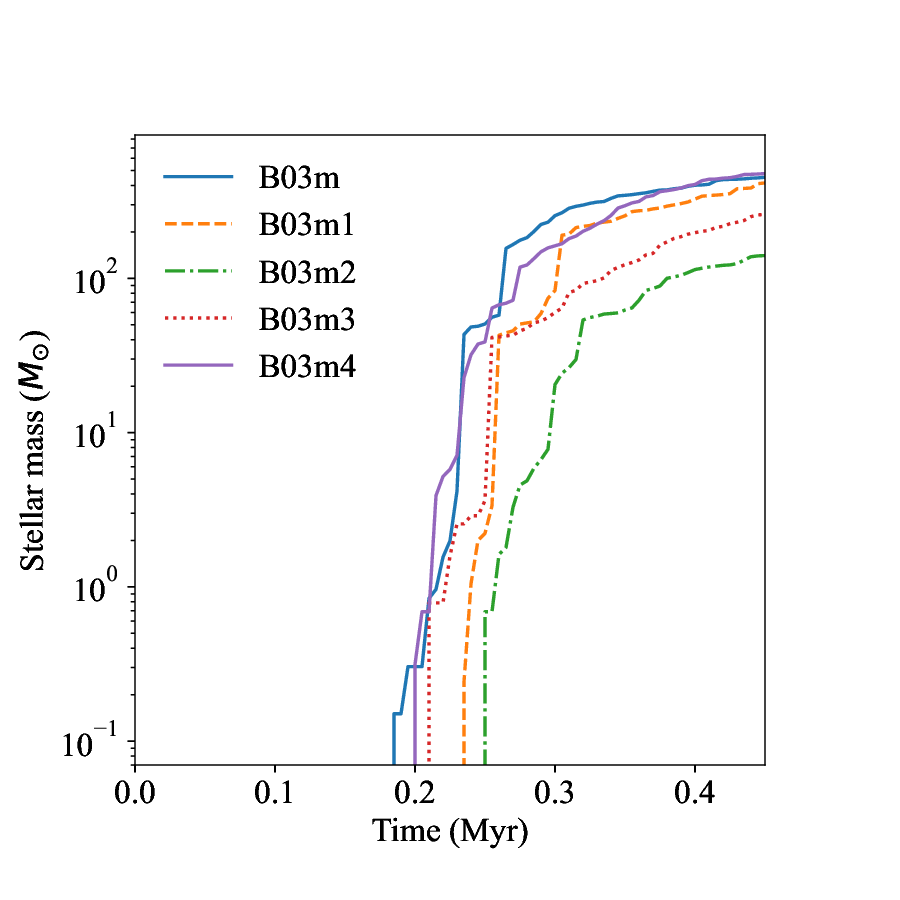}
 \end{center}
\caption{Similar to figure \ref{fig:B03massvstime_resolution}, but for models with different random number seeds of initial conditions. The different colors represent models with different random number seeds (color online).}\label{fig:massvstime_seed}
\end{figure}

When we set the turbulent velocity field for the initial conditions, we used a random number.
The randomness in the turbulence affects the shape of collapsing molecular clouds and the star clusters forming within them. To clarify run-to-run variations, we performed four additional runs for model B03m, but with different random seeds for the turbulent velocity field. Figure \ref{fig:massvstime_seed} shows the total stellar mass as a function of time in models B03m, {B03m1, B03m2, B03m3, and B03m4. These models adopt the same parameters except for random number seeds of the initial conditions.}  Owing to the randomness of the turbulent velocity field, the onset of star formation varies from 0.19 to 0.25 Myr. The final stellar mass is also different among the models. The lowest stellar mass at 0.45 Myr is 141\,$M_{\odot}$ whereas the highest is 478\,$M_{\odot}$. 

\begin{figure}[htbp]
 \begin{center}
  \includegraphics[width=8cm]{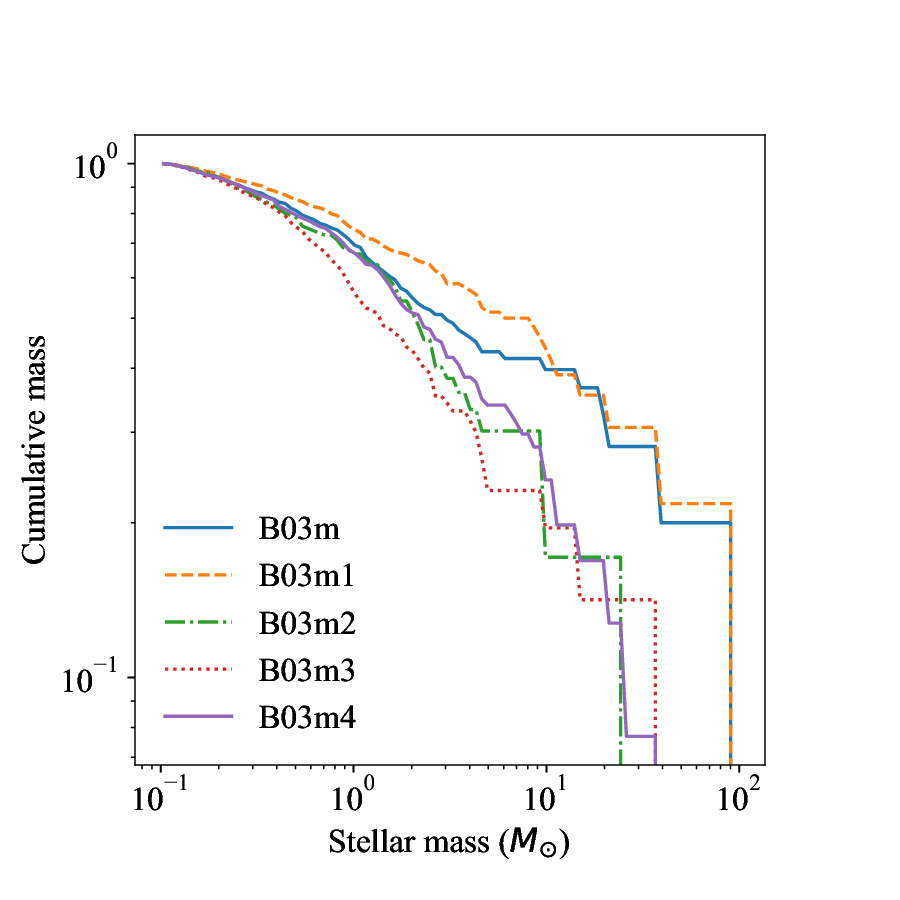}
 \end{center}
\caption{Similar to figure \ref{fig:B03m_rmax}c, but for models with a different random number seed. Different colors represent models with different random number seeds (color online).}\label{fig:CMF_seed}
\end{figure}

Molecular clouds of less than 1000\,$M_{\odot}$ do not have enough mass to adequately sample the IMF from 0.1 to 100\,$M_{\odot}$. The fraction of massive stars is only a small percentage in all stars. The random number seed for the initial conditions and star formation affect the formation of massive stars in these small clouds. Figure \ref{fig:CMF_seed} compares cumulative mass function in models with different random number seeds. Even if we assume the same initial gas mass, the masses of the most massive stars formed in these models vary from 23.5\,$M_{\odot}$ to 91.0\,$M_{\odot}$. Thus, the formation of massive stars in small molecular clouds is highly stochastic. This model is therefore not suitable for evaluating the effects of the value of $r_{\rm{max}}$ on the sampling of IMFs.

\subsection{Formation of star clusters from the turbulent molecular clouds with 4\,$\times$\,10$^4\,M_{\odot}$}\label{M40k}
\begin{figure*}[htbp]
 \begin{center}
   \includegraphics[width=5cm]{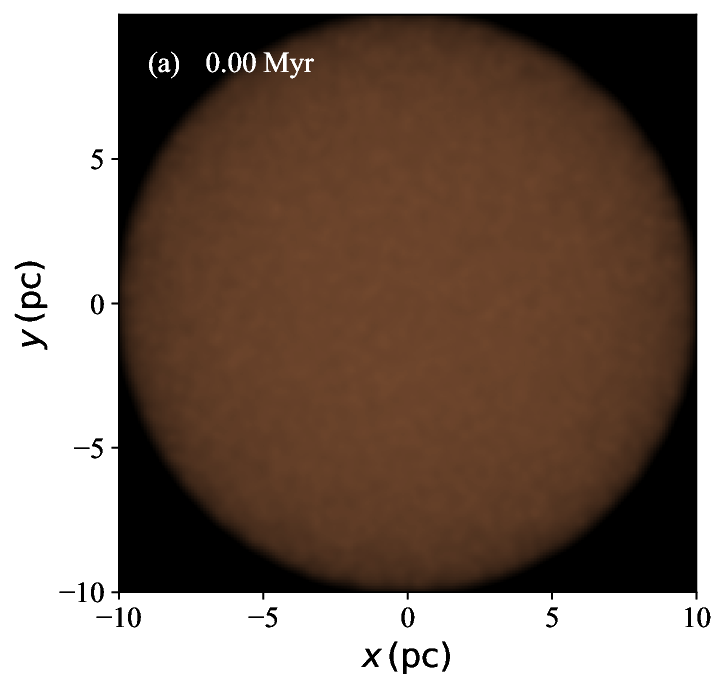}
  \includegraphics[width=5cm]{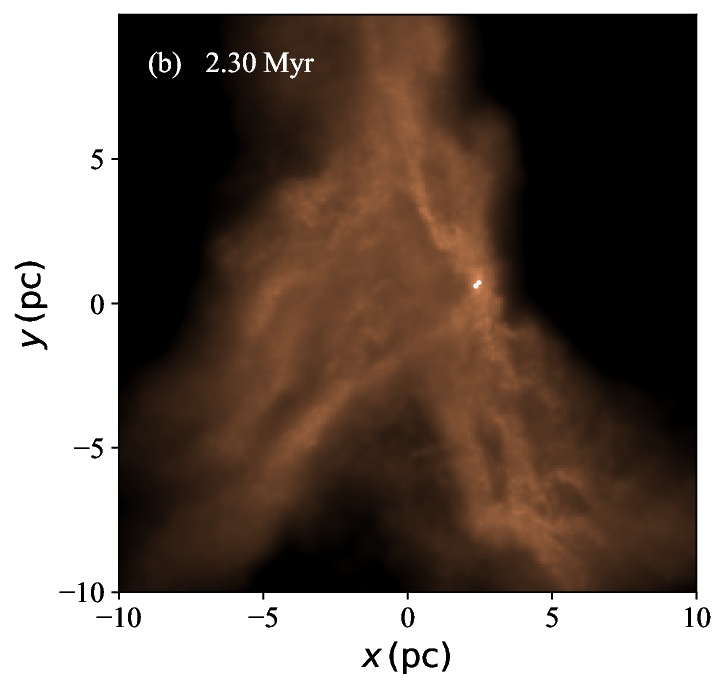}
  \includegraphics[width=5cm]{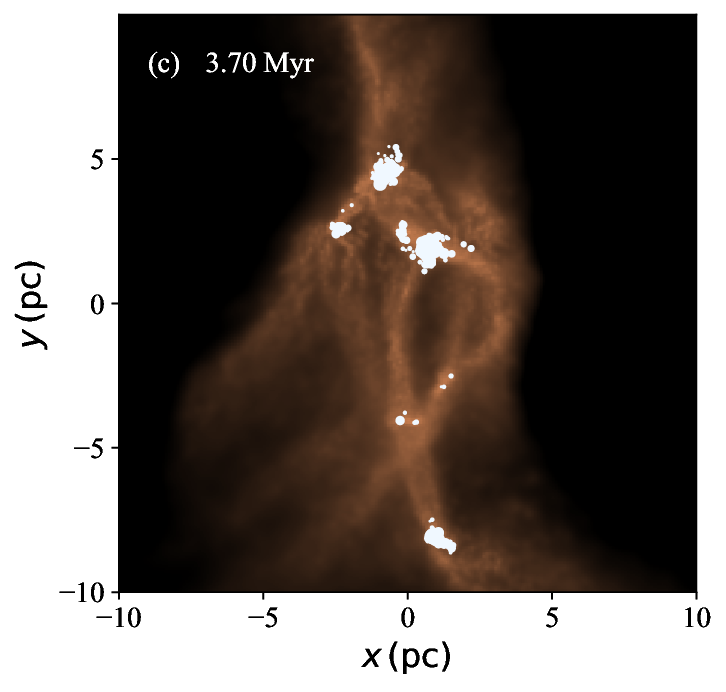}
 \end{center}
\caption{Similar to figure \ref{fig:snapshot}, but for model M40kl. Panels (a), (b), and (c) represent snapshots at 0.00, 2.30, and 3.70 Myr from the beginning of the simulation, respectively (color online).}\label{fig:snapshotM40k}
\end{figure*}

{The star formation in a larger cloud similarly behaves with that of models B03. Figure \ref{fig:snapshotM40k} shows snapshots of gas and stellar density distributions in model M40kl. The initial condition (figure \ref{fig:snapshotM40k}a) and the afterward evolution (figure \ref{fig:snapshotM40k}b) are induced with the same mechanism with models B03 (see section \ref{sec:B03}). This cloud makes several star clusters because of the large cloud's mass (figure \ref{fig:snapshotM40k}c).}

\subsubsection{Mass resolution}\label{sec:M40k_resolution}
\begin{figure}[htbp]
 \begin{center}
 \includegraphics[width=8cm]{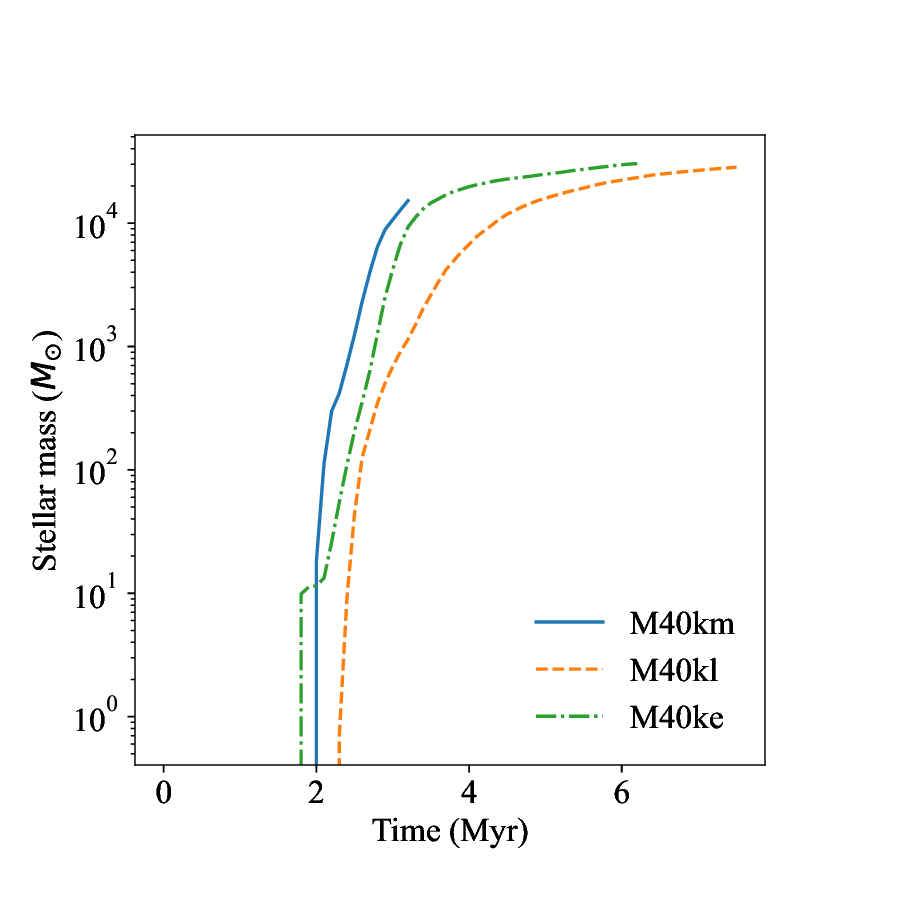}
 \end{center}
\caption{Similar to figure \ref{fig:B03massvstime_resolution}, but for models M40km (solid blue curve), M40kl (dashed orange curve), and M40ke (dash-dotted green curve, color online).}\label{fig:massvstime_M40k}
\end{figure}

In this subsection, we describe the formation of star clusters in a cloud with an initial gas mass of $4\,\times\,10^4\,M_{\odot}$. 
Figure \ref{fig:massvstime_M40k} shows the time evolution of stellar mass in models M40km and M40kl. Unlike model B03 (figure \ref{fig:B03massvstime_resolution}), the onset of star formation in model M40km is shifted to earlier phases than in model M40kl. The former model can resolve turbulent motions of gas more accurately than the latter. The chaotic nature of turbulence motion induces a high-density region locally. Models with a higher mass resolution have more chances to form a star-forming region, thus forming stars in an earlier phase. 

On the other hand, models B03 have a considerably high density (the free-fall time of 0.19 Myr) and compact clouds. In this model, the conditions for star formation can be easily satisfied. Therefore, the onset of star formation in models B03 weakly depends on the mass resolution. 

\subsubsection{Gravitational softening length}
The choice of gravitational softening length in M40k does not {largely} affect the evolution of the stellar mass {in the adopted range of $\epsilon_{\rm{g}}$, but the effect is similar to the} models B03. M40ke adopts the same gravitational softening length as in M40kl ($\epsilon_{\rm{g}}$\,=\,3.2\,$\times\,10^4$\,au), but the initial number of gas particles is the same as that in M40km ($N_{\rm{g}}$\,=\,4.0\,$\times\,10^6$). We have shown that star formation is significantly suppressed by increasing the gravitational softening length in B03. In contrast, the star formation is not suppressed in M40ke (the green dash-dotted curve in figure \ref{fig:massvstime_M40k}). This difference is caused by the size of the clouds. B03 has a radius of 0.5$\>$pc, which is comparable to the size of the softening length of B03l. However, the softening size is much smaller than the radius of M40k ($r_{\rm{t}}$\,=\,10$\>$pc). {Thanks to the large radius compared to the adopted gravitational softening length, M40kl can form stars even if the softening length is the same as B03l, which excessively suppresses star formation.}

\subsubsection{Maximum search radius}\label{sec:M40k_radius}
\begin{figure*}[htbp]
 \begin{center}
  \includegraphics[width=6cm]{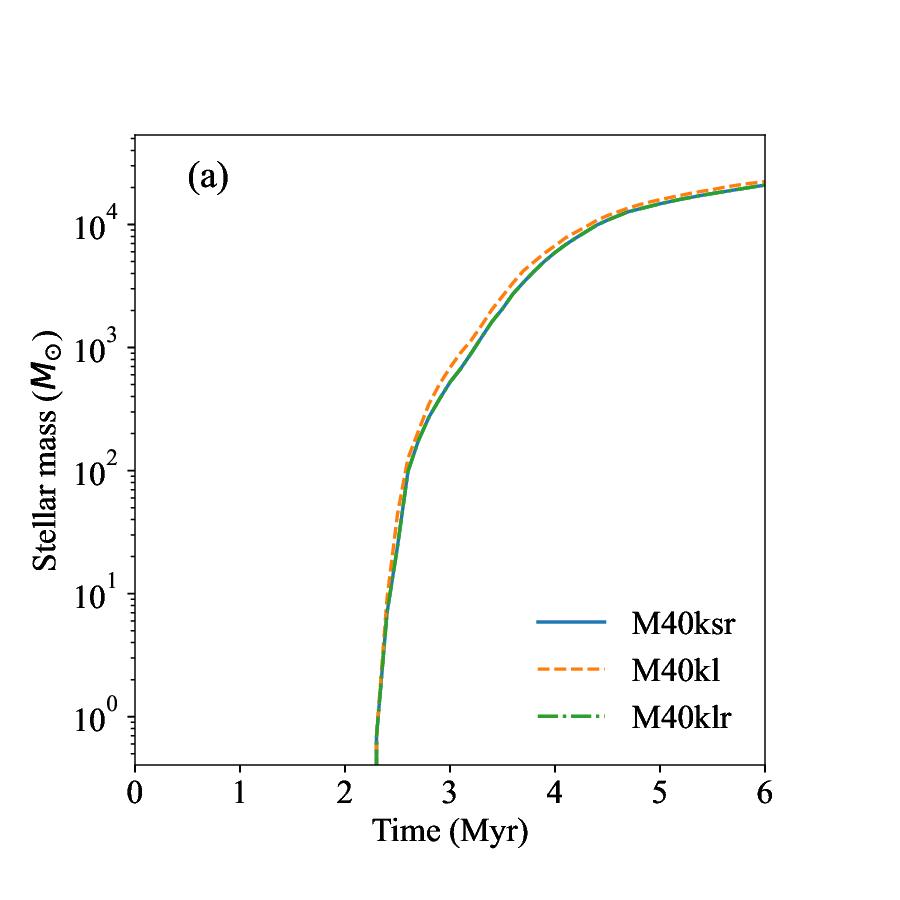}
  \hspace{-1cm}
  \includegraphics[width=6cm]{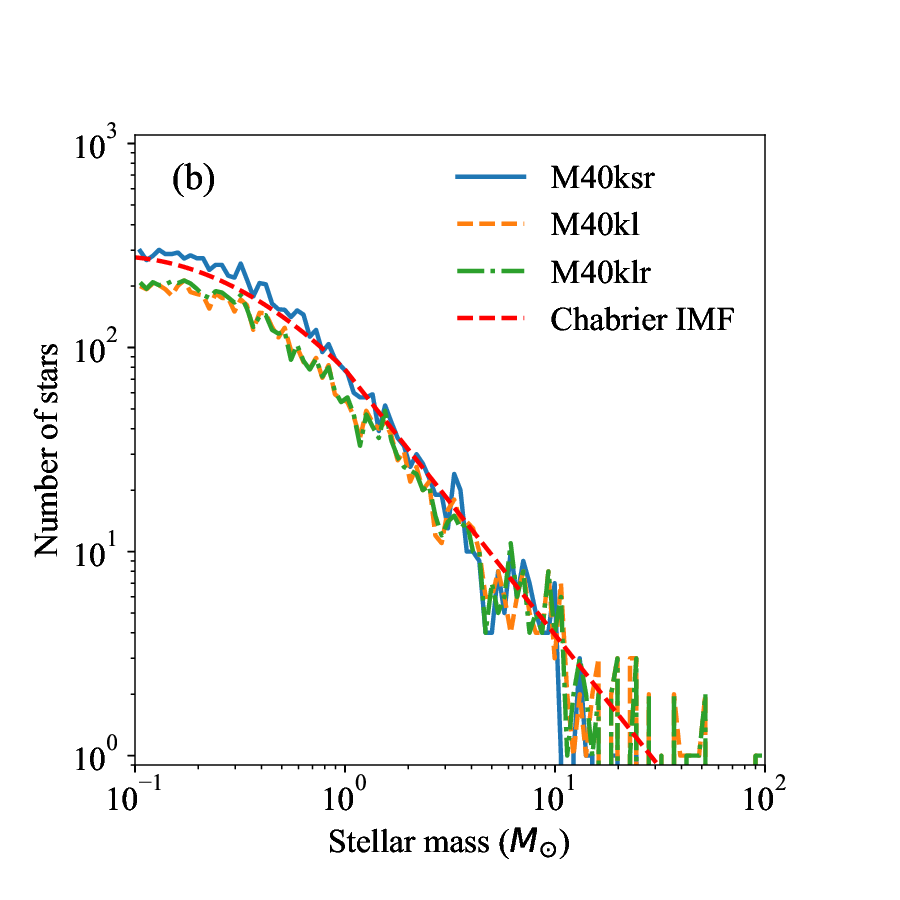}
  \hspace{-1cm}  
  \includegraphics[width=6cm]{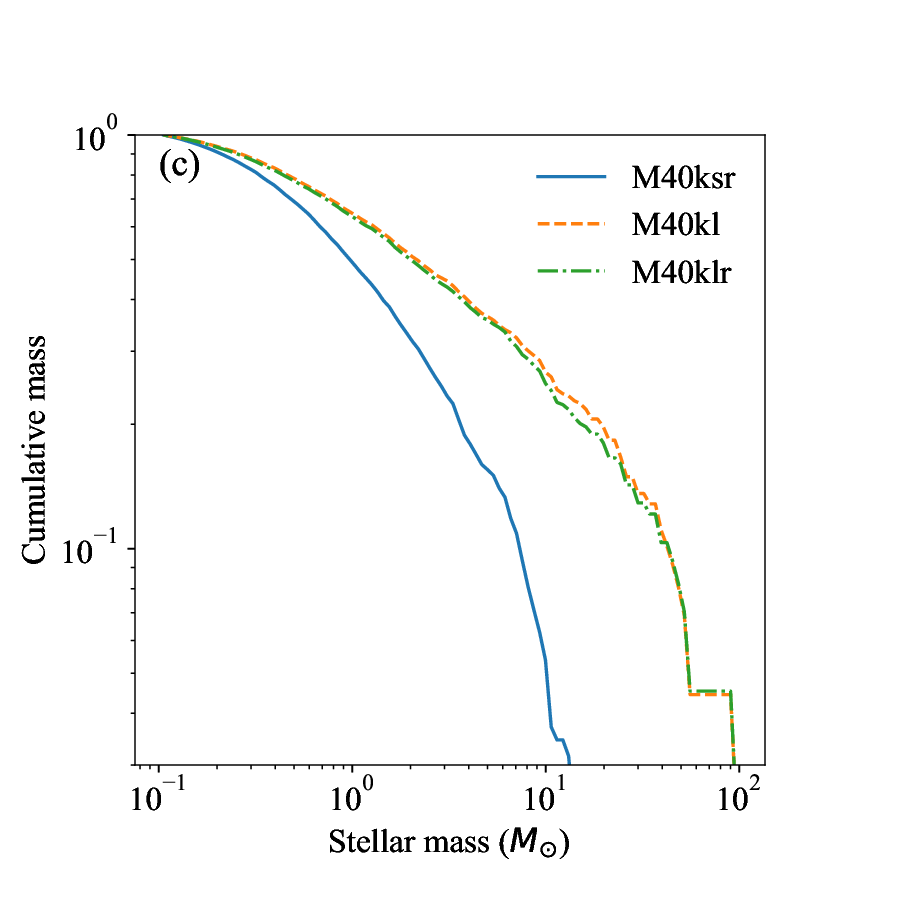}
 \end{center}
\caption{Similar to figure \ref{fig:B03m_rmax}, but for models M40ksr (solid blue curve), M40kl (dashed orange curve), and M40klr (dash-dotted green curve). Panels (b) and (c) are plotted at the time when the total stellar mass reaches 4000\,$M_{\odot}$ (3.8 Myr for M40ksr and 3.7 Myr for M40kl and M40klr, color online).}\label{fig:M40k_rmax}
\end{figure*}

The maximum search radius does not largely affect the evolution of the total stellar mass also in M40k. Figure \ref{fig:M40k_rmax}a represents the total stellar mass as a function of time in models M40ksr ($r_{\rm{max}}$\,=\,0.02$\>$pc), M40kl ($r_{\rm{max}}$\,=\,0.2$\>$pc), and M40klr ($r_{\rm{max}}$\,=\,2.0$\>$pc). As shown in this figure, the time evolution of the total stellar mass is the same in M40kl and M40klr. However, the total stellar mass in M40ksr is lower than the other models. This result is owed to the formation of massive stars being suppressed in this model.

Figures \ref{fig:M40k_rmax}b and \ref{fig:M40k_rmax}c show the stellar mass functions computed in M40ksr, M40kl, and M40klr. The cumulative functions in M40kl and M40klr are overlap. Thus, choosing a maximum search radius larger than $r_{\rm{th}}$ does not affect the shape of the stellar mass function.

Setting an exceedingly small search radius prevents forming massive stars. Figure \ref{fig:M40k_rmax}c clearly shows the lack of massive stars in M40ksr. The lack of massive stars in this model produces a larger number of low mass stars than those of other models. The most massive star formed in M40ksr is 23.9\,$M_{\odot}$, while M40kl and M40klr form stars with 94.7\,$M_{\odot}$ and 100.0\,$M_{\odot}$, respectively. 

Notably, massive stars can be formed in sufficiently high density regions even if a small search radius is chosen. However, most stars tend to form slightly above the threshold density for star formation (1.2\,$\times\,10^5\>$cm$^{-3}$ in this case). This case does not improve the sampling of the IMF in models with a small maximum search radius.

The choice of the maximum search radius affects the number of massive stars. The expected number of massive stars (10--100\,$M_{\odot}$) from the Chabrier IMF is $\approx$ 50 in a star cluster with 5000 stars. In M40kl and M40ksr, there are 43 and 41 massive stars, respectively. On the other hand, M40klr has only 13 massive stars. This result indicates that models with a maximum radius smaller than the estimated search radius (equation \ref{eq:radius}) tend to underestimate the fraction of massive stars.

\begin{figure}[htbp]
 \begin{center}
  \includegraphics[width=8cm]{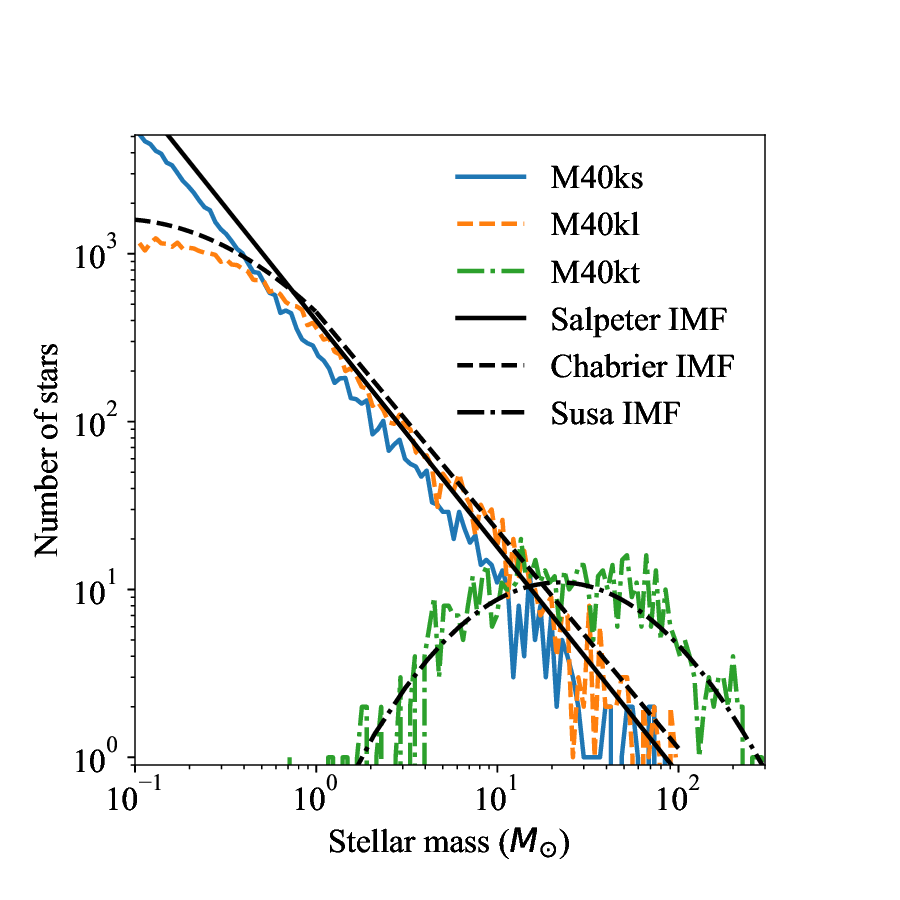}
\end{center}
\caption{Similar to figure \ref{fig:B03m_rmax}b, but for models adopting different IMFs. The solid blue, dashed orange, and dotted green curves represent M40ks (Salpeter IMF), M40kl (Chabrier IMF), and  M40kt (Susa IMF) at 6.0 Myr, respectively (color online).}\label{fig:MF_IMF}
\end{figure}

\begin{figure}[htbp]
 \begin{center}
  \includegraphics[width=8cm]{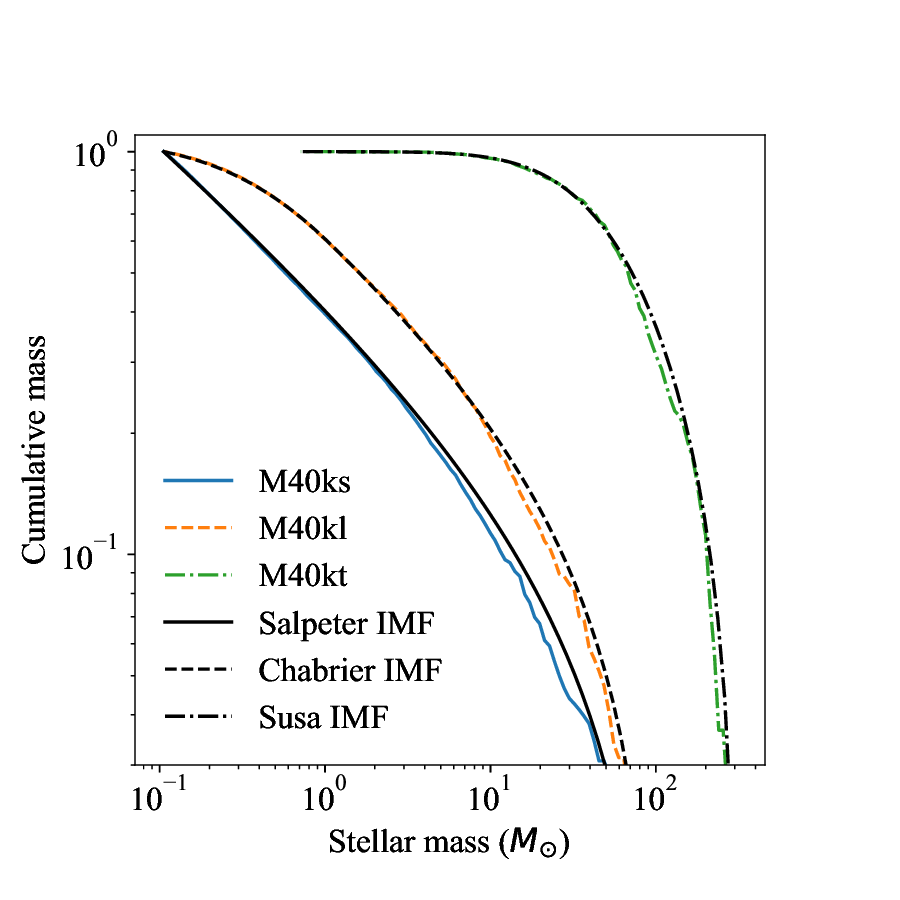}
 \end{center}
\caption{Similar to figure \ref{fig:B03m_rmax}c, but for models adopting different IMFs. The solid blue, dashed orange, and dotted green curves represent M40ks (Salpeter IMF) at 3.7 Myr, M40kl (Chabrier IMF) at 3.7 Myr, and  M40kt (Susa IMF) at 6.0 Myr, respectively(color online).}\label{fig:CMF_IMF}
\end{figure}

Models with the appropriate size of a maximum search radius can sample IMFs with a different shape. Figures \ref{fig:MF_IMF} and \ref{fig:CMF_IMF} show stellar mass functions computed in models with different IMFs. As shown in this figure, all models can fully sample the IMFs. Even if we assume Susa IMF, it is possible to create stars with a stellar mass of $\approx$ 300\,$M_{\odot}$. The required maximum search radius to form stars with 300\,$M_{\odot}$ in a region of $1.2$\,$\times$\,10$^5\>$cm$^{-3}$ is 0.29$\>$pc. In M40kt, we set $r_{\rm{max}}$\,=\,0.5$\>$pc. Thus, it is possible to fully sample any form of IMFs if a sufficiently large search radius is chosen. 

The maximum search radius should be adjusted depending on the threshold density of star formation ($n_{\rm{th}}$). If $n_{\rm{th}}$\,=\,$10^{4}\>$cm$^{-3}$ is chosen, $r_{\rm{max}}>0.48$$\>$pc must be set to allow a correct sampling of stars with 100\,$M_{\odot}$. However, if $n_{\rm{th}}$\,=\,$10^{7}\>$cm$^{-3}$ is adopted, the required value of $r_{\rm{max}}$ is only 0.05$\>$pc. In summary, it is necessary to set a maximum search radius larger than the value estimated from the threshold density for star formation to correctly sample the assumed IMF.

\section{Dwarf galaxy formation}\label{sec:UFD}
\begin{figure}[htbp]
 \begin{center}
  \includegraphics[width=8cm]{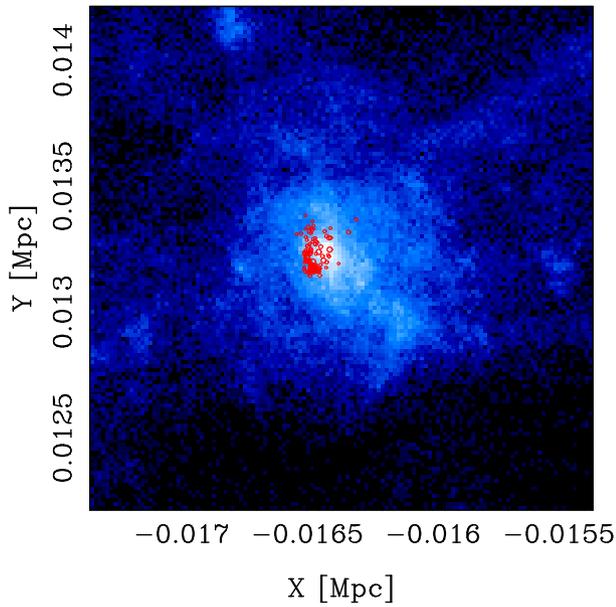}
 \end{center}
\caption{Spatial distribution of dark mater and stars in the simulated UFD at $z$\,=\,8.7. The color scale represents the log-scale surface density
of dark matter in each grid from 10$^{-0.9}\,M_{\odot}\>$pc$^{-2}$ (black) to 10$^{2.3}\,M_{\odot}\>$pc$^{-2}$ (white). Red circles show stars. More massive stars are drawn with larger plots (color online).}\label{fig:UFD}
\end{figure}

{In this section, we show that our star formation model can be applied to galaxy formation simulations. Figure \ref{fig:UFD} shows dark matter and stellar distribution of the simulated UFD. As shown in this figure, stars are formed at the center of the dark matter halo. The star formation was quenched at the redshift $z$\,=\,8.7 because the supernova feedback blow the gas away from the halo. At this redshift, total stellar and halo masses of this galaxy are 1.36\,$\times\,10^3$ $M_{\odot}$ and 7.46\,$\times\,10^6\,M_{\odot}$, respectively. The stellar mass-halo mass ratio is therefore 1.83\,$\times\,10^{-4}$, meaning that this galaxy is highly dark matter dominated. This result is consistent with the extrapolation from the abundance-matching results, predicting the stellar mass of less than $10^3\,M_{\odot}$ in a halo with $\sim\,10^7\,M_{\odot}$ \citep{2017MNRAS.467.2019R}. The half-mass radius and velocity dispersion of this UFD are 51\,pc and 2.0\,km\,s$^{-1}$. These values are consistent with the typical value of UFD around the Milky Way \citep{2019ARA&A..57..375S}.}

\begin{figure}[htbp]
 \begin{center}
  \includegraphics[width=8cm]{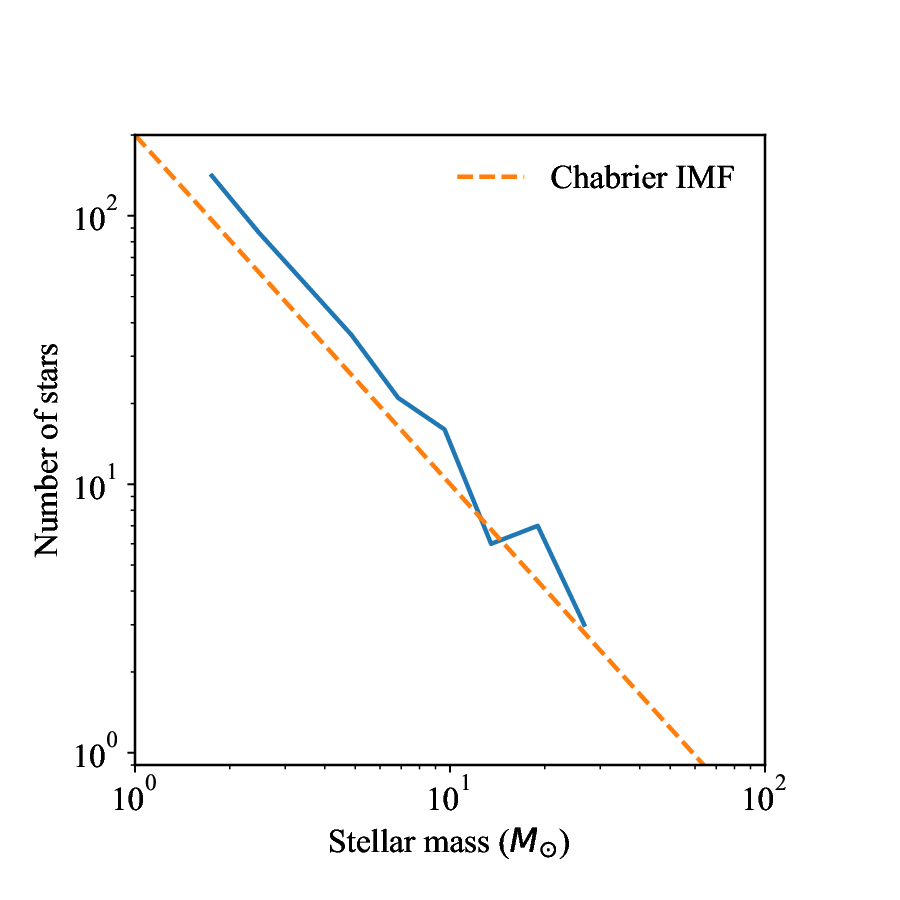}
\end{center}
\caption{Similar to figure \ref{fig:B03m_rmax}b, but for the simulated UFD at $z$\,=\,8.7. Plotted stellar mass is the initial stellar mass. The most massive star in this model (40\,$M_{\odot}$) has already exploded as supernovae at the time of this snapshot. Blue-solid and orange-dashed lines show the result of model UFD and the expected number of stars from the Chabrier IMF, respectively  (color online).}\label{fig:MF_UFD}
\end{figure}

{Expected stellar mass function is reproduced in this simulation. Figure \ref{fig:MF_UFD} shows masses of star particles formed in this model. As shown in this figure, the slope of the stellar mass function from 1.5\,$M_{\odot}$ to 40\,$M_{\odot}$ is consistent with the Chabrier IMF. The star with 40\,$M_{\odot}$ has already exploded as supernovae at the time of this snapshot. The lack of stars more massive than 40\,$M_{\odot}$ is owing to the small total stellar mass (1.36\,$\times\,10^3$ $M_{\odot}$). Models B03 (figure \ref{fig:B03m_rmax}) also lack stars with the high-mass end of the IMF. This result comes from the assumption that we restrict the gas mass within $r_{\rm{max}}$ to be larger than 2\,$m_{*}$ (see section \ref{sec:massive}). The cut-off in the stellar mass function at the lowest mass end is due to the lack of mass resolution. Since the initial gas particle mass is 18.5\,$M_{\odot}$ in this simulation, we set the minimum star particle mass to be 1.5\,$M_{\odot}$ in order not to overproduce stars. Note that if we have to treat feedback from low mass stars, we need to make compound star particles \citep{2017MNRAS.471.2151H,2020MNRAS.492....8A}.}

{Our scheme satisfies the Kennicutt-Schmidt relation. It is well know that the surface gas density and star formation rates are expressed with }
\begin{equation}
    \Sigma_{\rm{SFR}}\,=\,A \Sigma_{\rm{gas}}^N,
\end{equation}
{where $A$ and $N$ are the constants \citep{1959ApJ...129..243S,1989ApJ...344..685K,1998ApJ...498..541K}. \citet{1998ApJ...498..541K} has shown that the value of the power-law index is $N\,=\,1.4\,\pm{\,0.15}$. Recent surveys have found that this relation continues to dwarf galaxies \citep{2016ApJ...832...85T}. Star formation models for galaxy formation therefore needs to satisfy this relation.}

\begin{figure}[htbp]
 \begin{center}
  \includegraphics[width=8cm]{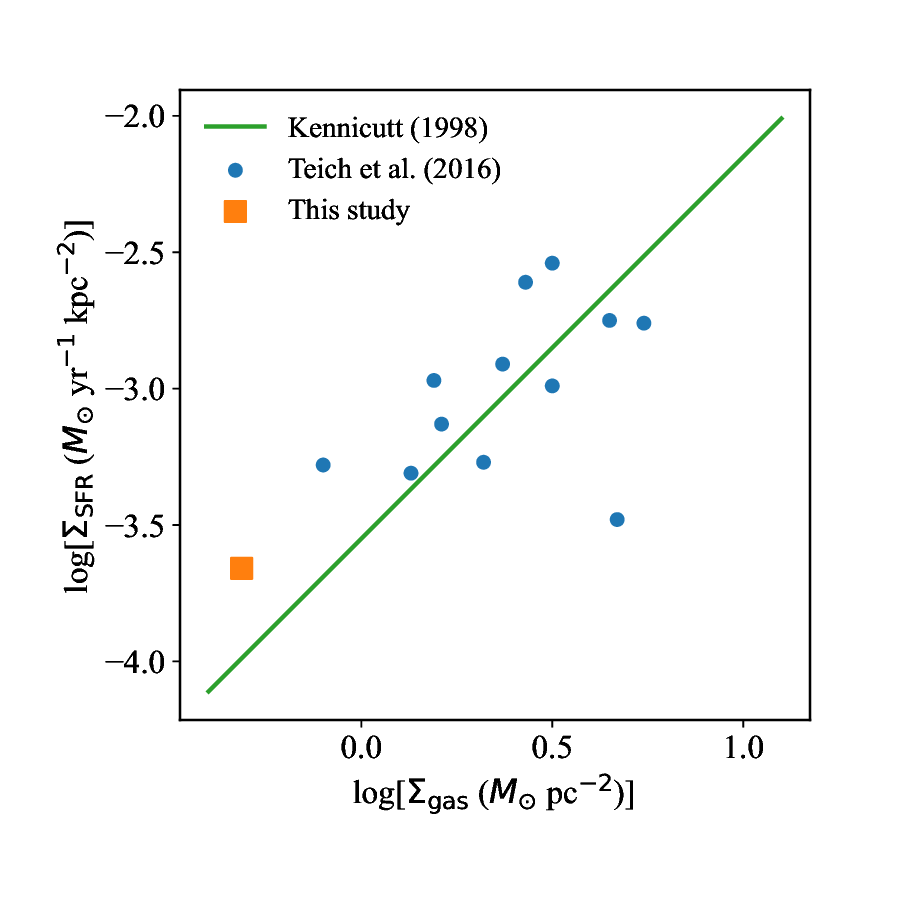}
\end{center}
\caption{The Kennicutt-Schmidt relation (color online). The orange square and blue filled-circles represent model UFD and observations by \citet{2016ApJ...832...85T}, respectively. Green line shows the fitted function in \citet{1998ApJ...498..541K}.}\label{fig:KS}
\end{figure}

{Figure \ref{fig:KS} shows the surface density of star formation rates and gas in our simulation and observations. In this simulation, we derive the surface density of gas within 100$\>$pc from the center of the galaxy ($\Sigma_{\rm{gas}}\,=\,10^{-0.31}\,M_{\odot}\,\rm{pc}^{-2}$) and star formation rates averaged in 100 Myr ($\Sigma_{\rm{SFR}}\,=\,10^{-3.66}\,M_{\odot}\,\rm{yr}^{-1}\,\rm{kpc}^{-2}$). These values are consistent with the Kennicutt-Schmidt relation. This result is also consistent with the value in simulations of dwarf galaxies, which adopts the star formation model similar to this study \citep{2020arXiv201007311G}. Since we computed the UFDs, both $\Sigma_{\rm{gas}}$ and $\Sigma_{\rm{SFR}}$ are located in the lowest values. As we assume the local Schmidt law in this simulation (equation \ref{eq:schmidt}), we can confirm that it is possible to adopt this star formation scheme to the simulations of galaxy formation.}

\section{Discussion}\label{sec:discussion}
\subsection{Comparison with other methods\label{sec:comparison}}
{In this subsection, we compare results computed in previous studies with different methods. We firstly discuss the gas density PDF presented in figure \ref{fig:DensityPDF}. The shape of the gas density PDF characterizes the evolution of the gas \citep{1994ApJ...423..681V}. Simulations of molecular clouds \citep[e.g.,][]{2001ApJ...546..980O, 2001ApJ...557..727V, 2005MNRAS.356..737S} and galaxies \citep[e.g.,][]{2001ApJ...559L..41W, 2007ApJ...660..276W,2003ApJ...590L...1K, 2008ApJ...673..810T, 2008ApJ...680.1083R} have shown that the PDF shows log-normal around a mean density. This feature is a characteristic of supersonic turbulence. Several studies have shown that there is a power-law tail at high-density region, which would be arisen from a balance between turbulence and gravity \citep{2020ApJ...903L...2J}. These features are also seen in observations \citep{2009A&A...508L..35K, 2011A&A...530A..64K, 2010A&A...512A..67L, 2010MNRAS.406.1350F, 2012A&A...540L..11S, 2013ApJ...766L..17S, 2015A&A...575A..79S, 2015A&A...578A..29S, 2015MNRAS.453L..41S, 2016A&A...587A..74S, 2015MNRAS.449.4465B, 2018A&A...617A.125C}. Our models also show the gas density PDF with the log-normal distribution around a mean density and the power-law tail at high-density ($n_{\rm{H}}\,\gtrsim\,10^5\>$cm$^{-3}$, figure \ref{fig:DensityPDF}). This result means that we can correctly compute the evolution of molecular clouds with supersonic turbulence and self-gravity even if we include the stochastically sampled star formation model.}

{Next, we discuss star formation efficiencies.} Since we adopt the initial conditions following \citet{2003MNRAS.343..413B} for models B03, we compare the total stellar mass in models B03h and B03m to those of \citet{2003MNRAS.343..413B}. The main difference between our models and those of \citet{2003MNRAS.343..413B} is the approach for the conversion of gas to stars. In \citet{2003MNRAS.343..413B}, they assume that stars are formed from sink particles, while our model stochastically converts gas particles to star particles. We need to confirm that our models do not largely alter the results obtained by the sink particle approach.  

As shown in section \ref{sec:B03_resolution}, the total masses in models B03h and B03m are 556\,$M_{\odot}$ and 452\,$M_{\odot}$, respectively. The average mass of all models with different random number seeds is 349\,$M_{\odot}$ (section \ref{sec:B03_rand}). These masses are roughly consistent with the stellar mass (415\,$M_{\odot}$) in the models of \citet{2003MNRAS.343..413B}. 

\begin{figure}[htbp]
 \begin{center}
  \includegraphics[width=8cm]{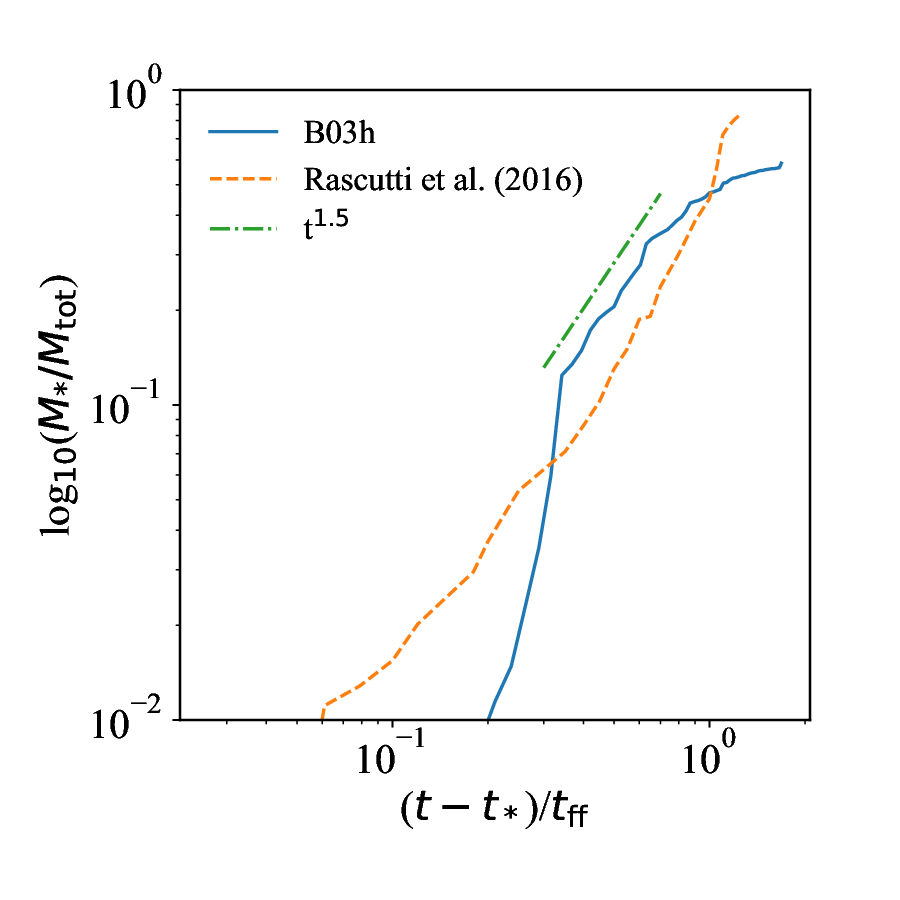}
 \end{center}
\caption{Time evolution of the stellar mass divided by the initial cloud mass ($M_{*}$: total stellar mass of the star cluster, $M_{\rm{tot}}$: initial total gas mass of the cloud, $t$: time from the beginning of the simulation, $t_{*}$: time when the first star formed, $t_{\rm{ff}}$: initial free-fall time of the cloud). The solid blue and dashed orange curves show model B03h and model $\Sigma$-M5E4-R15 without feedback in \citet{2016ApJ...829..130R}, respectively. The green dash-dotted line shows the $M_{*}\propto{t^{1.5}}$ fitted by \cite{2016ApJ...829..130R} (color online).\label{fig:SFE}}
\end{figure}

{Figure \ref{fig:SFE} compares the time evolution of the stellar mass divided by the initial cloud mass computed in models B03h and $\Sigma$-M5E4-R15 without feedback in \citet{2016ApJ...829..130R}. Both models assume uniform density sphere as the initial condition. Model B03h begins star formation later than that of the model in \citet{2016ApJ...829..130R}. This difference is caused by the assumed value of the threshold density for star formation. In model B03h, we assume $n_{\rm{th}}$\,=\,$3.0\,\times\,10^{8}$\,cm$^{-3}$ (table \ref{tab:model}) while the threshold density corresponds to $\sim\,1.5\,\times\,10^{4}$\,cm$^{-3}$ in model $\Sigma$-M5E4-R15. In fact, model B03n ($n_{\rm{th}}$\,=\,1.2\,$\times$\,10$^5$\,cm$^{-3}$) starts star formation $\sim$ 0.15 Myr earlier compared to model B03m ($n_{\rm{th}}$\,=\,1.2\,$\times$\,10$^7$\,cm$^{-3}$, figure \ref{fig:B03massvstime_softening}b).} 

{Early phases of the time evolution of stellar mass are different between models B03h and $\Sigma$-M5E4-R15. \citet{2016ApJ...829..130R} argued that there was a break of the power-law of the time evolution of star formation at around $M_{*}\,\sim\,0.1M_{\rm{cl},0}$, where $M_{\rm{cl},0}$ was the initial cloud mass. However, they confirmed that this was owing to the artificial outcome from their initial condition. This result means that the difference in the evolution of star formation could not come from the different star formation scheme but the assumption of the different initial conditions.} 

{For 0.2 $<\,(t-t_{*})/t_{\rm{ff}}\,<$ 1.0, star formation of both models shows similar slope. \citet{2016ApJ...829..130R} have shown that the evolution of the stellar mass can be fitted with $\sim\,t^{1.5}$ in this phase. As shown in figure \ref{fig:SFE}, time evolution of stellar mass can also be approximated with this power-law in model B03h. \citet{2015ApJ...800...49L} have argued the importance of the self-gravity. If they remove the self-gravity, the time evolution of the star formation efficiencies becomes slower compared to models with self-gravity. As both models B03h and $\Sigma$-M5E4-R15 in \citet{2016ApJ...829..130R} include self-gravity, this result suggests that the evolution of the SFE is not affected by the scheme of the star formation.}

Note that {none of the models compared here} adopt any form of feedback from massive stars. Because of this assumption, both models tend to convert a larger fraction of gas to stars ($\approx$ 40--50\%) than the observed inferred value \citep[10-30\%,][]{2003ARA&A..41...57L}. This issue is studied in our paper in this series \citep{2021arXiv210302829F}.

\subsection{Mass of gas particles}
\begin{figure}[htbp]
 \begin{center}
  \includegraphics[width=8cm]{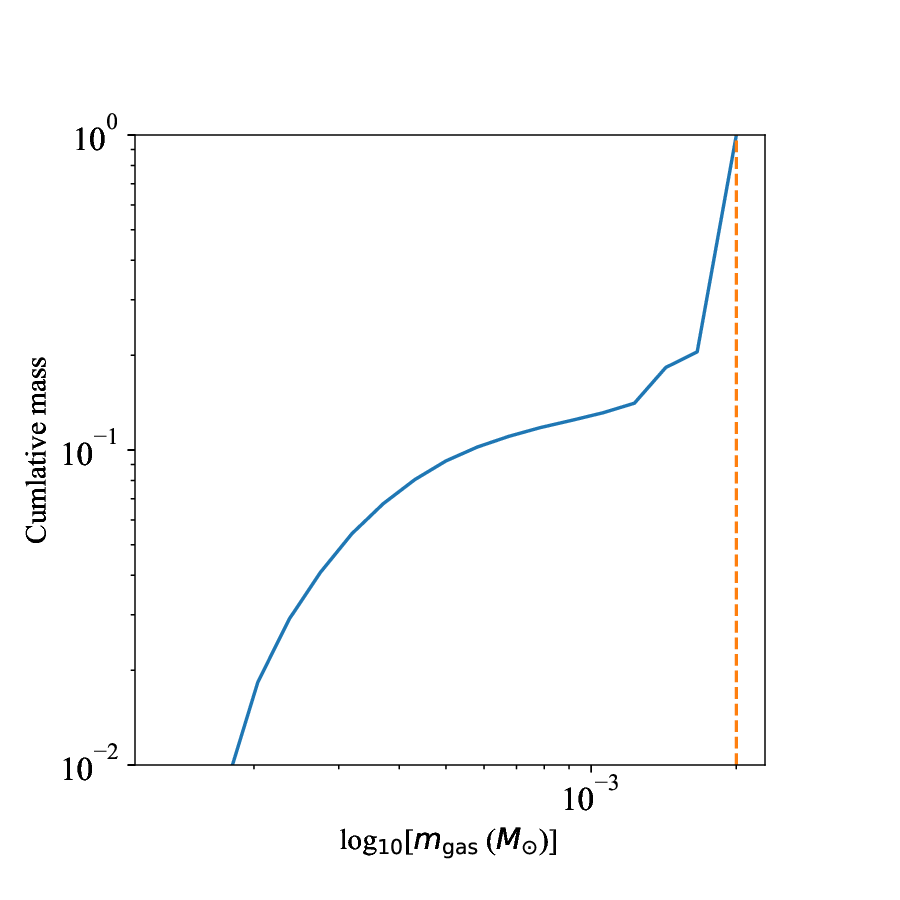}
 \end{center}
\caption{Gas particle mass distribution in model B03h at 0.45 Myr (blue solid curve). The orange-dashed curve shows the original gas particle mass (0.002\,$M_{\odot}$, color online).}\label{fig:GassMass}
\end{figure}

{We implemented the particle merging algorithm (section \ref{sec:sf}) to reduce the calculation cost. Figure \ref{fig:GassMass} shows the distribution of gas particle masses in model B03h at 0.45 Myr from the beginning of the simulation. Thanks to the particle merging algorithm, few particles are less than the original mass (= 0.002\,$M_{\odot}$). Mass fractions of gas particles less than 0.5, 0.25, and 0.125 times the initial mass are 0.031, 0.015, 0.0026, respectively.}

{The DISPH is insensitive to the contamination of gas particles with different masses. \citet{2013ApJ...768...44S} tested the evaluation of the density and pressure around the contact discontinuity of eight times different density with equal separation. This set-up results in the use of eight times lower mass particles in less dense regions. They have shown that the pressure is evaluated correctly under this assumption. These results mean that contamination of low-mass particles does not largely affect the results.}

\subsection{Massive star formation}\label{sec:massive}
Massive stars play an important role in the evolution of galaxies and star clusters. Photoionization, radiation pressure, and supernova explosions can heat the ISM \citep{2019ARA&A..57..227K}. Production of elements from massive stars is a source of galactic chemical evolution \citep[e.g.,][]{2018ApJ...855...63H, 2019ApJ...885...33H, 2020MNRAS.491.1832P}. Thus, massive stars in star-by-star simulations of galaxies must be assessed properly.

In this model, it is possible to fully sample the initial mass function by setting the maximum search radius appropriately. As shown in section \ref{M40k}, the formation of massive stars is suppressed if an excessively small search radius is set. Conversely, no discernible problems occur when a sufficiently large search radius is used. However, the typical size of the self-gravitating cores in giant molecular clouds is $\sim$ 0.1$\>$pc \citep[e.g.,][]{2007ARA&A..45..565M}. It is therefore unphysical to set an excessively large search radius compared to the size of the self-gravitating cores. In section \ref{M40k}, we confirmed that the sampling of the IMF was not different if we chose a search radius larger than $r_{\rm{th}}$ (equation \ref{eq:radius}). This result was because all stars were formed in a region with a density larger than the threshold density for star formation. Thus, we recommend setting the maximum search radius as small as possible but larger than $r_{\rm{th}}$ in a star-by-star simulation.

\begin{figure}[htbp]
 \begin{center}
  \includegraphics[width=8cm]{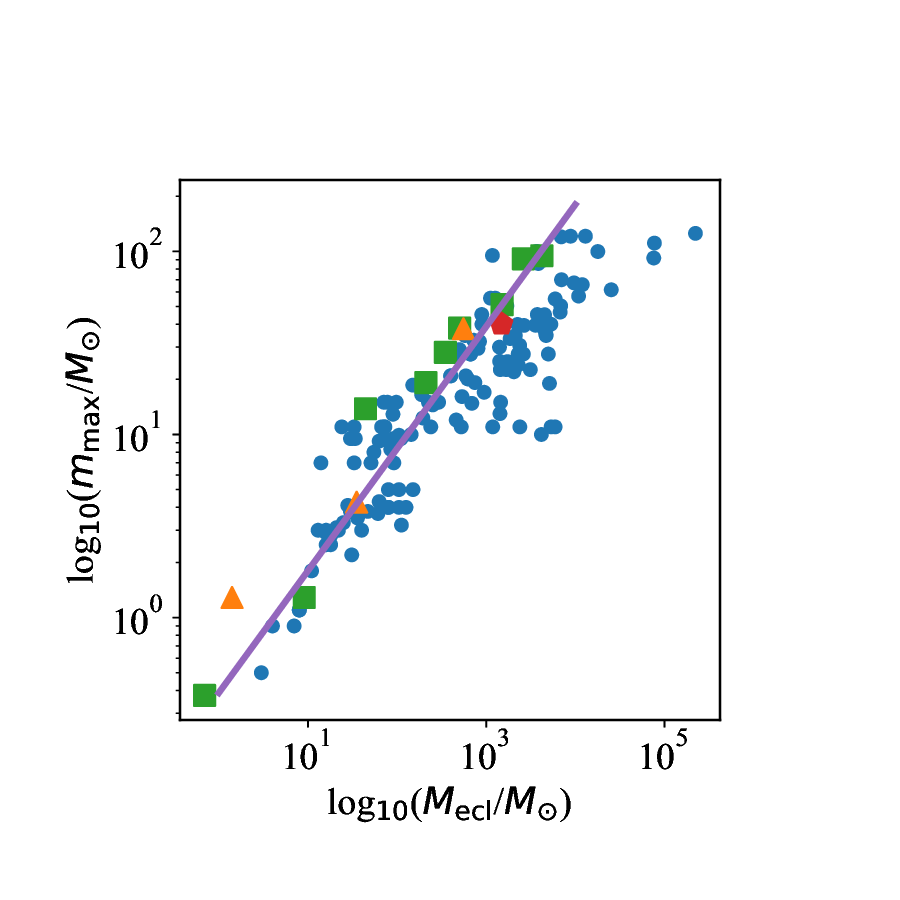}
 \end{center}
\caption{The mass of the most massive stars ($m_{\rm{max}}$) versus the total stellar mass ($M_{\rm{ecl}}$) of star clusters. Orange triangles represent model B03h at 0.185, 0.225, and 0.450 Myr from left to right. Green squares show model M40kl at 2.3, 2.4, 2.5, 2.7, 2.8, 2.9, 3.3, 3.5, and 3.7 Myr from left to right. The red-pentagon depicts model UFD at $z$\,=\,8.7. The purple-solid line is the computed relation in sink particle simulations \citep{2004MNRAS.349..735B}. Blue plots are observed data \citep{1959ApJ...130...57W, 1983AJ.....88..985S, 1984A&AS...57..205H, 1984ApJS...56..211F, 1985ApJ...291..571S, 
1985ApJ...292..148T, 
1986BAAS...18..910S, 
1987PASP...99..240G, 
1989MNRAS.236..263P, 1989ApJ...340..823W,
1989AJ.....97..107M, 
1995ApJ...454..151M, 1993AJ....105..980M, 
1990ApJ...356L..55D, 1990ApJ...362..147C, 1993ApJ...407..657C, 1997AJ....114..198C, 1991ApJ...371..171L, 1993PASP..105..588P, 1997A&A...320..159T, 1998A&AS..133...81T,  
1998ApJ...492..540H, 1998AJ....116.1816H, 1998ApJ...493..180M,  
1999AstL...25...10D, 1999A&A...347..532S, 1999A&A...342..515T, 1999MNRAS.302..714N, 1999A&A...349..825V, 2000A&A...360..539K, 2000AJ....120.1396H, 2001AJ....121.1050M, 2001AJ....122..866P, 
2002AJ....123..892P, 
2002A&A...388..158L, 2002ApJ...575..354G, 2002A&A...392..945P, 2002ApJ...581..258F, 2003AJ....125.1480A, 2003MNRAS.341..805P, 2003ARA&A..41...57L, 2003A&A...407..925R, 2003AJ....126.1861B, 2003A&A...411...83B, 2005A&A...435...95B, 2006A&A...455..923B, 2004ApJ...608..781D, 2006A&A...459..477D, 2004A&A...421..623K, 2004AJ....128..323N, 2004AJ....127.2826B, 2004AJ....128.2316S, 
2004NewAR..48..727N, 2004ApJ...614..818B, 2006AJ....132.1692B,
2004ApJ...611L..33B, 2004AJ....128.1684S, 2004AJ....128.2364R, 2004A&A...425..937D, 2005A&A...437..247D, 2004ApJS..154..374G, 2008ApJ...674..336G, 2005ApJ...620L..43O, 2005A&A...430..941P, 2006AJ....132.1100W, 2008hsf2.book..351W, 2007MNRAS.374..399F, 2007A&A...462..157P, 
2007AJ....133.1092W, 
2007MNRAS.375.1220M, 2007ApJ...659.1360W, 2007ApJ...660.1480M, 2007AJ....133.2072D, 2007A&A...471..813R, 2007AJ....134.1368C, 2007ApJ...667..963S, 2007A&A...474..515M, 2007A&A...476..199A, 2008A&A...483..209Y, 2008hsf1.book..390A, 2008hsf1.book..966D, 2008hsf2.book..124W, 2008hsf2.book..169L, 2008hsf2.book..497R, 2008hsf2.book..683R,  2008hsf2.book..735N, 2008A&A...478..219M, 2008ApJ...675.1319H, 
2008A&A...483..171N, 
2008NewA...13..508O, 2008MNRAS.389L..38S, 2009MNRAS.397.2049S, 2008MNRAS.390.1598F, 2008Ap&SS.318...25H, 2008A&A...492..441N, 2008ApJ...688.1142K, 2009A&A...493...79M, 2009MNRAS.394..900C, 2009A&A...497..195K, 2009A&A...493..931B, 2009MNRAS.397.1915B, 2009A&A...503..801F, 2010ApJ...715..671W, 2010ApJ...722..971C, 2011ApJ...727...64K, 2012MNRAS.419.1871D, 2013AJ....145..125V}.}\label{fig:Mmax-Mecl}
\end{figure}

{Although this model adopts stochastic approach for the star formation, it can reproduce the relationship between the most massive stars and enclosed mass in star clusters. Figure \ref{fig:Mmax-Mecl} compares this relation in our simulations and observations. As shown in this figure, our simulations and observations are nicely overlapped. This result is because we require gas mass within the search radius to be larger than $2\,m_{*}$ for the formation of stars with $m_{*}$ (section \ref{sec:sf}). In a small cloud (B03h) or in the early phases of the simulations, the mass in the search radius is not enough to form massive stars.}

{The relationship between the most massive stars and total stellar mass in star clusters cannot be reproduced with a random sampling of stellar mass from the IMF. \citet{2006MNRAS.365.1333W} investigated the maximum stellar mass in star clusters using Monte Carlo experiments. They have rejected the possibility of random sampling with confidence larger than 0.9999. This result is further confirmed with the newer data \citep{2013MNRAS.434...84W}. The most massive stars in models B03h, M40kl, and UFD suggest that the choice of the value of $r_{\rm{max}}$ and the minimum mass in $r_{\rm{max}}$ (2\,$m_{*}$) is appropriate.}

Dynamical evolution of star clusters with massive stars within the context of galaxy evolution is one of the challenging issues in astronomy. Because of dynamical friction, massive stars sink to the center of star clusters \citep[e.g.,][]{2010ARA&A..48..431P}. {Dynamical interactions within star clusters form compact binaries} \citep{2000ApJ...528L..17P,2006ApJ...637..937O,2015PhRvL.115e1101R,2014MNRAS.441.3703Z,2017MNRAS.464L..36A,2017PASJ...69...94F,2019MNRAS.487.2947D,2019MNRAS.486.3942K}, which can be detected by the gravitational wave observations \citep{2019PhRvX...9c1040A}. Because star clusters have crossing times of $\sim$ 1 Myr {and close encounters of stars often happen}, it is necessary to perform $N$-body simulations with a direct $N$-body integration scheme. The BRIDGE algorithm, {in which a direct integration scheme can be combined with the tree algorithm} \citep{2007PASJ...59.1095F}, makes it possible to perform simulations of star clusters within a parent galaxy. In this study, we cannot evaluate the dynamical evolution of star clusters because we have not implemented the BRIDGE algorithm. Our next paper will intensively discuss the implementation of BRIDGE into ASURA and the dynamical evolution of star clusters \citep{FujiiASURABRIDGE}.

\subsection{Applications}\label{sec:application}
We can adapt this model to star-by-star simulations from the scale of star clusters to galaxies. In this study, we have computed the formation of star clusters with a mass of one gas particle at less than 0.1\,$M_{\odot}$. All these runs correspond to case 2 of figure \ref{fig:SF_scheme}. This model can be also applied to simulations of galaxies with $m_{\rm{gas}}>$ 0.1\,$M_{\odot}$ {(model UFD)}. In this case, it is necessary to carefully assign the stellar mass to each star particle.

Mass conservation in a star-forming region is one of the primary concerns in star-by-star simulations. \citet{2017MNRAS.471.2151H} have constructed a method to assign stellar mass from the assumed IMF to newly formed star particles. In their model, stellar masses from a given IMF are assigned to a star particle until the sum of the sampled mass exceeds the mass of the star particle. The residual between the assigned mass and the initial mass of the particle is transferred to another star particle to guarantee mass conservation. Their model can be applied to simulations with wide ranges of mass resolutions. However, their model requires further consideration of an unphysical mass transfer among star particles.    

The method used to convert gas particles into star particles as applied in this study can complement their model. The former's advantage is that it can avoid an unphysical mass transfer between star particles. When performing simulations with $m_{\rm{min,\,IMF}} < m_{\rm{gas}}$, there are two approaches to assign a stellar mass to a star particle. One method is the extension of the model of \citet{2017MNRAS.471.2151H}. In this case, star particles with $m_{*} > m_{\rm{gas}}$ are formed through a  mass assemblage from neighbor gas particles, whereas several stellar masses are assigned to a star particle if $m_{*} \leq m_{\rm{gas}}$. The residuals will be compensated by the mass assemblage from surrounding gas particles. This case is suitable for simulations of galaxy formation, which require high computational costs. 

The other approach is that the sampled stellar mass from the IMF is simply assigned to each star particle. Simulations for understanding star cluster formation within the context of galaxy evolution must incorporate each star with masses from $m_{\rm{min,\,IMF}}$ to $m_{\rm{max,\,IMF}}$. This method can be applied to such simulations. In this case, the computational costs become relatively high compared to the first method because several stars are formed.

Models in this study stochastically form stars using star particles instead of applying the sink particle approach \citep[e.g.,][]{2003MNRAS.343..413B,2018PASJ...70S..54S, 2019MNRAS.489.1880H,2020MNRAS.tmpL..65D}.  In the sink particle approach, large fractions of unresolved gas are locked in sink particles if the resolution is not high enough \citep{1997MNRAS.288.1060B, 1997ApJ...489L.179T}. {Even if the sink particle approach can reproduce the mass function of molecular cloud cores, it requires an assumption of models converting sink particles to stars.} The star particle approach applied in this study is more suitable in simulations that do not have a high enough resolution to resolve the fragmentation and form low mass stars.

One possible application of our model is in addressing the issue of mass segregation. There are several instances of observational evidence wherein massive stars tend to reside in the center of star clusters \citep[e.g.,][]{1998ApJ...492..540H, 2006ApJ...644..355H, 2006AJ....132..253S}. These studies concluded that mass segregation is primordial. An alternative explanation is that cluster formation via the merger of sub-clumps can produce early-stage mass segregation \citep{2007ApJ...655L..45M, 2009ApJ...700L..99A, 2009MNRAS.400..657M}. In this model, the formation of massive stars might be biased in a high-density region because the maximum search radius is adopted. Moreover, we did not assume any initial mass segregation for the formation of massive stars. Careful studies of the location of massive star formation and the afterward dynamical evolution of these stars can refine our collective understanding of these issues.

Another application of this study's proposed model is that it will be able to competently study chemical tagging \citep[e.g.,][]{2010ApJ...721..582B}. There are almost no star-to-star variations of the chemical abundances in open star clusters \citep[e.g.,][]{2007AJ....133.1161D, 2010A&A...511A..56P, 2013MNRAS.431.3338R}. Star-by-star simulations using this study's model can follow chemically homogeneous groups of stars from their formation to the present. The model can also be applied to find signatures of first stars \citep[e.g.,][]{2014Sci...345..912A}.

Our models rely on the assumption of the IMF. It is highly challenging to find suitable conditions to form stars in each mass range \citep[e.g.,][]{2007ARA&A..45..565M}. Thus, it is not possible to construct a self-consistent model of star formation for star-by-star simulations of galaxies and star clusters given humanity's current understanding of the topic. Determining the critical conditions to form stars in different mass ranges is vital in constructing self-consistent models of star formation for simulations at larger scales.   

\section{Conclusions}\label{sec:conclusions}
This study has developed and tested a model of star formation for star-by-star simulations. In this model, we stochastically sample the IMF and assign the sampled mass to a newly formed star particle. We have updated the Chemical Evolution Library (CELib) to easily sample the assumed IMFs. 

In this model, gas particles are converted into star particles in two ways. If a mass of one gas particle is more massive than that of a star particle, the gas particle is spawned to form the new star particle. In the opposite case, a star particle is formed by combining mass from surrounding gas particles. 

We have newly introduced a maximum search radius ($r_{\rm{max}}$) to avoid combining mass from an unrealistically large region. We found that the search radius should be larger than the value estimated from the threshold density for star formation (equation \ref{eq:radius}). If the threshold density is 1.2\,$\times$\,10$^7\>$cm$^{-3}$, the required maximum search radius corresponds to 0.04$\>$pc to sample a stellar mass of 100\,$M_{\odot}$. In this case, models with $r_{\rm{max}}$ $\geq$ 0.2$\>$pc can fully sample any form of IMFs, whereas models with $r_{\rm{max}}$\,=\,0.02$\>$pc cannot sample the most massive stars.

The total mass of the molecular cloud affects the sampling of the most massive stars. In small star systems, it is difficult to adequately sample the most massive stars even if a sufficiently large maximum search radius is set. We also found that the threshold densities for star formation and star formation efficiency do not significantly affect the results. {With our star formation scheme, the observed relationship between the most massive stars and total stellar mass in star clusters can naturally be reproduced.}

{In this study, we also show that this model can be applied to cosmological zoom-in simulations of UFDs. We confirm that this scheme can form stars with masses following the IMFs. The surface densities of star formation rates and gas are consistent with the Kennicutt-Schmidt relation. With this scheme, we can compute the evolution of star clusters and dwarf galaxies with a mass resolution from $\sim$\,0.001\,$M_{\odot}$ to $\sim$\,10\,$M_{\odot}$.}

\section*{Funding}
YH is supported by the Special Postdoctoral Researchers (SPDR) program at RIKEN. MF is supported by The University of Tokyo Excellent Young Researcher Program. This work was supported by JSPS KAKENHI Grant Numbers 20K14532, 19H01933, 21K03614, 21H04499, and by MEXT as ``Program for Promoting Researches on the Supercomputer Fugaku" (hp200124: Toward a unified view of the universe: from large scale structures to planets).

\begin{ack}
We are grateful for the fruitful discussions with Junichiro Makino. Numerical computations were carried out on the Cray XC50 at the Center for Computational Astrophysics (CfCA) in National Astronomical Observatory of Japan and Cray XC40 at YITP in Kyoto University. This research also utilized NASA's Astrophysics Data System. We would like to thank Editage (www.editage.com) for English language editing.
\end{ack}

\bibliography{sampleNotes}

\begin{thebibliography}{}
\expandafter\ifx\csname natexlab\endcsname\relax\def\natexlab#1{#1}\fi
\providecommand{\url}[1]{\href{#1}{#1}}

\bibitem[{{Abbott} {et~al.}(2019){Abbott}, {Abbott}, {Abbott}, {Abraham},
  {Acernese}, {Ackley}, {Adams}, {Adhikari}, {Adya}, {Affeldt}, \&
  et~al.}]{2019PhRvX...9c1040A}
{Abbott}, B.~P., {Abbott}, R., {Abbott}, T.~D., {et~al.} 2019, Physical Review
  X, 9, 031040

\bibitem[{{Agertz} {et~al.}(2020{\natexlab{a}}){Agertz}, {Renaud}, {Feltzing},
  {Read}, {Ryde}, {Andersson}, {Rey}, {Bensby}, \&
  {Feuillet}}]{2020arXiv200606008A}
{Agertz}, O., {Renaud}, F., {Feltzing}, S., {et~al.} 2020{\natexlab{a}},
  \mnras, submitted (arXiv:2006.06008)

\bibitem[{{Agertz} {et~al.}(2020{\natexlab{b}}){Agertz}, {Pontzen}, {Read},
  {Rey}, {Orkney}, {Rosdahl}, {Teyssier}, {Verbeke}, {Kretschmer}, \&
  {Nickerson}}]{2020MNRAS.491.1656A}
{Agertz}, O., {Pontzen}, A., {Read}, J.~I., {et~al.} 2020{\natexlab{b}},
  \mnras, 491, 1656

\bibitem[{{Allison} {et~al.}(2009){Allison}, {Goodwin}, {Parker}, {de Grijs},
  {Portegies Zwart}, \& {Kouwenhoven}}]{2009ApJ...700L..99A}
{Allison}, R.~J., {Goodwin}, S.~P., {Parker}, R.~J., {et~al.} 2009, \apjl, 700,
  L99

\bibitem[{{Andrews} \& {Wolk}(2008)}]{2008hsf1.book..390A}
{Andrews}, S.~M., \& {Wolk}, S.~J. 2008, {The LkH{\ensuremath{\alpha}} 101
  Cluster}, ed. B.~{Reipurth}, Vol.~4, 390

\bibitem[{{Aoki} {et~al.}(2014){Aoki}, {Tominaga}, {Beers}, {Honda}, \&
  {Lee}}]{2014Sci...345..912A}
{Aoki}, W., {Tominaga}, N., {Beers}, T.~C., {Honda}, S., \& {Lee}, Y.~S. 2014,
  Science, 345, 912

\bibitem[{{Applebaum} {et~al.}(2021){Applebaum}, {Brooks}, {Christensen},
  {Munshi}, {Quinn}, {Shen}, \& {Tremmel}}]{2020arXiv200811207A}
{Applebaum}, E., {Brooks}, A.~M., {Christensen}, C.~R., {et~al.} 2021, \apj,
  906, 96

\bibitem[{{Applebaum} {et~al.}(2020){Applebaum}, {Brooks}, {Quinn}, \&
  {Christensen}}]{2020MNRAS.492....8A}
{Applebaum}, E., {Brooks}, A.~M., {Quinn}, T.~R., \& {Christensen}, C.~R. 2020,
  \mnras, 492, 8

\bibitem[{{Ascenso} {et~al.}(2007){Ascenso}, {Alves}, {Vicente}, \&
  {Lago}}]{2007A&A...476..199A}
{Ascenso}, J., {Alves}, J., {Vicente}, S., \& {Lago}, M.~T.~V.~T. 2007, \aap,
  476, 199

\bibitem[{{Askar} {et~al.}(2017){Askar}, {Szkudlarek}, {Gondek-Rosi{\'n}ska},
  {Giersz}, \& {Bulik}}]{2017MNRAS.464L..36A}
{Askar}, A., {Szkudlarek}, M., {Gondek-Rosi{\'n}ska}, D., {Giersz}, M., \&
  {Bulik}, T. 2017, \mnras, 464, L36

\bibitem[{{Aspin}(2003)}]{2003AJ....125.1480A}
{Aspin}, C. 2003, \aj, 125, 1480

\bibitem[{{Asplund} {et~al.}(2009){Asplund}, {Grevesse}, {Sauval}, \&
  {Scott}}]{2009ARA&A..47..481A}
{Asplund}, M., {Grevesse}, N., {Sauval}, A.~J., \& {Scott}, P. 2009, \araa, 47,
  481

\bibitem[{{Baba} {et~al.}(2004){Baba}, {Nagata}, {Nagayama}, {Nagashima},
  {Kato}, {Kurita}, {Sato}, {Nakajima}, {Tamura}, {Nakaya}, \&
  {Sugitani}}]{2004ApJ...614..818B}
{Baba}, D., {Nagata}, T., {Nagayama}, T., {et~al.} 2004, \apj, 614, 818

\bibitem[{{Baba} {et~al.}(2006){Baba}, {Sato}, {Nagashima}, {Nishiyama},
  {Kato}, {Haba}, {Nagata}, {Nagayama}, {Tamura}, \&
  {Sugitani}}]{2006AJ....132.1692B}
{Baba}, D., {Sato}, S., {Nagashima}, C., {et~al.} 2006, \aj, 132, 1692

\bibitem[{{Barnes} \& {Hut}(1986)}]{1986Natur.324..446B}
{Barnes}, J., \& {Hut}, P. 1986, \nat, 324, 446

\bibitem[{{Bate} \& {Bonnell}(2005)}]{2005MNRAS.356.1201B}
{Bate}, M.~R., \& {Bonnell}, I.~A. 2005, \mnras, 356, 1201

\bibitem[{{Bate} {et~al.}(1995){Bate}, {Bonnell}, \&
  {Price}}]{1995MNRAS.277..362B}
{Bate}, M.~R., {Bonnell}, I.~A., \& {Price}, N.~M. 1995, \mnras, 277, 362

\bibitem[{{Bate} \& {Burkert}(1997)}]{1997MNRAS.288.1060B}
{Bate}, M.~R., \& {Burkert}, A. 1997, \mnras, 288, 1060

\bibitem[{{B{\'e}dorf} {et~al.}(2014){B{\'e}dorf}, {Gaburov}, {Fujii},
  {Nitadori}, {Ishiyama}, \& {Portegies Zwart}}]{2014hpcn.conf...54B}
{B{\'e}dorf}, J., {Gaburov}, E., {Fujii}, M.~S., {et~al.} 2014, in Proceedings
  of the International Conference for High Performance Computing, 54--65

\bibitem[{{Bland-Hawthorn} {et~al.}(2010){Bland-Hawthorn}, {Karlsson},
  {Sharma}, {Krumholz}, \& {Silk}}]{2010ApJ...721..582B}
{Bland-Hawthorn}, J., {Karlsson}, T., {Sharma}, S., {Krumholz}, M., \& {Silk},
  J. 2010, \apj, 721, 582

\bibitem[{{Bleuler} \& {Teyssier}(2014)}]{2014MNRAS.445.4015B}
{Bleuler}, A., \& {Teyssier}, R. 2014, \mnras, 445, 4015

\bibitem[{{Bohigas} \& {Tapia}(2003)}]{2003AJ....126.1861B}
{Bohigas}, J., \& {Tapia}, M. 2003, \aj, 126, 1861

\bibitem[{{Bohigas} {et~al.}(2004){Bohigas}, {Tapia}, {Roth}, \&
  {Ruiz}}]{2004AJ....127.2826B}
{Bohigas}, J., {Tapia}, M., {Roth}, M., \& {Ruiz}, M.~T. 2004, \aj, 127, 2826

\bibitem[{{Bonanos} {et~al.}(2004){Bonanos}, {Stanek}, {Udalski},
  {Wyrzykowski}, {{\.Z}ebru{\'n}}, {Kubiak}, {Szyma{\'n}ski}, {Szewczyk},
  {Pietrzy{\'n}ski}, \& {Soszy{\'n}ski}}]{2004ApJ...611L..33B}
{Bonanos}, A.~Z., {Stanek}, K.~Z., {Udalski}, A., {et~al.} 2004, \apjl, 611,
  L33

\bibitem[{{Bonatto} \& {Bica}(2009)}]{2009MNRAS.397.1915B}
{Bonatto}, C., \& {Bica}, E. 2009, \mnras, 397, 1915

\bibitem[{{Bonnell} {et~al.}(2003){Bonnell}, {Bate}, \&
  {Vine}}]{2003MNRAS.343..413B}
{Bonnell}, I.~A., {Bate}, M.~R., \& {Vine}, S.~G. 2003, \mnras, 343, 413

\bibitem[{{Bonnell} {et~al.}(2004){Bonnell}, {Vine}, \&
  {Bate}}]{2004MNRAS.349..735B}
{Bonnell}, I.~A., {Vine}, S.~G., \& {Bate}, M.~R. 2004, \mnras, 349, 735

\bibitem[{{Borissova} {et~al.}(2006){Borissova}, {Ivanov}, {Minniti}, \&
  {Geisler}}]{2006A&A...455..923B}
{Borissova}, J., {Ivanov}, V.~D., {Minniti}, D., \& {Geisler}, D. 2006, \aap,
  455, 923

\bibitem[{{Borissova} {et~al.}(2005){Borissova}, {Ivanov}, {Minniti},
  {Geisler}, \& {Stephens}}]{2005A&A...435...95B}
{Borissova}, J., {Ivanov}, V.~D., {Minniti}, D., {Geisler}, D., \& {Stephens},
  A.~W. 2005, \aap, 435, 95

\bibitem[{{Borissova} {et~al.}(2003){Borissova}, {Pessev}, {Ivanov}, {Saviane},
  {Kurtev}, \& {Ivanov}}]{2003A&A...411...83B}
{Borissova}, J., {Pessev}, P., {Ivanov}, V.~D., {et~al.} 2003, \aap, 411, 83

\bibitem[{{Bouy} {et~al.}(2009){Bouy}, {Hu{\'e}lamo}, {Mart{\'\i}n}, {Marchis},
  {Barrado Y Navascu{\'e}s}, {Kolb}, {Marchetti}, {Petr-Gotzens}, {Sterzik},
  {Ivanov}, {K{\"o}hler}, \& {N{\"u}rnberger}}]{2009A&A...493..931B}
{Bouy}, H., {Hu{\'e}lamo}, N., {Mart{\'\i}n}, E.~L., {et~al.} 2009, \aap, 493,
  931

\bibitem[{{Bressan} {et~al.}(1993){Bressan}, {Fagotto}, {Bertelli}, \&
  {Chiosi}}]{1993A&AS..100..647B}
{Bressan}, A., {Fagotto}, F., {Bertelli}, G., \& {Chiosi}, C. 1993, \aaps, 100,
  647

\bibitem[{{Brunt}(2015)}]{2015MNRAS.449.4465B}
{Brunt}, C.~M. 2015, \mnras, 449, 4465

\bibitem[{{Carpenter} {et~al.}(1997){Carpenter}, {Meyer}, {Dougados}, {Strom},
  \& {Hillenbrand}}]{1997AJ....114..198C}
{Carpenter}, J.~M., {Meyer}, M.~R., {Dougados}, C., {Strom}, S.~E., \&
  {Hillenbrand}, L.~A. 1997, \aj, 114, 198

\bibitem[{{Carpenter} {et~al.}(1990){Carpenter}, {Snell}, \&
  {Schloerb}}]{1990ApJ...362..147C}
{Carpenter}, J.~M., {Snell}, R.~L., \& {Schloerb}, F.~P. 1990, \apj, 362, 147

\bibitem[{{Carpenter} {et~al.}(1993){Carpenter}, {Snell}, {Schloerb}, \&
  {Skrutskie}}]{1993ApJ...407..657C}
{Carpenter}, J.~M., {Snell}, R.~L., {Schloerb}, F.~P., \& {Skrutskie}, M.~F.
  1993, \apj, 407, 657

\bibitem[{{Chabrier}(2003)}]{2003PASP..115..763C}
{Chabrier}, G. 2003, \pasp, 115, 763

\bibitem[{{Chen} {et~al.}(2007){Chen}, {de Grijs}, \&
  {Zhao}}]{2007AJ....134.1368C}
{Chen}, L., {de Grijs}, R., \& {Zhao}, J.~L. 2007, \aj, 134, 1368

\bibitem[{{Cichowolski} {et~al.}(2009){Cichowolski}, {Romero}, {Ortega},
  {Cappa}, \& {Vasquez}}]{2009MNRAS.394..900C}
{Cichowolski}, S., {Romero}, G.~A., {Ortega}, M.~E., {Cappa}, C.~E., \&
  {Vasquez}, J. 2009, \mnras, 394, 900

\bibitem[{{Clark} {et~al.}(2005){Clark}, {Bonnell}, {Zinnecker}, \&
  {Bate}}]{2005MNRAS.359..809C}
{Clark}, P.~C., {Bonnell}, I.~A., {Zinnecker}, H., \& {Bate}, M.~R. 2005,
  \mnras, 359, 809

\bibitem[{{Col{\'\i}n} {et~al.}(2013){Col{\'\i}n}, {V{\'a}zquez-Semadeni}, \&
  {G{\'o}mez}}]{2013MNRAS.435.1701C}
{Col{\'\i}n}, P., {V{\'a}zquez-Semadeni}, E., \& {G{\'o}mez}, G.~C. 2013,
  \mnras, 435, 1701

\bibitem[{{Corbelli} {et~al.}(2018){Corbelli}, {Elmegreen}, {Braine}, \&
  {Thilker}}]{2018A&A...617A.125C}
{Corbelli}, E., {Elmegreen}, B.~G., {Braine}, J., \& {Thilker}, D. 2018, \aap,
  617, A125

\bibitem[{{Covey} {et~al.}(2010){Covey}, {Lada}, {Rom{\'a}n-Z{\'u}{\~n}iga},
  {Muench}, {Forbrich}, \& {Ascenso}}]{2010ApJ...722..971C}
{Covey}, K.~R., {Lada}, C.~J., {Rom{\'a}n-Z{\'u}{\~n}iga}, C., {et~al.} 2010,
  \apj, 722, 971

\bibitem[{{Dahm}(2008)}]{2008hsf1.book..966D}
{Dahm}, S.~E. 2008, {The Young Cluster and Star Forming Region NGC 2264}, ed.
  B.~{Reipurth}, Vol.~4, 966

\bibitem[{{Dahm} \& {Hillenbrand}(2007)}]{2007AJ....133.2072D}
{Dahm}, S.~E., \& {Hillenbrand}, L.~A. 2007, \aj, 133, 2072

\bibitem[{{Dale} {et~al.}(2012){Dale}, {Ercolano}, \&
  {Bonnell}}]{2012MNRAS.424..377D}
{Dale}, J.~E., {Ercolano}, B., \& {Bonnell}, I.~A. 2012, \mnras, 424, 377

\bibitem[{{Dale} {et~al.}(2014){Dale}, {Ngoumou}, {Ercolano}, \&
  {Bonnell}}]{2014MNRAS.442..694D}
{Dale}, J.~E., {Ngoumou}, J., {Ercolano}, B., \& {Bonnell}, I.~A. 2014, \mnras,
  442, 694

\bibitem[{{Dalla Vecchia} \& {Schaye}(2012)}]{2012MNRAS.426..140D}
{Dalla Vecchia}, C., \& {Schaye}, J. 2012, \mnras, 426, 140

\bibitem[{{Dambis}(1999)}]{1999AstL...25...10D}
{Dambis}, A.~K. 1999, Astronomy Letters, 25, 10

\bibitem[{{Damiani} {et~al.}(2004){Damiani}, {Flaccomio}, {Micela},
  {Sciortino}, {Harnden}, \& {Murray}}]{2004ApJ...608..781D}
{Damiani}, F., {Flaccomio}, E., {Micela}, G., {et~al.} 2004, \apj, 608, 781

\bibitem[{{Damiani} {et~al.}(2006){Damiani}, {Prisinzano}, {Micela}, \&
  {Sciortino}}]{2006A&A...459..477D}
{Damiani}, F., {Prisinzano}, L., {Micela}, G., \& {Sciortino}, S. 2006, \aap,
  459, 477

\bibitem[{{Davies} {et~al.}(2012){Davies}, {Clark}, {Trombley}, {Figer},
  {Najarro}, {Crowther}, {Kudritzki}, {Thompson}, {Urquhart}, \&
  {Hindson}}]{2012MNRAS.419.1871D}
{Davies}, B., {Clark}, J.~S., {Trombley}, C., {et~al.} 2012, \mnras, 419, 1871

\bibitem[{{De Silva} {et~al.}(2007){De Silva}, {Freeman}, {Asplund},
  {Bland-Hawthorn}, {Bessell}, \& {Collet}}]{2007AJ....133.1161D}
{De Silva}, G.~M., {Freeman}, K.~C., {Asplund}, M., {et~al.} 2007, \aj, 133,
  1161

\bibitem[{{de Wit} {et~al.}(2004){de Wit}, {Testi}, {Palla}, {Vanzi}, \&
  {Zinnecker}}]{2004A&A...425..937D}
{de Wit}, W.~J., {Testi}, L., {Palla}, F., {Vanzi}, L., \& {Zinnecker}, H.
  2004, \aap, 425, 937

\bibitem[{{de Wit} {et~al.}(2005){de Wit}, {Testi}, {Palla}, \&
  {Zinnecker}}]{2005A&A...437..247D}
{de Wit}, W.~J., {Testi}, L., {Palla}, F., \& {Zinnecker}, H. 2005, \aap, 437,
  247

\bibitem[{{Depoy} {et~al.}(1990){Depoy}, {Lada}, {Gatley}, \&
  {Probst}}]{1990ApJ...356L..55D}
{Depoy}, D.~L., {Lada}, E.~A., {Gatley}, I., \& {Probst}, R. 1990, \apjl, 356,
  L55

\bibitem[{{Di Carlo} {et~al.}(2019){Di Carlo}, {Giacobbo}, {Mapelli},
  {Pasquato}, {Spera}, {Wang}, \& {Haardt}}]{2019MNRAS.487.2947D}
{Di Carlo}, U.~N., {Giacobbo}, N., {Mapelli}, M., {et~al.} 2019, \mnras, 487,
  2947

\bibitem[{{Dobbs} {et~al.}(2020){Dobbs}, {Liow}, \&
  {Rieder}}]{2020MNRAS.tmpL..65D}
{Dobbs}, C.~L., {Liow}, K.~Y., \& {Rieder}, S. 2020, \mnras, 496, L1

\bibitem[{{Emerick} {et~al.}(2019){Emerick}, {Bryan}, \& {Mac
  Low}}]{2019MNRAS.482.1304E}
{Emerick}, A., {Bryan}, G.~L., \& {Mac Low}, M.-M. 2019, \mnras, 482, 1304

\bibitem[{{Fagotto} {et~al.}(1994{\natexlab{a}}){Fagotto}, {Bressan},
  {Bertelli}, \& {Chiosi}}]{1994A&AS..105...29F}
{Fagotto}, F., {Bressan}, A., {Bertelli}, G., \& {Chiosi}, C.
  1994{\natexlab{a}}, \aaps, 105, 29

\bibitem[{{Fagotto} {et~al.}(1994{\natexlab{b}}){Fagotto}, {Bressan},
  {Bertelli}, \& {Chiosi}}]{1994A&AS..104..365F}
---. 1994{\natexlab{b}}, \aaps, 104, 365

\bibitem[{{Faustini} {et~al.}(2009){Faustini}, {Molinari}, {Testi}, \&
  {Brand}}]{2009A&A...503..801F}
{Faustini}, F., {Molinari}, S., {Testi}, L., \& {Brand}, J. 2009, \aap, 503,
  801

\bibitem[{{Federrath} {et~al.}(2010){Federrath}, {Banerjee}, {Clark}, \&
  {Klessen}}]{2010ApJ...713..269F}
{Federrath}, C., {Banerjee}, R., {Clark}, P.~C., \& {Klessen}, R.~S. 2010,
  \apj, 713, 269

\bibitem[{{Ferland} {et~al.}(1998){Ferland}, {Korista}, {Verner}, {Ferguson},
  {Kingdon}, \& {Verner}}]{1998PASP..110..761F}
{Ferland}, G.~J., {Korista}, K.~T., {Verner}, D.~A., {et~al.} 1998, \pasp, 110,
  761

\bibitem[{{Ferland} {et~al.}(2013){Ferland}, {Porter}, {van Hoof}, {Williams},
  {Abel}, {Lykins}, {Shaw}, {Henney}, \& {Stancil}}]{2013RMxAA..49..137F}
{Ferland}, G.~J., {Porter}, R.~L., {van Hoof}, P.~A.~M., {et~al.} 2013, Rev.
  Mex. Astron. Astrofis., 49, 137

\bibitem[{{Ferland} {et~al.}(2017){Ferland}, {Chatzikos}, {Guzm{\'a}n},
  {Lykins}, {van Hoof}, {Williams}, {Abel}, {Badnell}, {Keenan}, {Porter}, \&
  {Stancil}}]{2017RMxAA..53..385F}
{Ferland}, G.~J., {Chatzikos}, M., {Guzm{\'a}n}, F., {et~al.} 2017, Rev. Mex.
  Astron. Astrofis., 53, 385

\bibitem[{{Figer} {et~al.}(2002){Figer}, {Najarro}, {Gilmore}, {Morris}, {Kim},
  {Serabyn}, {McLean}, {Gilbert}, {Graham}, {Larkin}, {Levenson}, \&
  {Teplitz}}]{2002ApJ...581..258F}
{Figer}, D.~F., {Najarro}, F., {Gilmore}, D., {et~al.} 2002, \apj, 581, 258

\bibitem[{{Font} {et~al.}(2020){Font}, {McCarthy}, {Poole-Mckenzie},
  {Stafford}, {Brown}, {Schaye}, {Crain}, {Theuns}, \&
  {Schaller}}]{2020MNRAS.498.1765F}
{Font}, A.~S., {McCarthy}, I.~G., {Poole-Mckenzie}, R., {et~al.} 2020, \mnras,
  498, 1765

\bibitem[{{Forte} \& {Orsatti}(1984)}]{1984ApJS...56..211F}
{Forte}, J.~C., \& {Orsatti}, A.~M. 1984, \apjs, 56, 211

\bibitem[{{Froebrich} {et~al.}(2008){Froebrich}, {Meusinger}, \&
  {Scholz}}]{2008MNRAS.390.1598F}
{Froebrich}, D., {Meusinger}, H., \& {Scholz}, A. 2008, \mnras, 390, 1598

\bibitem[{{Froebrich} \& {Rowles}(2010)}]{2010MNRAS.406.1350F}
{Froebrich}, D., \& {Rowles}, J. 2010, \mnras, 406, 1350

\bibitem[{{Froebrich} {et~al.}(2007){Froebrich}, {Scholz}, \&
  {Raftery}}]{2007MNRAS.374..399F}
{Froebrich}, D., {Scholz}, A., \& {Raftery}, C.~L. 2007, \mnras, 374, 399

\bibitem[{{Fujii} {et~al.}(2007){Fujii}, {Iwasawa}, {Funato}, \&
  {Makino}}]{2007PASJ...59.1095F}
{Fujii}, M., {Iwasawa}, M., {Funato}, Y., \& {Makino}, J. 2007, \pasj, 59, 1095

\bibitem[{{Fujii}(2015)}]{2015PASJ...67...59F}
{Fujii}, M.~S. 2015, \pasj, 67, 59

\bibitem[{{Fujii} \& {Portegies Zwart}(2015)}]{2015MNRAS.449..726F}
{Fujii}, M.~S., \& {Portegies Zwart}, S. 2015, \mnras, 449, 726

\bibitem[{{Fujii} \& {Portegies Zwart}(2016)}]{2016ApJ...817....4F}
---. 2016, \apj, 817, 4

\bibitem[{{Fujii} {et~al.}(2021{\natexlab{a}}){Fujii}, {Saitoh}, {Hirai}, \&
  {Wang}}]{2021arXiv210302829F}
{Fujii}, M.~S., {Saitoh}, T.~R., {Hirai}, Y., \& {Wang}, L. 2021{\natexlab{a}},
  \pasj, submitted (arXiv:2103.02829)

\bibitem[{{Fujii} {et~al.}(2021{\natexlab{b}}){Fujii}, {Saitoh}, {Wang}, \&
  {Hirai}}]{FujiiASURABRIDGE}
{Fujii}, M.~S., {Saitoh}, T.~R., {Wang}, L., \& {Hirai}, Y. 2021{\natexlab{b}},
  \pasj, in press (arXiv:2101.05934)

\bibitem[{{Fujii} {et~al.}(2017){Fujii}, {Tanikawa}, \&
  {Makino}}]{2017PASJ...69...94F}
{Fujii}, M.~S., {Tanikawa}, A., \& {Makino}, J. 2017, \pasj, 69, 94

\bibitem[{{Fukushima} {et~al.}(2020){Fukushima}, {Yajima}, {Sugimura},
  {Hosokawa}, {Omukai}, \& {Matsumoto}}]{2020MNRAS.497.3830F}
{Fukushima}, H., {Yajima}, H., {Sugimura}, K., {et~al.} 2020, \mnras, 497, 3830

\bibitem[{{Gaia Collaboration} {et~al.}(2018){Gaia Collaboration}, {Brown},
  {Vallenari}, {Prusti}, {de Bruijne}, {Babusiaux}, {Bailer-Jones}, {Biermann},
  {Evans}, {Eyer}, {Jansen}, {Jordi}, {Klioner}, {Lammers}, {Lindegren},
  {Luri}, {Mignard}, {Panem}, {Pourbaix}, {Randich}, {Sartoretti}, {Siddiqui},
  {Soubiran}, {van Leeuwen}, {Walton}, {Arenou}, {Bastian}, {Cropper},
  {Drimmel}, {Katz}, {Lattanzi}, {Bakker}, {Cacciari}, {Casta{\~n}eda},
  {Chaoul}, {Cheek}, {De Angeli}, {Fabricius}, {Guerra}, {Holl}, {Masana},
  {Messineo}, {Mowlavi}, {Nienartowicz}, {Panuzzo}, {Portell}, {Riello},
  {Seabroke}, {Tanga}, {Th{\'e}venin}, {Gracia-Abril}, {Comoretto},
  {Garcia-Reinaldos}, {Teyssier}, {Altmann}, {Andrae}, {Audard},
  {Bellas-Velidis}, {Benson}, {Berthier}, {Blomme}, {Burgess}, {Busso},
  {Carry}, {Cellino}, {Clementini}, {Clotet}, {Creevey}, {Davidson}, {De
  Ridder}, {Delchambre}, {Dell'Oro}, {Ducourant},
  {Fern{\'a}ndez-Hern{\'a}ndez}, {Fouesneau}, {Fr{\'e}mat}, {Galluccio},
  {Garc{\'\i}a-Torres}, {Gonz{\'a}lez-N{\'u}{\~n}ez}, {Gonz{\'a}lez-Vidal},
  {Gosset}, {Guy}, {Halbwachs}, {Hambly}, {Harrison}, {Hern{\'a}ndez},
  {Hestroffer}, {Hodgkin}, {Hutton}, {Jasniewicz}, {Jean-Antoine-Piccolo},
  {Jordan}, {Korn}, {Krone-Martins}, {Lanzafame}, {Lebzelter}, {L{\"o}ffler},
  {Manteiga}, {Marrese}, {Mart{\'\i}n-Fleitas}, {Moitinho}, {Mora}, {Muinonen},
  {Osinde}, {Pancino}, {Pauwels}, {Petit}, {Recio-Blanco}, {Richards},
  {Rimoldini}, {Robin}, {Sarro}, {Siopis}, {Smith}, {Sozzetti}, {S{\"u}veges},
  {Torra}, {van Reeven}, {Abbas}, {Abreu Aramburu}, {Accart}, {Aerts},
  {Altavilla}, {{\'A}lvarez}, {Alvarez}, {Alves}, {Anderson}, {Andrei},
  {Anglada Varela}, {Antiche}, {Antoja}, {Arcay}, {Astraatmadja}, {Bach},
  {Baker}, {Balaguer-N{\'u}{\~n}ez}, {Balm}, {Barache}, {Barata}, {Barbato},
  {Barblan}, {Barklem}, {Barrado}, {Barros}, {Barstow}, {Bartholom{\'e}
  Mu{\~n}oz}, {Bassilana}, {Becciani}, {Bellazzini}, {Berihuete}, {Bertone},
  {Bianchi}, {Bienaym{\'e}}, {Blanco-Cuaresma}, {Boch}, {Boeche}, {Bombrun},
  {Borrachero}, {Bossini}, {Bouquillon}, {Bourda}, {Bragaglia}, {Bramante},
  {Breddels}, {Bressan}, {Brouillet}, {Br{\"u}semeister}, {Brugaletta},
  {Bucciarelli}, {Burlacu}, {Busonero}, {Butkevich}, {Buzzi}, {Caffau},
  {Cancelliere}, {Cannizzaro}, {Cantat-Gaudin}, {Carballo}, {Carlucci},
  {Carrasco}, {Casamiquela}, {Castellani}, {Castro-Ginard}, {Charlot},
  {Chemin}, {Chiavassa}, {Cocozza}, {Costigan}, {Cowell}, {Crifo}, {Crosta},
  {Crowley}, {Cuypers}, {Dafonte}, {Damerdji}, {Dapergolas}, {David}, {David},
  {de Laverny}, {De Luise}, {De March}, {de Martino}, {de Souza}, {de Torres},
  {Debosscher}, {del Pozo}, {Delbo}, {Delgado}, {Delgado}, {Di Matteo},
  {Diakite}, {Diener}, {Distefano}, {Dolding}, {Drazinos}, {Dur{\'a}n},
  {Edvardsson}, {Enke}, {Eriksson}, {Esquej}, {Eynard Bontemps}, {Fabre},
  {Fabrizio}, {Faigler}, {Falc{\~a}o}, {Farr{\`a}s Casas}, {Federici},
  {Fedorets}, {Fernique}, {Figueras}, {Filippi}, {Findeisen}, {Fonti},
  {Fraile}, {Fraser}, {Fr{\'e}zouls}, {Gai}, {Galleti}, {Garabato},
  {Garc{\'\i}a-Sedano}, {Garofalo}, {Garralda}, {Gavel}, {Gavras}, {Gerssen},
  {Geyer}, {Giacobbe}, {Gilmore}, {Girona}, {Giuffrida}, {Glass}, {Gomes},
  {Granvik}, {Gueguen}, {Guerrier}, {Guiraud}, {Guti{\'e}rrez-S{\'a}nchez},
  {Haigron}, {Hatzidimitriou}, {Hauser}, {Haywood}, {Heiter}, {Helmi}, {Heu},
  {Hilger}, {Hobbs}, {Hofmann}, {Holland}, {Huckle}, {Hypki}, {Icardi},
  {Jan{\ss}en}, {Jevardat de Fombelle}, {Jonker}, {Juh{\'a}sz}, {Julbe},
  {Karampelas}, {Kewley}, {Klar}, {Kochoska}, {Kohley}, {Kolenberg},
  {Kontizas}, {Kontizas}, {Koposov}, {Kordopatis}, {Kostrzewa-Rutkowska},
  {Koubsky}, {Lambert}, {Lanza}, {Lasne}, {Lavigne}, {Le Fustec}, {Le
  Poncin-Lafitte}, {Lebreton}, {Leccia}, {Leclerc}, {Lecoeur-Taibi},
  {Lenhardt}, {Leroux}, {Liao}, {Licata}, {Lindstr{\o}m}, {Lister}, {Livanou},
  {Lobel}, {L{\'o}pez}, {Managau}, {Mann}, {Mantelet}, {Marchal}, {Marchant},
  {Marconi}, {Marinoni}, {Marschalk{\'o}}, {Marshall}, {Martino}, {Marton},
  {Mary}, {Massari}, {Matijevi{\v{c}}}, {Mazeh}, {McMillan}, {Messina},
  {Michalik}, {Millar}, {Molina}, {Molinaro}, {Moln{\'a}r}, {Montegriffo},
  {Mor}, {Morbidelli}, {Morel}, {Morris}, {Mulone}, {Muraveva}, {Musella},
  {Nelemans}, {Nicastro}, {Noval}, {O'Mullane}, {Ord{\'e}novic},
  {Ord{\'o}{\~n}ez-Blanco}, {Osborne}, {Pagani}, {Pagano}, {Pailler},
  {Palacin}, {Palaversa}, {Panahi}, {Pawlak}, {Piersimoni}, {Pineau}, {Plachy},
  {Plum}, {Poggio}, {Poujoulet}, {Pr{\v{s}}a}, {Pulone}, {Racero}, {Ragaini},
  {Rambaux}, {Ramos-Lerate}, {Regibo}, {Reyl{\'e}}, {Riclet}, {Ripepi}, {Riva},
  {Rivard}, {Rixon}, {Roegiers}, {Roelens}, {Romero-G{\'o}mez}, {Rowell},
  {Royer}, {Ruiz-Dern}, {Sadowski}, {Sagrist{\`a} Sell{\'e}s}, {Sahlmann},
  {Salgado}, {Salguero}, {Sanna}, {Santana-Ros}, {Sarasso}, {Savietto},
  {Schultheis}, {Sciacca}, {Segol}, {Segovia}, {S{\'e}gransan}, {Shih},
  {Siltala}, {Silva}, {Smart}, {Smith}, {Solano}, {Solitro}, {Sordo}, {Soria
  Nieto}, {Souchay}, {Spagna}, {Spoto}, {Stampa}, {Steele},
  {Steidelm{\"u}ller}, {Stephenson}, {Stoev}, {Suess}, {Surdej}, {Szabados},
  {Szegedi-Elek}, {Tapiador}, {Taris}, {Tauran}, {Taylor}, {Teixeira},
  {Terrett}, {Teyssand ier}, {Thuillot}, {Titarenko}, {Torra Clotet}, {Turon},
  {Ulla}, {Utrilla}, {Uzzi}, {Vaillant}, {Valentini}, {Valette}, {van Elteren},
  {Van Hemelryck}, {van Leeuwen}, {Vaschetto}, {Vecchiato}, {Veljanoski},
  {Viala}, {Vicente}, {Vogt}, {von Essen}, {Voss}, {Votruba}, {Voutsinas},
  {Walmsley}, {Weiler}, {Wertz}, {Wevers}, {Wyrzykowski}, {Yoldas},
  {{\v{Z}}erjal}, {Ziaeepour}, {Zorec}, {Zschocke}, {Zucker}, {Zurbach}, \&
  {Zwitter}}]{2018A&A...616A...1G}
{Gaia Collaboration}, {Brown}, A.~G.~A., {Vallenari}, A., {et~al.} 2018, \aap,
  616, A1

\bibitem[{{Garmany} \& {Walborn}(1987)}]{1987PASP...99..240G}
{Garmany}, C.~D., \& {Walborn}, N.~R. 1987, \pasp, 99, 240

\bibitem[{{Gatto} {et~al.}(2017){Gatto}, {Walch}, {Naab}, {Girichidis},
  {W{\"u}nsch}, {Glover}, {Klessen}, {Clark}, {Peters}, {Derigs}, {Baczynski},
  \& {Puls}}]{2017MNRAS.466.1903G}
{Gatto}, A., {Walch}, S., {Naab}, T., {et~al.} 2017, \mnras, 466, 1903

\bibitem[{{Getman} {et~al.}(2002){Getman}, {Feigelson}, {Townsley}, {Bally},
  {Lada}, \& {Reipurth}}]{2002ApJ...575..354G}
{Getman}, K.~V., {Feigelson}, E.~D., {Townsley}, L., {et~al.} 2002, \apj, 575,
  354

\bibitem[{{Gill} {et~al.}(2004){Gill}, {Knebe}, \&
  {Gibson}}]{2004MNRAS.351..399G}
{Gill}, S.~P.~D., {Knebe}, A., \& {Gibson}, B.~K. 2004, \mnras, 351, 399

\bibitem[{{Grand} {et~al.}(2017){Grand}, {G{\'o}mez}, {Marinacci}, {Pakmor},
  {Springel}, {Campbell}, {Frenk}, {Jenkins}, \& {White}}]{Auriga}
{Grand}, R. J.~J., {G{\'o}mez}, F.~A., {Marinacci}, F., {et~al.} 2017, \mnras,
  467, 179

\bibitem[{{Gutcke} {et~al.}(2021){Gutcke}, {Pakmor}, {Naab}, \&
  {Springel}}]{2020arXiv201007311G}
{Gutcke}, T.~A., {Pakmor}, R., {Naab}, T., \& {Springel}, V. 2021, \mnras, 501,
  5597

\bibitem[{{Gutermuth} {et~al.}(2004){Gutermuth}, {Megeath}, {Muzerolle},
  {Allen}, {Pipher}, {Myers}, \& {Fazio}}]{2004ApJS..154..374G}
{Gutermuth}, R.~A., {Megeath}, S.~T., {Muzerolle}, J., {et~al.} 2004, \apjs,
  154, 374

\bibitem[{{Gutermuth} {et~al.}(2008){Gutermuth}, {Myers}, {Megeath}, {Allen},
  {Pipher}, {Muzerolle}, {Porras}, {Winston}, \& {Fazio}}]{2008ApJ...674..336G}
{Gutermuth}, R.~A., {Myers}, P.~C., {Megeath}, S.~T., {et~al.} 2008, \apj, 674,
  336

\bibitem[{{Haardt} \& {Madau}(2012)}]{2012ApJ...746..125H}
{Haardt}, F., \& {Madau}, P. 2012, \apj, 746, 125

\bibitem[{{Hahn} \& {Abel}(2011)}]{2011MNRAS.415.2101H}
{Hahn}, O., \& {Abel}, T. 2011, \mnras, 415, 2101

\bibitem[{{Haisch} {et~al.}(2000){Haisch}, {Lada}, \&
  {Lada}}]{2000AJ....120.1396H}
{Haisch}, Karl~E., J., {Lada}, E.~A., \& {Lada}, C.~J. 2000, \aj, 120, 1396

\bibitem[{{Harayama} {et~al.}(2008){Harayama}, {Eisenhauer}, \&
  {Martins}}]{2008ApJ...675.1319H}
{Harayama}, Y., {Eisenhauer}, F., \& {Martins}, F. 2008, \apj, 675, 1319

\bibitem[{{Hasan} {et~al.}(2008){Hasan}, {Hasan}, \&
  {Shah}}]{2008Ap&SS.318...25H}
{Hasan}, P., {Hasan}, S.~N., \& {Shah}, U. 2008, \apss, 318, 25

\bibitem[{{He} {et~al.}(2019){He}, {Ricotti}, \& {Geen}}]{2019MNRAS.489.1880H}
{He}, C.-C., {Ricotti}, M., \& {Geen}, S. 2019, \mnras, 489, 1880

\bibitem[{{Heske} \& {Wendker}(1984)}]{1984A&AS...57..205H}
{Heske}, A., \& {Wendker}, H.~J. 1984, \aaps, 57, 205

\bibitem[{{Hillenbrand} \& {Hartmann}(1998)}]{1998ApJ...492..540H}
{Hillenbrand}, L.~A., \& {Hartmann}, L.~W. 1998, \apj, 492, 540

\bibitem[{{Hillenbrand} {et~al.}(1998){Hillenbrand}, {Strom}, {Calvet},
  {Merrill}, {Gatley}, {Makidon}, {Meyer}, \&
  {Skrutskie}}]{1998AJ....116.1816H}
{Hillenbrand}, L.~A., {Strom}, S.~E., {Calvet}, N., {et~al.} 1998, \aj, 116,
  1816

\bibitem[{{Hirai} {et~al.}(2015){Hirai}, {Ishimaru}, {Saitoh}, {Fujii},
  {Hidaka}, \& {Kajino}}]{2015ApJ...814...41H}
{Hirai}, Y., {Ishimaru}, Y., {Saitoh}, T.~R., {et~al.} 2015, \apj, 814, 41

\bibitem[{{Hirai} {et~al.}(2017){Hirai}, {Ishimaru}, {Saitoh}, {Fujii},
  {Hidaka}, \& {Kajino}}]{2017MNRAS.466.2474H}
---. 2017, \mnras, 466, 2474

\bibitem[{{Hirai} \& {Saitoh}(2017)}]{2017ApJ...838L..23H}
{Hirai}, Y., \& {Saitoh}, T.~R. 2017, \apjl, 838, L23

\bibitem[{{Hirai} {et~al.}(2018){Hirai}, {Saitoh}, {Ishimaru}, \&
  {Wanajo}}]{2018ApJ...855...63H}
{Hirai}, Y., {Saitoh}, T.~R., {Ishimaru}, Y., \& {Wanajo}, S. 2018, \apj, 855,
  63

\bibitem[{{Hirai} {et~al.}(2019){Hirai}, {Wanajo}, \&
  {Saitoh}}]{2019ApJ...885...33H}
{Hirai}, Y., {Wanajo}, S., \& {Saitoh}, T.~R. 2019, \apj, 885, 33

\bibitem[{{Hopkins} {et~al.}(2011){Hopkins}, {Quataert}, \&
  {Murray}}]{2011MNRAS.417..950H}
{Hopkins}, P.~F., {Quataert}, E., \& {Murray}, N. 2011, \mnras, 417, 950

\bibitem[{{Hopkins} {et~al.}(2018{\natexlab{a}}){Hopkins}, {Wetzel},
  {Kere{\v{s}}}, {Faucher-Gigu{\`e}re}, {Quataert}, {Boylan-Kolchin}, {Murray},
  {Hayward}, \& {El-Badry}}]{2018MNRAS.477.1578H}
{Hopkins}, P.~F., {Wetzel}, A., {Kere{\v{s}}}, D., {et~al.} 2018{\natexlab{a}},
  \mnras, 477, 1578

\bibitem[{{Hopkins} {et~al.}(2018{\natexlab{b}}){Hopkins}, {Wetzel},
  {Kere{\v{s}}}, {Faucher-Gigu{\`e}re}, {Quataert}, {Boylan-Kolchin}, {Murray},
  {Hayward}, {Garrison-Kimmel}, {Hummels}, {Feldmann}, {Torrey}, {Ma},
  {Angl{\'e}s-Alc{\'a}zar}, {Su}, {Orr}, {Schmitz}, {Escala}, {Sanderson},
  {Grudi{\'c}}, {Hafen}, {Kim}, {Fitts}, {Bullock}, {Wheeler}, {Chan},
  {Elbert}, \& {Narayanan}}]{FIRE2}
---. 2018{\natexlab{b}}, \mnras, 480, 800

\bibitem[{{Howard} {et~al.}(2014){Howard}, {Pudritz}, \&
  {Harris}}]{2014MNRAS.438.1305H}
{Howard}, C.~S., {Pudritz}, R.~E., \& {Harris}, W.~E. 2014, \mnras, 438, 1305

\bibitem[{{Howard} {et~al.}(2018){Howard}, {Pudritz}, \&
  {Harris}}]{2018NatAs...2..725H}
---. 2018, Nature Astronomy, 2, 725

\bibitem[{{Hu}(2019)}]{2019MNRAS.483.3363H}
{Hu}, C.-Y. 2019, \mnras, 483, 3363

\bibitem[{{Hu} {et~al.}(2017){Hu}, {Naab}, {Glover}, {Walch}, \&
  {Clark}}]{2017MNRAS.471.2151H}
{Hu}, C.-Y., {Naab}, T., {Glover}, S. C.~O., {Walch}, S., \& {Clark}, P.~C.
  2017, \mnras, 471, 2151

\bibitem[{{Hubber} {et~al.}(2013{\natexlab{a}}){Hubber}, {Allison}, {Smith}, \&
  {Goodwin}}]{2013MNRAS.430.1599H}
{Hubber}, D.~A., {Allison}, R.~J., {Smith}, R., \& {Goodwin}, S.~P.
  2013{\natexlab{a}}, \mnras, 430, 1599

\bibitem[{{Hubber} {et~al.}(2013{\natexlab{b}}){Hubber}, {Walch}, \&
  {Whitworth}}]{2013MNRAS.430.3261H}
{Hubber}, D.~A., {Walch}, S., \& {Whitworth}, A.~P. 2013{\natexlab{b}}, \mnras,
  430, 3261

\bibitem[{{Huff} \& {Stahler}(2006)}]{2006ApJ...644..355H}
{Huff}, E.~M., \& {Stahler}, S.~W. 2006, \apj, 644, 355

\bibitem[{{Jappsen} {et~al.}(2005){Jappsen}, {Klessen}, {Larson}, {Li}, \& {Mac
  Low}}]{2005A&A...435..611J}
{Jappsen}, A.~K., {Klessen}, R.~S., {Larson}, R.~B., {Li}, Y., \& {Mac Low},
  M.~M. 2005, \aap, 435, 611

\bibitem[{{Jaupart} \& {Chabrier}(2020)}]{2020ApJ...903L...2J}
{Jaupart}, E., \& {Chabrier}, G. 2020, \apjl, 903, L2

\bibitem[{{Kaas} {et~al.}(2004){Kaas}, {Olofsson}, {Bontemps}, {Andr{\'e}},
  {Nordh}, {Huldtgren}, {Prusti}, {Persi}, {Delgado}, {Motte}, {Abergel},
  {Boulanger}, {Burgdorf}, {Casali}, {Cesarsky}, {Davies}, {Falgarone},
  {Montmerle}, {Perault}, {Puget}, \& {Sibille}}]{2004A&A...421..623K}
{Kaas}, A.~A., {Olofsson}, G., {Bontemps}, S., {et~al.} 2004, \aap, 421, 623

\bibitem[{{Kainulainen} {et~al.}(2011){Kainulainen}, {Beuther}, {Banerjee},
  {Federrath}, \& {Henning}}]{2011A&A...530A..64K}
{Kainulainen}, J., {Beuther}, H., {Banerjee}, R., {Federrath}, C., \&
  {Henning}, T. 2011, \aap, 530, A64

\bibitem[{{Kainulainen} {et~al.}(2009){Kainulainen}, {Beuther}, {Henning}, \&
  {Plume}}]{2009A&A...508L..35K}
{Kainulainen}, J., {Beuther}, H., {Henning}, T., \& {Plume}, R. 2009, \aap,
  508, L35

\bibitem[{{Katz}(1992)}]{1992ApJ...391..502K}
{Katz}, N. 1992, \apj, 391, 502

\bibitem[{{Kennicutt}(1989)}]{1989ApJ...344..685K}
{Kennicutt}, Robert~C., J. 1989, \apj, 344, 685

\bibitem[{{Kennicutt}(1998)}]{1998ApJ...498..541K}
---. 1998, \apj, 498, 541

\bibitem[{{Kim} \& {Ostriker}(2017)}]{2017ApJ...846..133K}
{Kim}, C.-G., \& {Ostriker}, E.~C. 2017, \apj, 846, 133

\bibitem[{{Kim} {et~al.}(2018{\natexlab{a}}){Kim}, {Kim}, \&
  {Ostriker}}]{2018ApJ...859...68K}
{Kim}, J.-G., {Kim}, W.-T., \& {Ostriker}, E.~C. 2018{\natexlab{a}}, \apj, 859,
  68

\bibitem[{{Kim} {et~al.}(2018{\natexlab{b}}){Kim}, {Ma}, {Grudi{\'c}},
  {Hopkins}, {Hayward}, {Wetzel}, {Faucher-Gigu{\`e}re}, {Kere{\v{s}}},
  {Garrison-Kimmel}, \& {Murray}}]{2018MNRAS.474.4232K}
{Kim}, J.-h., {Ma}, X., {Grudi{\'c}}, M.~Y., {et~al.} 2018{\natexlab{b}},
  \mnras, 474, 4232

\bibitem[{{Kirk} \& {Myers}(2011)}]{2011ApJ...727...64K}
{Kirk}, H., \& {Myers}, P.~C. 2011, \apj, 727, 64

\bibitem[{{Klassen} {et~al.}(2016){Klassen}, {Pudritz}, {Kuiper}, {Peters}, \&
  {Banerjee}}]{2016ApJ...823...28K}
{Klassen}, M., {Pudritz}, R.~E., {Kuiper}, R., {Peters}, T., \& {Banerjee}, R.
  2016, \apj, 823, 28

\bibitem[{{Kn{\"o}dlseder}(2000)}]{2000A&A...360..539K}
{Kn{\"o}dlseder}, J. 2000, \aap, 360, 539

\bibitem[{{Knollmann} \& {Knebe}(2009)}]{2009ApJS..182..608K}
{Knollmann}, S.~R., \& {Knebe}, A. 2009, \apjs, 182, 608

\bibitem[{{Koenig} {et~al.}(2008){Koenig}, {Allen}, {Gutermuth}, {Hora},
  {Brunt}, \& {Muzerolle}}]{2008ApJ...688.1142K}
{Koenig}, X.~P., {Allen}, L.~E., {Gutermuth}, R.~A., {et~al.} 2008, \apj, 688,
  1142

\bibitem[{{Kraus} {et~al.}(2009){Kraus}, {Weigelt}, {Balega}, {Docobo},
  {Hofmann}, {Preibisch}, {Schertl}, {Tamazian}, {Driebe}, {Ohnaka}, {Petrov},
  {Sch{\"o}ller}, \& {Smith}}]{2009A&A...497..195K}
{Kraus}, S., {Weigelt}, G., {Balega}, Y.~Y., {et~al.} 2009, \aap, 497, 195

\bibitem[{{Kravtsov}(2003)}]{2003ApJ...590L...1K}
{Kravtsov}, A.~V. 2003, \apjl, 590, L1

\bibitem[{{Kroupa}(2001)}]{2001MNRAS.322..231K}
{Kroupa}, P. 2001, \mnras, 322, 231

\bibitem[{{Krumholz} {et~al.}(2019){Krumholz}, {McKee}, \& {Bland
  -Hawthorn}}]{2019ARA&A..57..227K}
{Krumholz}, M.~R., {McKee}, C.~F., \& {Bland -Hawthorn}, J. 2019, \araa, 57,
  227

\bibitem[{{Krumholz} {et~al.}(2004){Krumholz}, {McKee}, \&
  {Klein}}]{2004ApJ...611..399K}
{Krumholz}, M.~R., {McKee}, C.~F., \& {Klein}, R.~I. 2004, \apj, 611, 399

\bibitem[{{Kumamoto} {et~al.}(2019){Kumamoto}, {Fujii}, \&
  {Tanikawa}}]{2019MNRAS.486.3942K}
{Kumamoto}, J., {Fujii}, M.~S., \& {Tanikawa}, A. 2019, \mnras, 486, 3942

\bibitem[{{Lada} \& {Lada}(2003)}]{2003ARA&A..41...57L}
{Lada}, C.~J., \& {Lada}, E.~A. 2003, \araa, 41, 57

\bibitem[{{Lada} {et~al.}(1991){Lada}, {Depoy}, {Evans}, \&
  {Gatley}}]{1991ApJ...371..171L}
{Lada}, E.~A., {Depoy}, D.~L., {Evans}, Neal~J., I., \& {Gatley}, I. 1991,
  \apj, 371, 171

\bibitem[{{Lah{\'e}n} {et~al.}(2019){Lah{\'e}n}, {Naab}, {Johansson},
  {Elmegreen}, {Hu}, \& {Walch}}]{2019ApJ...879L..18L}
{Lah{\'e}n}, N., {Naab}, T., {Johansson}, P.~H., {et~al.} 2019, \apjl, 879, L18

\bibitem[{{Lah{\'e}n} {et~al.}(2020){Lah{\'e}n}, {Naab}, {Johansson},
  {Elmegreen}, {Hu}, {Walch}, {Steinwand el}, \&
  {Moster}}]{2020ApJ...891....2L}
---. 2020, \apj, 891, 2

\bibitem[{{Lata} {et~al.}(2002){Lata}, {Pandey}, {Sagar}, \&
  {Mohan}}]{2002A&A...388..158L}
{Lata}, S., {Pandey}, A.~K., {Sagar}, R., \& {Mohan}, V. 2002, \aap, 388, 158

\bibitem[{{Lee} {et~al.}(2015){Lee}, {Chang}, \&
  {Murray}}]{2015ApJ...800...49L}
{Lee}, E.~J., {Chang}, P., \& {Murray}, N. 2015, \apj, 800, 49

\bibitem[{{Lombardi} {et~al.}(2010){Lombardi}, {Lada}, \&
  {Alves}}]{2010A&A...512A..67L}
{Lombardi}, M., {Lada}, C.~J., \& {Alves}, J. 2010, \aap, 512, A67

\bibitem[{{Luhman}(2008)}]{2008hsf2.book..169L}
{Luhman}, K.~L. 2008, {Chamaeleon}, ed. B.~{Reipurth}, Vol.~5, 169

\bibitem[{{Ma} {et~al.}(2020){Ma}, {Grudi{\'c}}, {Quataert}, {Hopkins},
  {Faucher-Gigu{\`e}re}, {Boylan-Kolchin}, {Wetzel}, {Kim}, {Murray}, \&
  {Kere{\v{s}}}}]{2020MNRAS.493.4315M}
{Ma}, X., {Grudi{\'c}}, M.~Y., {Quataert}, E., {et~al.} 2020, \mnras, 493, 4315

\bibitem[{{Ma{\'\i}z Apell{\'a}niz} {et~al.}(2007){Ma{\'\i}z Apell{\'a}niz},
  {Walborn}, {Morrell}, {Niemela}, \& {Nelan}}]{2007ApJ...660.1480M}
{Ma{\'\i}z Apell{\'a}niz}, J., {Walborn}, N.~R., {Morrell}, N.~I., {Niemela},
  V.~S., \& {Nelan}, E.~P. 2007, \apj, 660, 1480

\bibitem[{{Marco} \& {Negueruela}(2009)}]{2009A&A...493...79M}
{Marco}, A., \& {Negueruela}, I. 2009, \aap, 493, 79

\bibitem[{{Martins} {et~al.}(2008){Martins}, {Hillier}, {Paumard},
  {Eisenhauer}, {Ott}, \& {Genzel}}]{2008A&A...478..219M}
{Martins}, F., {Hillier}, D.~J., {Paumard}, T., {et~al.} 2008, \aap, 478, 219

\bibitem[{{Massey} {et~al.}(2001){Massey}, {DeGioia-Eastwood}, \&
  {Waterhouse}}]{2001AJ....121.1050M}
{Massey}, P., {DeGioia-Eastwood}, K., \& {Waterhouse}, E. 2001, \aj, 121, 1050

\bibitem[{{Massey} {et~al.}(1989){Massey}, {Garmany}, {Silkey}, \&
  {Degioia-Eastwood}}]{1989AJ.....97..107M}
{Massey}, P., {Garmany}, C.~D., {Silkey}, M., \& {Degioia-Eastwood}, K. 1989,
  \aj, 97, 107

\bibitem[{{Massey} \& {Hunter}(1998)}]{1998ApJ...493..180M}
{Massey}, P., \& {Hunter}, D.~A. 1998, \apj, 493, 180

\bibitem[{{Massey} \& {Johnson}(1993)}]{1993AJ....105..980M}
{Massey}, P., \& {Johnson}, J. 1993, \aj, 105, 980

\bibitem[{{Massey} {et~al.}(1995){Massey}, {Johnson}, \&
  {Degioia-Eastwood}}]{1995ApJ...454..151M}
{Massey}, P., {Johnson}, K.~E., \& {Degioia-Eastwood}, K. 1995, \apj, 454, 151

\bibitem[{{Mayne} {et~al.}(2007){Mayne}, {Naylor}, {Littlefair}, {Saunders}, \&
  {Jeffries}}]{2007MNRAS.375.1220M}
{Mayne}, N.~J., {Naylor}, T., {Littlefair}, S.~P., {Saunders}, E.~S., \&
  {Jeffries}, R.~D. 2007, \mnras, 375, 1220

\bibitem[{{McKee} \& {Ostriker}(2007)}]{2007ARA&A..45..565M}
{McKee}, C.~F., \& {Ostriker}, E.~C. 2007, \araa, 45, 565

\bibitem[{{McMillan} {et~al.}(2007){McMillan}, {Vesperini}, \& {Portegies
  Zwart}}]{2007ApJ...655L..45M}
{McMillan}, S. L.~W., {Vesperini}, E., \& {Portegies Zwart}, S.~F. 2007, \apjl,
  655, L45

\bibitem[{{Menten} {et~al.}(2007){Menten}, {Reid}, {Forbrich}, \&
  {Brunthaler}}]{2007A&A...474..515M}
{Menten}, K.~M., {Reid}, M.~J., {Forbrich}, J., \& {Brunthaler}, A. 2007, \aap,
  474, 515

\bibitem[{{Moeckel} \& {Bonnell}(2009)}]{2009MNRAS.400..657M}
{Moeckel}, N., \& {Bonnell}, I.~A. 2009, \mnras, 400, 657

\bibitem[{{Morris} \& {Monaghan}(1997)}]{1997JCoPh.136...41M}
{Morris}, J.~P., \& {Monaghan}, J.~J. 1997, Journal of Computational Physics,
  136, 41

\bibitem[{{Navarro} \& {White}(1993)}]{1993MNRAS.265..271N}
{Navarro}, J.~F., \& {White}, S.~D.~M. 1993, \mnras, 265, 271

\bibitem[{{Naylor} \& {Fabian}(1999)}]{1999MNRAS.302..714N}
{Naylor}, T., \& {Fabian}, A.~C. 1999, \mnras, 302, 714

\bibitem[{{Naz{\'e}} {et~al.}(2008){Naz{\'e}}, {Rauw}, \&
  {Manfroid}}]{2008A&A...483..171N}
{Naz{\'e}}, Y., {Rauw}, G., \& {Manfroid}, J. 2008, \aap, 483, 171

\bibitem[{{Negueruela} \& {Marco}(2008)}]{2008A&A...492..441N}
{Negueruela}, I., \& {Marco}, A. 2008, \aap, 492, 441

\bibitem[{{Nelan} {et~al.}(2004){Nelan}, {Walborn}, {Wallace}, {Moffat},
  {Makidon}, {Gies}, \& {Panagia}}]{2004AJ....128..323N}
{Nelan}, E.~P., {Walborn}, N.~R., {Wallace}, D.~J., {et~al.} 2004, \aj, 128,
  323

\bibitem[{{Neuh{\"a}user} \& {Forbrich}(2008)}]{2008hsf2.book..735N}
{Neuh{\"a}user}, R., \& {Forbrich}, J. 2008, {The Corona Australis Star Forming
  Region}, ed. B.~{Reipurth}, Vol.~5, 735

\bibitem[{{Niemela} \& {Gamen}(2004)}]{2004NewAR..48..727N}
{Niemela}, V., \& {Gamen}, R. 2004, New Astronomy Reviews, 48, 727

\bibitem[{{Nomoto} {et~al.}(2013){Nomoto}, {Kobayashi}, \&
  {Tominaga}}]{2013ARA&A..51..457N}
{Nomoto}, K., {Kobayashi}, C., \& {Tominaga}, N. 2013, \araa, 51, 457

\bibitem[{{Oey} \& {Clarke}(2005)}]{2005ApJ...620L..43O}
{Oey}, M.~S., \& {Clarke}, C.~J. 2005, \apjl, 620, L43

\bibitem[{{Okamoto} {et~al.}(2003){Okamoto}, {Jenkins}, {Eke}, {Quilis}, \&
  {Frenk}}]{2003MNRAS.345..429O}
{Okamoto}, T., {Jenkins}, A., {Eke}, V.~R., {Quilis}, V., \& {Frenk}, C.~S.
  2003, \mnras, 345, 429

\bibitem[{{O'Leary} {et~al.}(2006){O'Leary}, {Rasio}, {Fregeau}, {Ivanova}, \&
  {O'Shaughnessy}}]{2006ApJ...637..937O}
{O'Leary}, R.~M., {Rasio}, F.~A., {Fregeau}, J.~M., {Ivanova}, N., \&
  {O'Shaughnessy}, R. 2006, \apj, 637, 937

\bibitem[{{Ortolani} {et~al.}(2008){Ortolani}, {Bonatto}, {Bica}, {Momany}, \&
  {Barbuy}}]{2008NewA...13..508O}
{Ortolani}, S., {Bonatto}, C., {Bica}, E., {Momany}, Y., \& {Barbuy}, B. 2008,
  New Astronomy, 13, 508

\bibitem[{{Ostriker} {et~al.}(2001){Ostriker}, {Stone}, \&
  {Gammie}}]{2001ApJ...546..980O}
{Ostriker}, E.~C., {Stone}, J.~M., \& {Gammie}, C.~F. 2001, \apj, 546, 980

\bibitem[{{Pancino} {et~al.}(2010){Pancino}, {Carrera}, {Rossetti}, \&
  {Gallart}}]{2010A&A...511A..56P}
{Pancino}, E., {Carrera}, R., {Rossetti}, E., \& {Gallart}, C. 2010, \aap, 511,
  A56

\bibitem[{{Pandey} {et~al.}(1989){Pandey}, {Bhatt}, {Mahra}, \&
  {Sagar}}]{1989MNRAS.236..263P}
{Pandey}, A.~K., {Bhatt}, B.~C., {Mahra}, H.~S., \& {Sagar}, R. 1989, \mnras,
  236, 263

\bibitem[{{Park} \& {Sung}(2002)}]{2002AJ....123..892P}
{Park}, B.-G., \& {Sung}, H. 2002, \aj, 123, 892

\bibitem[{{Paunzen} {et~al.}(2007){Paunzen}, {Netopil}, \&
  {Zwintz}}]{2007A&A...462..157P}
{Paunzen}, E., {Netopil}, M., \& {Zwintz}, K. 2007, \aap, 462, 157

\bibitem[{{Pelupessy} {et~al.}(2013){Pelupessy}, {van Elteren}, {de Vries},
  {McMillan}, {Drost}, \& {Portegies Zwart}}]{2013A&A...557A..84P}
{Pelupessy}, F.~I., {van Elteren}, A., {de Vries}, N., {et~al.} 2013, \aap,
  557, A84

\bibitem[{{Penny} {et~al.}(1993){Penny}, {Gies}, {Hartkopf}, {Mason}, \&
  {Turner}}]{1993PASP..105..588P}
{Penny}, L.~R., {Gies}, D.~R., {Hartkopf}, W.~I., {Mason}, B.~D., \& {Turner},
  N.~H. 1993, \pasp, 105, 588

\bibitem[{{Portegies Zwart} \& {McMillan}(2018)}]{2018araa.book.....P}
{Portegies Zwart}, S., \& {McMillan}, S. 2018, {Astrophysical Recipes; The art
  of AMUSE (Bristol: IOP Publishing)}, doi:10.1088/978-0-7503-1320-9

\bibitem[{{Portegies Zwart} {et~al.}(2013){Portegies Zwart}, {McMillan}, {van
  Elteren}, {Pelupessy}, \& {de Vries}}]{2013CoPhC.184..456P}
{Portegies Zwart}, S., {McMillan}, S.~L.~W., {van Elteren}, E., {Pelupessy},
  I., \& {de Vries}, N. 2013, Computer Physics Communications, 184, 456

\bibitem[{{Portegies Zwart} {et~al.}(2009){Portegies Zwart}, {McMillan},
  {Harfst}, {Groen}, {Fujii}, {Nuall{\'a}in}, {Glebbeek}, {Heggie}, {Lombardi},
  {Hut}, {Angelou}, {Banerjee}, {Belkus}, {Fragos}, {Fregeau}, {Gaburov},
  {Izzard}, {Juri{\'c}}, {Justham}, {Sottoriva}, {Teuben}, {van Bever},
  {Yaron}, \& {Zemp}}]{2009NewA...14..369P}
{Portegies Zwart}, S., {McMillan}, S., {Harfst}, S., {et~al.} 2009, New
  Astronomy, 14, 369

\bibitem[{{Portegies Zwart} \& {McMillan}(2000)}]{2000ApJ...528L..17P}
{Portegies Zwart}, S.~F., \& {McMillan}, S.~L.~W. 2000, \apjl, 528, L17

\bibitem[{{Portegies Zwart} {et~al.}(2010){Portegies Zwart}, {McMillan}, \&
  {Gieles}}]{2010ARA&A..48..431P}
{Portegies Zwart}, S.~F., {McMillan}, S. L.~W., \& {Gieles}, M. 2010, \araa,
  48, 431

\bibitem[{{Portinari} {et~al.}(1998){Portinari}, {Chiosi}, \&
  {Bressan}}]{1998A&A...334..505P}
{Portinari}, L., {Chiosi}, C., \& {Bressan}, A. 1998, \aap, 334, 505

\bibitem[{{Pozzo} {et~al.}(2003){Pozzo}, {Naylor}, {Jeffries}, \&
  {Drew}}]{2003MNRAS.341..805P}
{Pozzo}, M., {Naylor}, T., {Jeffries}, R.~D., \& {Drew}, J.~E. 2003, \mnras,
  341, 805

\bibitem[{{Prantzos} {et~al.}(2020){Prantzos}, {Abia}, {Cristallo}, {Limongi},
  \& {Chieffi}}]{2020MNRAS.491.1832P}
{Prantzos}, N., {Abia}, C., {Cristallo}, S., {Limongi}, M., \& {Chieffi}, A.
  2020, \mnras, 491, 1832

\bibitem[{{Preibisch} {et~al.}(2002){Preibisch}, {Balega}, {Schertl}, \&
  {Weigelt}}]{2002A&A...392..945P}
{Preibisch}, T., {Balega}, Y.~Y., {Schertl}, D., \& {Weigelt}, G. 2002, \aap,
  392, 945

\bibitem[{{Preibisch} \& {Zinnecker}(2001)}]{2001AJ....122..866P}
{Preibisch}, T., \& {Zinnecker}, H. 2001, \aj, 122, 866

\bibitem[{{Prisinzano} {et~al.}(2005){Prisinzano}, {Damiani}, {Micela}, \&
  {Sciortino}}]{2005A&A...430..941P}
{Prisinzano}, L., {Damiani}, F., {Micela}, G., \& {Sciortino}, S. 2005, \aap,
  430, 941

\bibitem[{{Rahmati} {et~al.}(2013){Rahmati}, {Pawlik}, {Rai{\v c}evi{\` c}}, \&
  {Schaye}}]{2013MNRAS.430.2427R}
{Rahmati}, A., {Pawlik}, A.~H., {Rai{\v c}evi{\` c}}, M., \& {Schaye}, J. 2013,
  \mnras, 430, 2427

\bibitem[{{Raskutti} {et~al.}(2016){Raskutti}, {Ostriker}, \&
  {Skinner}}]{2016ApJ...829..130R}
{Raskutti}, S., {Ostriker}, E.~C., \& {Skinner}, M.~A. 2016, \apj, 829, 130

\bibitem[{{Rauw} \& {De Becker}(2008)}]{2008hsf2.book..497R}
{Rauw}, G., \& {De Becker}, M. 2008, {The Multiwavelength Picture of Star
  Formation in the Very Young Open Cluster NGC 6383}, ed. B.~{Reipurth},
  Vol.~5, 497

\bibitem[{{Rauw} {et~al.}(2003){Rauw}, {De Becker}, {Gosset}, {Pittard}, \&
  {Stevens}}]{2003A&A...407..925R}
{Rauw}, G., {De Becker}, M., {Gosset}, E., {Pittard}, J.~M., \& {Stevens},
  I.~R. 2003, \aap, 407, 925

\bibitem[{{Read} {et~al.}(2017){Read}, {Iorio}, {Agertz}, \&
  {Fraternali}}]{2017MNRAS.467.2019R}
{Read}, J.~I., {Iorio}, G., {Agertz}, O., \& {Fraternali}, F. 2017, \mnras,
  467, 2019

\bibitem[{{Reddy} {et~al.}(2013){Reddy}, {Giridhar}, \&
  {Lambert}}]{2013MNRAS.431.3338R}
{Reddy}, A.~B.~S., {Giridhar}, S., \& {Lambert}, D.~L. 2013, \mnras, 431, 3338

\bibitem[{{Revaz} {et~al.}(2016){Revaz}, {Arnaudon}, {Nichols}, {Bonvin}, \&
  {Jablonka}}]{2016A&A...588A..21R}
{Revaz}, Y., {Arnaudon}, A., {Nichols}, M., {Bonvin}, V., \& {Jablonka}, P.
  2016, \aap, 588, A21

\bibitem[{{Revaz} \& {Jablonka}(2012)}]{2012A&A...538A..82R}
{Revaz}, Y., \& {Jablonka}, P. 2012, \aap, 538, A82

\bibitem[{{Rey} {et~al.}(2019){Rey}, {Pontzen}, {Agertz}, {Orkney}, {Read},
  {Saintonge}, \& {Pedersen}}]{2019ApJ...886L...3R}
{Rey}, M.~P., {Pontzen}, A., {Agertz}, O., {et~al.} 2019, \apjl, 886, L3

\bibitem[{{Robertson} \& {Kravtsov}(2008)}]{2008ApJ...680.1083R}
{Robertson}, B.~E., \& {Kravtsov}, A.~V. 2008, \apj, 680, 1083

\bibitem[{{Rodney} \& {Reipurth}(2008)}]{2008hsf2.book..683R}
{Rodney}, S.~A., \& {Reipurth}, B. 2008, {The W40 Cloud Complex}, ed.
  B.~{Reipurth}, Vol.~5, 683

\bibitem[{{Rodriguez} {et~al.}(2015){Rodriguez}, {Morscher}, {Pattabiraman},
  {Chatterjee}, {Haster}, \& {Rasio}}]{2015PhRvL.115e1101R}
{Rodriguez}, C.~L., {Morscher}, M., {Pattabiraman}, B., {et~al.} 2015, \prl,
  115, 051101

\bibitem[{{Roman-Lopes}(2007)}]{2007A&A...471..813R}
{Roman-Lopes}, A. 2007, \aap, 471, 813

\bibitem[{{Roman-Lopes} \& {Abraham}(2004)}]{2004AJ....128.2364R}
{Roman-Lopes}, A., \& {Abraham}, Z. 2004, \aj, 128, 2364

\bibitem[{{Rosswog}(2009)}]{2009NewAR..53...78R}
{Rosswog}, S. 2009, New Astronomy, 53, 78

\bibitem[{{Saitoh}(2017)}]{2017AJ....153...85S}
{Saitoh}, T.~R. 2017, \aj, 153, 85

\bibitem[{{Saitoh} {et~al.}(2008){Saitoh}, {Daisaka}, {Kokubo}, {Makino},
  {Okamoto}, {Tomisaka}, {Wada}, \& {Yoshida}}]{2008PASJ...60..667S}
{Saitoh}, T.~R., {Daisaka}, H., {Kokubo}, E., {et~al.} 2008, \pasj, 60, 667

\bibitem[{{Saitoh} {et~al.}(2009){Saitoh}, {Daisaka}, {Kokubo}, {Makino},
  {Okamoto}, {Tomisaka}, {Wada}, \& {Yoshida}}]{2009PASJ...61..481S}
---. 2009, \pasj, 61, 481

\bibitem[{{Saitoh} \& {Makino}(2009)}]{2009ApJ...697L..99S}
{Saitoh}, T.~R., \& {Makino}, J. 2009, \apjl, 697, L99

\bibitem[{{Saitoh} \& {Makino}(2010)}]{2010PASJ...62..301S}
---. 2010, \pasj, 62, 301

\bibitem[{{Saitoh} \& {Makino}(2013)}]{2013ApJ...768...44S}
---. 2013, \apj, 768, 44

\bibitem[{{Salpeter}(1955)}]{1955ApJ...121..161S}
{Salpeter}, E.~E. 1955, \apj, 121, 161

\bibitem[{{Sanchawala} {et~al.}(2007){Sanchawala}, {Chen}, {Ojha}, {Ghosh},
  {Nakajima}, {Tamura}, {Baba}, {Sato}, \& {Tsujimoto}}]{2007ApJ...667..963S}
{Sanchawala}, K., {Chen}, W.-P., {Ojha}, D., {et~al.} 2007, \apj, 667, 963

\bibitem[{{Schaerer}(2002)}]{2002A&A...382...28S}
{Schaerer}, D. 2002, \aap, 382, 28

\bibitem[{{Schmidt}(1959)}]{1959ApJ...129..243S}
{Schmidt}, M. 1959, \apj, 129, 243

\bibitem[{{Schneider} {et~al.}(2012){Schneider}, {Csengeri}, {Hennemann},
  {Motte}, {Didelon}, {Federrath}, {Bontemps}, {Di Francesco}, {Arzoumanian},
  {Minier}, {Andr{\'e}}, {Hill}, {Zavagno}, {Nguyen-Luong}, {Attard},
  {Bernard}, {Elia}, {Fallscheer}, {Griffin}, {Kirk}, {Klessen}, {K{\"o}nyves},
  {Martin}, {Men'shchikov}, {Palmeirim}, {Peretto}, {Pestalozzi}, {Russeil},
  {Sadavoy}, {Sousbie}, {Testi}, {Tremblin}, {Ward-Thompson}, \&
  {White}}]{2012A&A...540L..11S}
{Schneider}, N., {Csengeri}, T., {Hennemann}, M., {et~al.} 2012, \aap, 540, L11

\bibitem[{{Schneider} {et~al.}(2013){Schneider}, {Andr{\'e}}, {K{\"o}nyves},
  {Bontemps}, {Motte}, {Federrath}, {Ward-Thompson}, {Arzoumanian},
  {Benedettini}, {Bressert}, {Didelon}, {Di Francesco}, {Griffin}, {Hennemann},
  {Hill}, {Palmeirim}, {Pezzuto}, {Peretto}, {Roy}, {Rygl}, {Spinoglio}, \&
  {White}}]{2013ApJ...766L..17S}
{Schneider}, N., {Andr{\'e}}, P., {K{\"o}nyves}, V., {et~al.} 2013, \apjl, 766,
  L17

\bibitem[{{Schneider} {et~al.}(2015{\natexlab{a}}){Schneider}, {Ossenkopf},
  {Csengeri}, {Klessen}, {Federrath}, {Tremblin}, {Girichidis}, {Bontemps}, \&
  {Andr{\'e}}}]{2015A&A...575A..79S}
{Schneider}, N., {Ossenkopf}, V., {Csengeri}, T., {et~al.} 2015{\natexlab{a}},
  \aap, 575, A79

\bibitem[{{Schneider} {et~al.}(2015{\natexlab{b}}){Schneider}, {Csengeri},
  {Klessen}, {Tremblin}, {Ossenkopf}, {Peretto}, {Simon}, {Bontemps}, \&
  {Federrath}}]{2015A&A...578A..29S}
{Schneider}, N., {Csengeri}, T., {Klessen}, R.~S., {et~al.} 2015{\natexlab{b}},
  \aap, 578, A29

\bibitem[{{Schneider} {et~al.}(2015{\natexlab{c}}){Schneider}, {Bontemps},
  {Girichidis}, {Rayner}, {Motte}, {Andr{\'e}}, {Russeil}, {Abergel},
  {Anderson}, {Arzoumanian}, {Benedettini}, {Csengeri}, {Didelon}, {di},
  {Griffin}, {Hill}, {Klessen}, {Ossenkopf}, {Pezzuto}, {Rivera-Ingraham},
  {Spinoglio}, {Tremblin}, \& {Zavagno}}]{2015MNRAS.453L..41S}
{Schneider}, N., {Bontemps}, S., {Girichidis}, P., {et~al.} 2015{\natexlab{c}},
  \mnras, 453, L41

\bibitem[{{Schneider} {et~al.}(2016){Schneider}, {Bontemps}, {Motte},
  {Ossenkopf}, {Klessen}, {Simon}, {Fechtenbaum}, {Herpin}, {Tremblin},
  {Csengeri}, {Myers}, {Hill}, {Cunningham}, \&
  {Federrath}}]{2016A&A...587A..74S}
{Schneider}, N., {Bontemps}, S., {Motte}, F., {et~al.} 2016, \aap, 587, A74

\bibitem[{{Schnurr} {et~al.}(2008){Schnurr}, {Casoli}, {Chen{\'e}}, {Moffat},
  \& {St-Louis}}]{2008MNRAS.389L..38S}
{Schnurr}, O., {Casoli}, J., {Chen{\'e}}, A.~N., {Moffat}, A.~F.~J., \&
  {St-Louis}, N. 2008, \mnras, 389, L38

\bibitem[{{Schnurr} {et~al.}(2009){Schnurr}, {Chen{\'e}}, {Casoli}, {Moffat},
  \& {St-Louis}}]{2009MNRAS.397.2049S}
{Schnurr}, O., {Chen{\'e}}, A.~N., {Casoli}, J., {Moffat}, A.~F.~J., \&
  {St-Louis}, N. 2009, \mnras, 397, 2049

\bibitem[{{Sellgren}(1983)}]{1983AJ.....88..985S}
{Sellgren}, K. 1983, \aj, 88, 985

\bibitem[{{Selman} {et~al.}(1999){Selman}, {Melnick}, {Bosch}, \&
  {Terlevich}}]{1999A&A...347..532S}
{Selman}, F., {Melnick}, J., {Bosch}, G., \& {Terlevich}, R. 1999, \aap, 347,
  532

\bibitem[{{Shen} {et~al.}(2010){Shen}, {Wadsley}, \&
  {Stinson}}]{2010MNRAS.407.1581S}
{Shen}, S., {Wadsley}, J., \& {Stinson}, G. 2010, \mnras, 407, 1581

\bibitem[{{Sherry} {et~al.}(2004){Sherry}, {Walter}, \&
  {Wolk}}]{2004AJ....128.2316S}
{Sherry}, W.~H., {Walter}, F.~M., \& {Wolk}, S.~J. 2004, \aj, 128, 2316

\bibitem[{{Shima} {et~al.}(2018){Shima}, {Tasker}, {Federrath}, \&
  {Habe}}]{2018PASJ...70S..54S}
{Shima}, K., {Tasker}, E.~J., {Federrath}, C., \& {Habe}, A. 2018, \pasj, 70,
  S54

\bibitem[{{Silkey} \& {Massey}(1986)}]{1986BAAS...18..910S}
{Silkey}, M., \& {Massey}, P. 1986, in Bulletin of the American Astronomical
  Society, Vol.~18, 910

\bibitem[{{Simon}(2019)}]{2019ARA&A..57..375S}
{Simon}, J.~D. 2019, \araa, 57, 375

\bibitem[{{Slyz} {et~al.}(2005){Slyz}, {Devriendt}, {Bryan}, \&
  {Silk}}]{2005MNRAS.356..737S}
{Slyz}, A.~D., {Devriendt}, J. E.~G., {Bryan}, G., \& {Silk}, J. 2005, \mnras,
  356, 737

\bibitem[{{Smith} {et~al.}(1985){Smith}, {Bentley}, {Castelaz}, {Gehrz},
  {Grasdalen}, \& {Hackwell}}]{1985ApJ...291..571S}
{Smith}, J., {Bentley}, A., {Castelaz}, M., {et~al.} 1985, \apj, 291, 571

\bibitem[{{Smith}(2021)}]{2020arXiv201010533S}
{Smith}, M.~C. 2021, \mnras, 502, 5417

\bibitem[{{Sormani} {et~al.}(2017){Sormani}, {Tre{\ss}}, {Klessen}, \&
  {Glover}}]{2017MNRAS.466..407S}
{Sormani}, M.~C., {Tre{\ss}}, R.~G., {Klessen}, R.~S., \& {Glover}, S. C.~O.
  2017, \mnras, 466, 407

\bibitem[{{Springel}(2005)}]{2005MNRAS.364.1105S}
{Springel}, V. 2005, \mnras, 364, 1105

\bibitem[{{Springel} \& {Hernquist}(2003)}]{2003MNRAS.339..289S}
{Springel}, V., \& {Hernquist}, L. 2003, \mnras, 339, 289

\bibitem[{{Steinmetz} \& {Mueller}(1994)}]{1994A&A...281L..97S}
{Steinmetz}, M., \& {Mueller}, E. 1994, \aap, 281, L97

\bibitem[{{Stinson} {et~al.}(2006){Stinson}, {Seth}, {Katz}, {Wadsley},
  {Governato}, \& {Quinn}}]{2006MNRAS.373.1074S}
{Stinson}, G., {Seth}, A., {Katz}, N., {et~al.} 2006, \mnras, 373, 1074

\bibitem[{{Stolte} {et~al.}(2006){Stolte}, {Brandner}, {Brandl}, \&
  {Zinnecker}}]{2006AJ....132..253S}
{Stolte}, A., {Brandner}, W., {Brandl}, B., \& {Zinnecker}, H. 2006, \aj, 132,
  253

\bibitem[{{Sung} {et~al.}(2004){Sung}, {Bessell}, \&
  {Chun}}]{2004AJ....128.1684S}
{Sung}, H., {Bessell}, M.~S., \& {Chun}, M.-Y. 2004, \aj, 128, 1684

\bibitem[{{Susa} {et~al.}(2014){Susa}, {Hasegawa}, \&
  {Tominaga}}]{2014ApJ...792...32S}
{Susa}, H., {Hasegawa}, K., \& {Tominaga}, N. 2014, \apj, 792, 32

\bibitem[{{Tasker} \& {Bryan}(2008)}]{2008ApJ...673..810T}
{Tasker}, E.~J., \& {Bryan}, G.~L. 2008, \apj, 673, 810

\bibitem[{{Teich} {et~al.}(2016){Teich}, {McNichols}, {Nims}, {Cannon},
  {Adams}, {Giovanelli}, {Haynes}, {McQuinn}, {Salzer}, {Skillman},
  {Bernstein-Cooper}, {Dolphin}, {Elson}, {Haurberg}, {J{\'o}zsa}, {Ott},
  {Saintonge}, {Warren}, {Cave}, {Hagen}, {Huang}, {Janowiecki}, {Marshall},
  {Thomann}, \& {Van Sistine}}]{2016ApJ...832...85T}
{Teich}, Y.~G., {McNichols}, A.~T., {Nims}, E., {et~al.} 2016, \apj, 832, 85

\bibitem[{{Testi} {et~al.}(1998){Testi}, {Palla}, \&
  {Natta}}]{1998A&AS..133...81T}
{Testi}, L., {Palla}, F., \& {Natta}, A. 1998, \aaps, 133, 81

\bibitem[{{Testi} {et~al.}(1999){Testi}, {Palla}, \&
  {Natta}}]{1999A&A...342..515T}
---. 1999, \aap, 342, 515

\bibitem[{{Testi} {et~al.}(1997){Testi}, {Palla}, {Prusti}, {Natta}, \&
  {Maltagliati}}]{1997A&A...320..159T}
{Testi}, L., {Palla}, F., {Prusti}, T., {Natta}, A., \& {Maltagliati}, S. 1997,
  \aap, 320, 159

\bibitem[{{Truelove} {et~al.}(1997){Truelove}, {Klein}, {McKee}, {Holliman},
  {Howell}, \& {Greenough}}]{1997ApJ...489L.179T}
{Truelove}, J.~K., {Klein}, R.~I., {McKee}, C.~F., {et~al.} 1997, \apjl, 489,
  L179

\bibitem[{{Turner}(1985)}]{1985ApJ...292..148T}
{Turner}, D.~G. 1985, \apj, 292, 148

\bibitem[{{Vallenari} {et~al.}(1999){Vallenari}, {Richichi}, {Carraro}, \&
  {Girardi}}]{1999A&A...349..825V}
{Vallenari}, A., {Richichi}, A., {Carraro}, G., \& {Girardi}, L. 1999, \aap,
  349, 825

\bibitem[{{Vargas {\'A}lvarez} {et~al.}(2013){Vargas {\'A}lvarez},
  {Kobulnicky}, {Bradley}, {Kannappan}, {Norris}, {Cool}, \&
  {Miller}}]{2013AJ....145..125V}
{Vargas {\'A}lvarez}, C.~A., {Kobulnicky}, H.~A., {Bradley}, D.~R., {et~al.}
  2013, \aj, 145, 125

\bibitem[{{Vazquez-Semadeni}(1994)}]{1994ApJ...423..681V}
{Vazquez-Semadeni}, E. 1994, \apj, 423, 681

\bibitem[{{V{\'a}zquez-Semadeni} \& {Garc{\'\i}a}(2001)}]{2001ApJ...557..727V}
{V{\'a}zquez-Semadeni}, E., \& {Garc{\'\i}a}, N. 2001, \apj, 557, 727

\bibitem[{{Vogelsberger} {et~al.}(2020){Vogelsberger}, {Marinacci}, {Torrey},
  \& {Puchwein}}]{2020NatRP...2...42V}
{Vogelsberger}, M., {Marinacci}, F., {Torrey}, P., \& {Puchwein}, E. 2020,
  Nature Reviews Physics, 2, 42

\bibitem[{{Wada}(2001)}]{2001ApJ...559L..41W}
{Wada}, K. 2001, \apjl, 559, L41

\bibitem[{{Wada} \& {Norman}(2007)}]{2007ApJ...660..276W}
{Wada}, K., \& {Norman}, C.~A. 2007, \apj, 660, 276

\bibitem[{{Walker}(1959)}]{1959ApJ...130...57W}
{Walker}, M.~F. 1959, \apj, 130, 57

\bibitem[{{Wall} {et~al.}(2019){Wall}, {McMillan}, {Mac Low}, {Klessen}, \&
  {Portegies Zwart}}]{2019ApJ...887...62W}
{Wall}, J.~E., {McMillan}, S. L.~W., {Mac Low}, M.-M., {Klessen}, R.~S., \&
  {Portegies Zwart}, S. 2019, \apj, 887, 62

\bibitem[{{Wang} \& {Looney}(2007)}]{2007ApJ...659.1360W}
{Wang}, S., \& {Looney}, L.~W. 2007, \apj, 659, 1360

\bibitem[{{Weidner} \& {Kroupa}(2006)}]{2006MNRAS.365.1333W}
{Weidner}, C., \& {Kroupa}, P. 2006, \mnras, 365, 1333

\bibitem[{{Weidner} {et~al.}(2013){Weidner}, {Kroupa}, \&
  {Pflamm-Altenburg}}]{2013MNRAS.434...84W}
{Weidner}, C., {Kroupa}, P., \& {Pflamm-Altenburg}, J. 2013, \mnras, 434, 84

\bibitem[{{Wheeler} {et~al.}(2019){Wheeler}, {Hopkins}, {Pace},
  {Garrison-Kimmel}, {Boylan-Kolchin}, {Wetzel}, {Bullock}, {Kere{\v{s}}},
  {Faucher-Gigu{\`e}re}, \& {Quataert}}]{2019MNRAS.490.4447W}
{Wheeler}, C., {Hopkins}, P.~F., {Pace}, A.~B., {et~al.} 2019, \mnras, 490,
  4447

\bibitem[{{Wilking} {et~al.}(2008){Wilking}, {Gagn{\'e}}, \&
  {Allen}}]{2008hsf2.book..351W}
{Wilking}, B.~A., {Gagn{\'e}}, M., \& {Allen}, L.~E. 2008, {Star Formation in
  the {\ensuremath{\rho}} Ophiuchi Molecular Cloud}, ed. B.~{Reipurth}, Vol.~5,
  351

\bibitem[{{Wilking} {et~al.}(1989){Wilking}, {Lada}, \&
  {Young}}]{1989ApJ...340..823W}
{Wilking}, B.~A., {Lada}, C.~J., \& {Young}, E.~T. 1989, \apj, 340, 823

\bibitem[{{Wolff} {et~al.}(2007){Wolff}, {Strom}, {Dror}, \&
  {Venn}}]{2007AJ....133.1092W}
{Wolff}, S.~C., {Strom}, S.~E., {Dror}, D., \& {Venn}, K. 2007, \aj, 133, 1092

\bibitem[{{Wolk} {et~al.}(2008){Wolk}, {Bourke}, \&
  {Vigil}}]{2008hsf2.book..124W}
{Wolk}, S.~J., {Bourke}, T.~L., \& {Vigil}, M. 2008, {The Embedded Massive Star
  Forming Region RCW 38}, ed. B.~{Reipurth}, Vol.~5, 124

\bibitem[{{Wolk} {et~al.}(2006){Wolk}, {Spitzbart}, {Bourke}, \&
  {Alves}}]{2006AJ....132.1100W}
{Wolk}, S.~J., {Spitzbart}, B.~D., {Bourke}, T.~L., \& {Alves}, J. 2006, \aj,
  132, 1100

\bibitem[{{Wolk} {et~al.}(2010){Wolk}, {Winston}, {Bourke}, {Gutermuth},
  {Megeath}, {Spitzbart}, \& {Osten}}]{2010ApJ...715..671W}
{Wolk}, S.~J., {Winston}, E., {Bourke}, T.~L., {et~al.} 2010, \apj, 715, 671

\bibitem[{{Yun} {et~al.}(2008){Yun}, {Djupvik}, {Delgado}, \&
  {Alfaro}}]{2008A&A...483..209Y}
{Yun}, J.~L., {Djupvik}, A.~A., {Delgado}, A.~J., \& {Alfaro}, E.~J. 2008,
  \aap, 483, 209

\bibitem[{{Ziosi} {et~al.}(2014){Ziosi}, {Mapelli}, {Branchesi}, \&
  {Tormen}}]{2014MNRAS.441.3703Z}
{Ziosi}, B.~M., {Mapelli}, M., {Branchesi}, M., \& {Tormen}, G. 2014, \mnras,
  441, 3703

\end{thebibliography}

\end{document}